\documentclass[trackchanges,twocolumn]{aastex701} %
\usepackage{amsmath}
\usepackage{amssymb}
\usepackage{xspace,xcolor}
\usepackage{natbib}
\usepackage{outlines}
\usepackage{savesym}
\usepackage{booktabs}
\usepackage{longtable}
\usepackage{tikz}
\usepackage{graphicx}
\usepackage{hyperref}
\usepackage{siunitx}
\usepackage{comment}
\usepackage{float}
\usepackage{braket}

\newcommand{\beq}{\begin{equation}}
\newcommand{\eeq}{\end{equation}}
\definecolor{blazeorange}{rgb}{1.0, 0.4, 0.0}
\definecolor{seagreen}{rgb}{0.18, 0.55, 0.34}
\definecolor{rufous}{rgb}{0.66, 0.11, 0.03}
\definecolor{royalfuchsia}{rgb}{0.79, 0.17, 0.57}
\definecolor{scarlet}{rgb}{1.0, 0.13, 0.0}
\definecolor{royalpurple}{rgb}{0.47, 0.32, 0.66}
\definecolor{darkblue}{rgb}{0, 0, 0.66}

\newcommand{\JHU}{William H. Miller III Department of Physics \& Astronomy, Johns Hopkins University, 3400 N Charles St, Baltimore, MD 21218, USA}
\newcommand{\ARCO}{Astrophysics Research Center of the Open University (ARCO), Department of Natural Sciences, Ra’anana 4353701, Israel}
\newcommand{\GWDC}{Department of Physics, The George Washington University, Washington, DC 20052, USA}
\newcommand{\Bunkyo}{Research Center for the Early Universe, Graduate School of Science, The University of Tokyo, Bunkyo, Tokyo 113-0033, Japan}

\begin{document}

\title{Distinct First-to-Second Peak Yield Ratios and Timescales Reveal a Sub-dominant Prompt Channel}


\correspondingauthor{yzenati1@jhu.edu}

\author[0000-0002-0632-8897]{Yossef Zenati}
\affiliation{\ARCO}
\affiliation{\JHU}
\email{yzenati1@jhu.edu}

\author[0000-0001-7833-1043]{Paz Beniamini}
\affiliation{\ARCO}
\affiliation{\GWDC}
\email{pazb@openu.ac.il}

\author{Kenta Hotokezaka}
\affiliation{\Bunkyo}
\email{kentah@g.ecc.u-tokyo.ac.jp}

\begin{abstract}

Stellar abundances reveal non-monotonic [Y/Eu] and [Sr/Eu] evolution, a systematic decline with increasing [Eu/H] at low metallicity, a minimum at $[\rm{Eu/H}] \sim -0.3$ and then a rise at high metallicity. This behavior requires at least three distinct neutron-capture sources operating on different timescales. We develop a one-zone chemical-evolution model constraining their typical delay-times, rates, and yield ratios. Reproducing the observed $\rm{[Y/Eu]}$ and $\rm{[Sr/Eu]}$ sequences requires, a delayed $r$-process channel (most likely binary neutron-star mergers) dominating Eu production ($\gtrsim 95\%$ of total Eu). A prompt channel preferentially producing first-peak elements with minimal Eu, explaining the increasing [Y/Eu] at decreasing [Eu/H] below $[\rm{Eu/H}] \lesssim -2.5$; and delayed AGB $s$-process enrichment with delays greater than $t_{min} = 0.3-0.6$\,Gyr reproducing the late-time upturn in Y (Sr). Our model quantitatively reproduces all constraints, including the large $\Delta[\rm{Y/Eu}] \approx 0.6$ dex variation between the late-time rise [Eu/H] and the minimum value, the location of the minimum at [Eu/H] $\sim -0.3$ and late-time rise. The first-to-second peak yield ratios correspond to $[\rm{Y/Eu}] \approx -0.3$ (prompt) and $\approx -0.8$ (BNS mergers). The observed $\Delta[\rm{Y/Eu}]$ amplitude establishes a model-independent lower limit on the first to second peak yield ratio $\gtrsim 3.4$ between the prompt and delayed channels, ruling out models with similar prompt and delayed yield ratios. These results demonstrate that explaining the observed heavy-element abundance patterns requires multiple channels with distinct nucleosynthetic signatures and operational timescales, providing constraints on the relative rates, delay times, and yield patterns of candidate production sites.
\end{abstract}

\keywords{
Supernovae: individual: SNe -- Binary Neutron stars: individual: BNS -- r-process: chemical evolution}

\section{Introduction} 
\label{sec:intro}

The astrophysical origin of the heaviest elements produced by rapid neutron capture ($r$-process) remains one of the central questions in nuclear astrophysics \citep{Lattimer_Schramm74,ThielemannF+17,CowanJ+21}. This process accounts for approximately half of all elements heavier than iron \citep{BurbidgeM+57}. While second- and third-peak element abundances in metal-poor stars, such as Ba and Eu, are largely universal and consistent with the solar $r$-process pattern \citep[e.g,][]{FrebelA08}. The slow neutron-capture process ($s$-process) operates in asymptotic giant branch (AGB) stars \citep[e.g.,][]{BussoM+99,BussoM+01,KarakasA_LattanzioJ14,PalmeriniS+21}, with barium receiving $\sim 15\%$--$85\%$ $s$-process contributions depending on metallicity \citep{TravaglioC+04, BisterzoS+14,BisterzoS+15}. The $r$-process occurs in high-neutron-density environments ($n_n \gtrsim 10^{20}$ cm$^{-3}$) where neutron captures proceed faster than beta decays \citep{BurbidgeM+57, CowanJ+21}. The ratio of first-peak elements like Sr, Y, and Zr to second- and third-peak elements varies substantially \citep{CoteB+18,Frebel2018,CowanJ+21,ArconesA+23}. A notable example is provided by HD~122563-like stars, which show an enhancement of light neutron-capture elements relative to heavy species \citep{Honda2006ApJ,OkadaH+25}. Such abundance patterns indicate that Galactic heavy element production may require more than a classical $r$- and $s$-processes framework \citep{Hansen2014ApJ}. The fundamental question is what astrophysical channels dominate the production of different heavy elements and how their contributions evolve across cosmic time. Previous studies invoked an additional contribution to account for the evolution of Sr, Y, and Zr \citep{TravaglioC+04,Ishimaru2005,Montes2007ApJ,Vincenzo2021MNRAS,KobayashiC+23,Molero2023MNRAS}. Multiple astrophysical channels have been proposed as potential $r$-process sites, distinguished primarily by their delay times between star formation and nucleosynthetic injection. Binary neutron-star mergers have long been considered prime candidates due to their neutron-rich ejecta \citep{Lattimer_Schramm74,LiX_PaczynskiB98, KasenD+17_Nat,HotokezakaH+16}. Collapsars, the collapse of rapidly rotating massive stars into black holes accompanied by energetic disk outflows, represent a potential prompt channel \citep{MacFadyen_Woosley99,Surman2008ApJ,Siegel+19_Nat,Gottlieb2025,Shibata2025PhRvD}. Additional proposed channels include magnetorotational supernovae \citep{Nishimura2015ApJ,MostaP+18} and magnetar-powered explosions \citep{Thompson2003ApJ,MetzgerB+08,ThompsonT_udDoula18}. Different lines of evidence constrain the primary $r$-process channel occurring in nature.

Milk-Way (MW) stellar abundances provide a direct measure of the time-integrated product of event rate and ejected mass, implying a Galactic Eu production rate of approximately $10^{-7}$~M$_\odot$~yr$^{-1}$ \citep{Bauswein2014}. However, this measurement alone cannot distinguish frequent low-yield events from rare high-yield events. Independent observations break this degeneracy. Ultra-faint dwarf (UFD) galaxies, with much lower stellar mass, can break this degeneracy. \cite{JiA+16_RetII,Roederer2016} discovered $r$-process enhanced stars in Reticulum~II with [Eu/H]$\approx-1$ while in other UFDs, with comparable number of stars (and [Fe/H]) only upper limits on the relative Eu abundance were obtained, and were lower by up to two orders of magnitude \citep{Frebel2010,Frebel2014,Roederer2014} as compared to Reticulum~II. These large fluctuations suggest that UFDs resolve single enrichment events (Reticulum~II experienced one major $r$-process event during its star formation phase, while other UFDs experienced none) and allow one to independently infer the rate and synthesized mass per event \citep{BHP2016a}.

An independent way to break the rate-mass per event degeneracy relies on measurement of radioactive isotopes in the early solar system as compared with present-day values \citep{Hotokezaka2015,Bartos2019Natur,Cote2019ApJ} and the variance of radioactively stable elements at a given [Fe/H] \citep{Beniamini_Hotokezaka20}. The rate and mass per event of the main $r$-process channel inferred by these considerations were consistent with binary neutron star (BNS) merger estimates based on theoretical calculations as well as observations of short GRB rates and the luminosities of kilonovae candidates \citep{Hotokezaka+18}. 
The 2017 detection of gravitational waves and electromagnetic counterparts from GW170817 provided direct empirical confirmation that BNS mergers eject substantial $r$-process material \citep{Abbott+17, VillarV+17,CoulterD+17}. Multi-wavelength kilonova observations constrain the ejecta mass to approximately $0.03-0.06$~M$_\odot$ with $r$-process mass fraction of $0.1-0.3$, implying roughly $1-3 \times 10^{-5}$~M$_\odot$ of Eu per event \citep{KasenD+17_Nat,DomotoN+21}. 
Interestingly, the initial BNS rate estimates, based on the discovery of GW170817, implied an overproduction of $r$-process by BNS relative to the total required rate described above. The lack of similar events, in the last years, has led to a reduction of the rate (approximately $320^{+490}_{-240}$~Gpc$^{-3}$~yr$^{-1}$ \citealt{LIGO+23_Rates}), which is now in good agreement with the requirement based on $r$-process abundances (see figures 2 and 3 in \citealt{Hotokezaka+18}).

Despite this convergent evidence favoring BNS mergers as a dominant source, three observational features may apriori suggest that additional channels may contribute. First, low-metallicity stars exhibit significant $r$-process abundances, which might naively suggest a prompt metallicity-independent source operating at early times. However, \citet{Tarumi_Hotokezaka_Beniamini21} demonstrated that the observed abundances of second- and third-peak elements in such stars are consistent with a delayed channel operating on timescales of approximately $100$ Myr to $1$ Gyr, as expected for BNS mergers. For instance, Collapsars, which precede the average star formation and have an effective negative delay, would overproduce $r$-process material at the lowest metallicities, thereby constraining their contribution to second- and third-peak elements. 

Second, in MW stars, [Eu/Fe] declines at [Fe/H]$\gtrsim -1$. This suggests that the typical delay time associated with Eu production is significantly shorter than that of Fe production (dominated by Ia SNe at this range of [Fe/H]).
Quantitatively, reproducing the observed feature requires that the bulk of Eu be produced with delay times of order several hundred Myr or less and a delay-time distribution (DTD) declining steeper than $\sim t^{-1.5}$ \citep{Hotokezaka+18,CoteB+19_NotBNS,Beniamini_Tsvi19}. The required steepness is marginally consistent with the demographics of tight Galactic BNS systems that merge within a Hubble time \citep{Beniamini_Tsvi16,Maoz_Nakar25}. Third, and perhaps most critically, systematic variations in the ratio of first-peak to second-peak abundances provide compelling evidence for multiple $r$-process channels with distinct nucleosynthetic signatures. We investigate this in the present work and aim to constrain the existence and contribution of additional enrichment channels.

Galactic chemical evolution (GCE) models provide a systematic framework for translating these observational constraints into quantitative estimates of event rates, DTD, and nucleosynthetic yields \citep{Matteucci2014,CoteB+18,KobayashiC+23}. One-zone models treat the interstellar medium as a well-mixed reservoir evolving under star formation, inflows, outflows, and enrichment from sources characterized by specified DTDs and yields. The time evolution of an element is governed by the convolution of the star-formation history (SFH) with the source DTD and element-specific yield. This framework isolates how different DTDs and yield patterns map onto observable abundance trends. 

We employ a one-zone GCE framework to systematically investigate the relative contributions of prompt and delayed $r$-process sources. We confront our models with stellar abundance data from the SAGA database \cite[][]{SudaT+08_SAGA,SudaT+11_SAGA,SudaT+17_SAGA}, focusing on first-peak tracers Sr and Y, second-peak tracers Ba and Eu, and their correlations in both the MW and dwarf galaxies. 

We demonstrate that reproducing the observed trends requires at least one main $r$-process channel which is delayed relative to the SFR and is the main producer of second and third peak elements, a prompt channel that contributes mainly first peak elements, and a further delayed (compared to the primary $r$-process channel) $s$-process enrichment from asymptotic giant branch stars. We begin with an overview of the core arguments in \S \ref {sec:highlight}. Then, in \S\ref{sec:onezone} we present our one-zone chemical evolution framework and the analytic expectations it yields. \S\ref{sec:results} confronts our models with observations. A discussion and summary are provided in \S\ref{sec:conclusions}.

\begin{figure*}[t]
    \centering
    \includegraphics[width=0.49\linewidth]{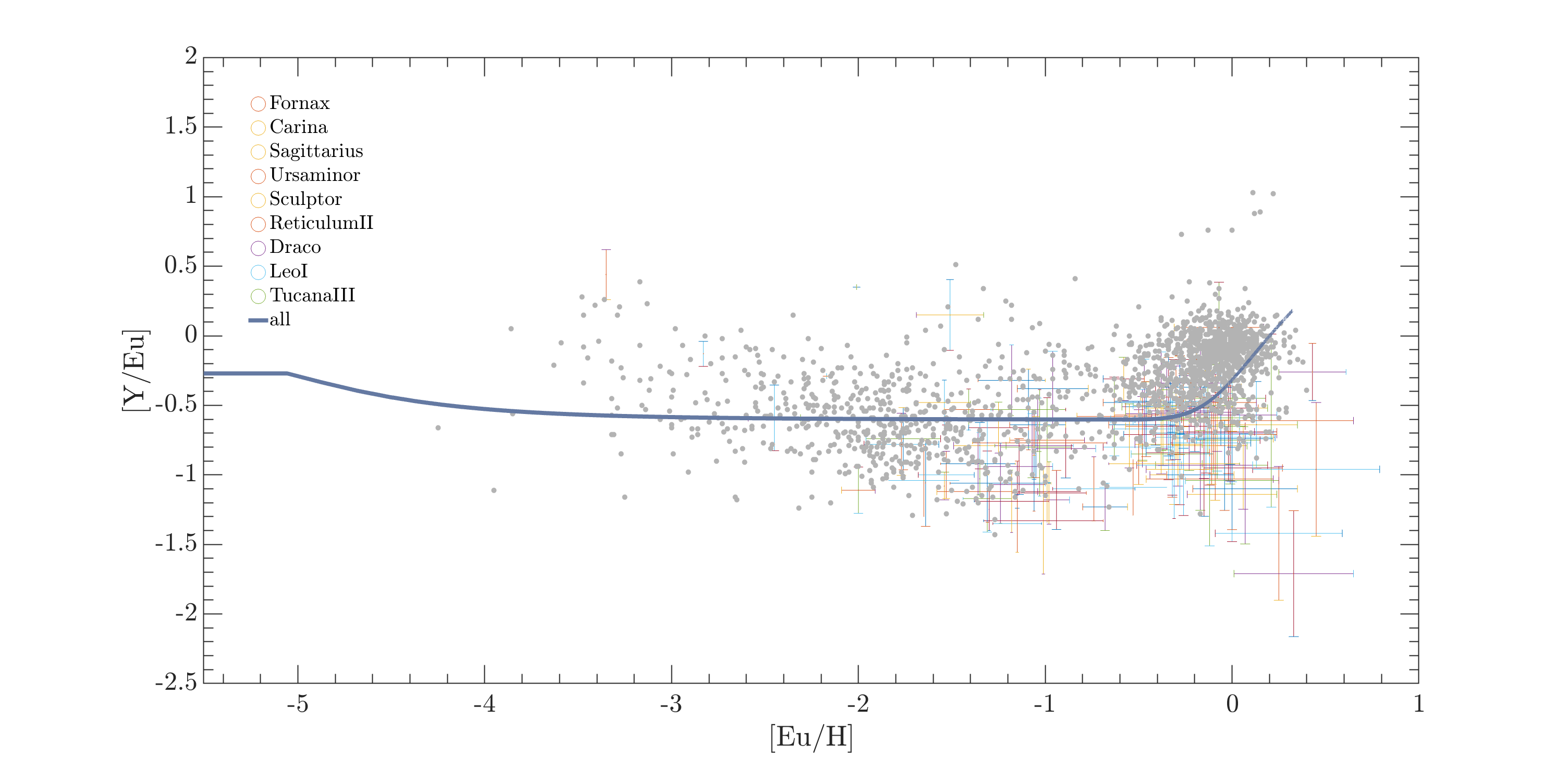}
    \includegraphics[width=0.49\linewidth]{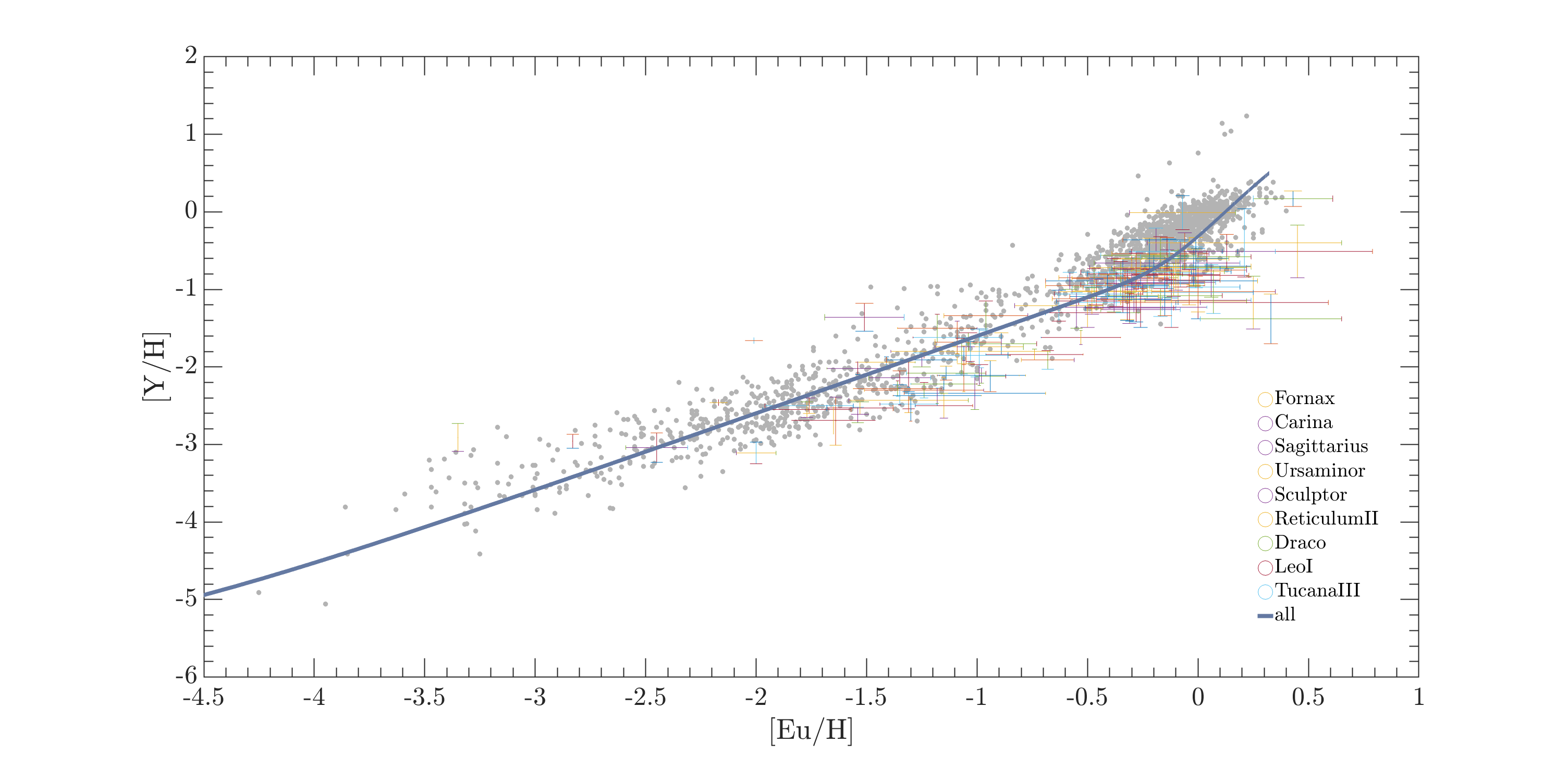}
    \includegraphics[width=0.49\linewidth]{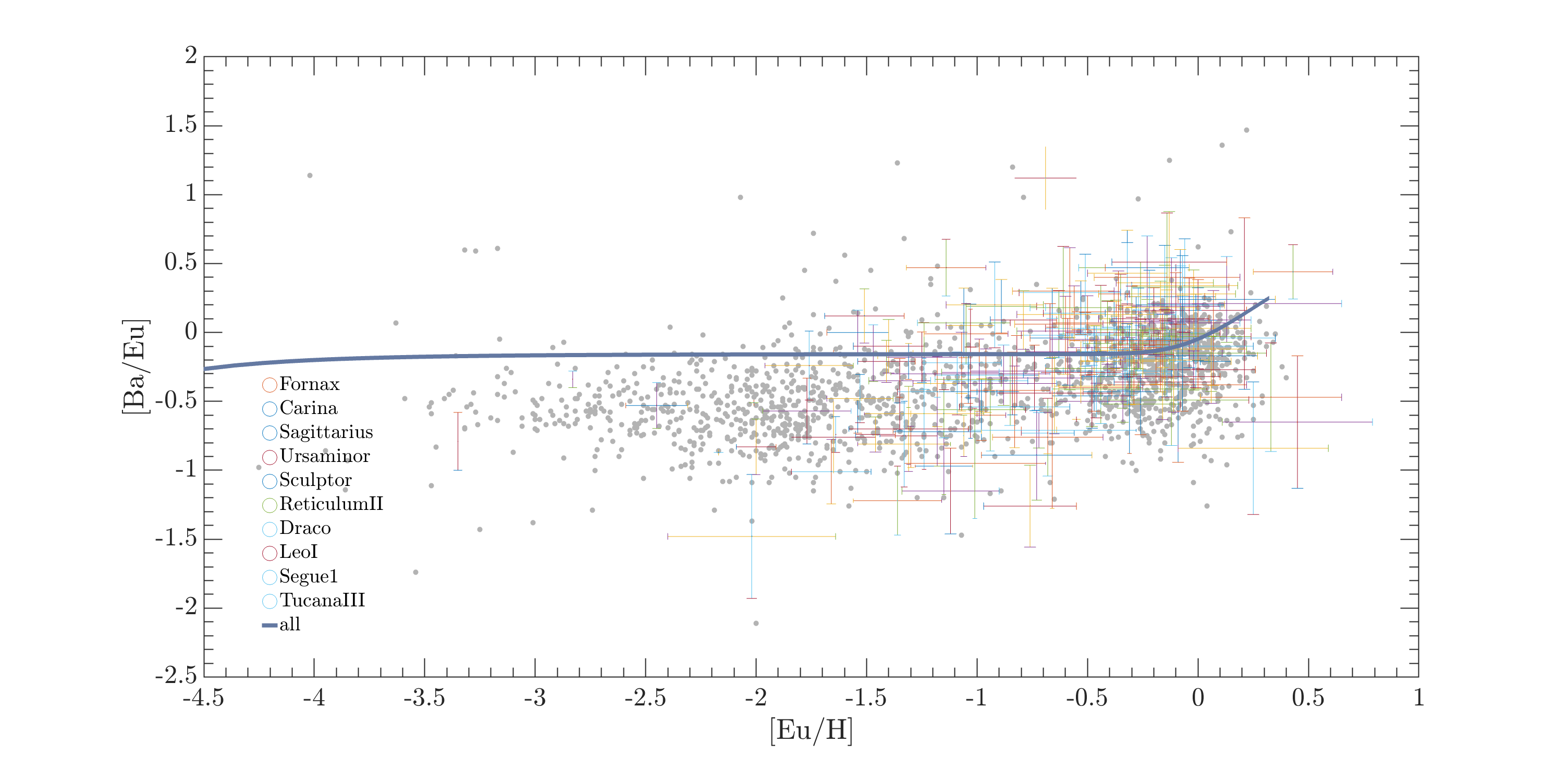}
    \includegraphics[width=0.49\linewidth]{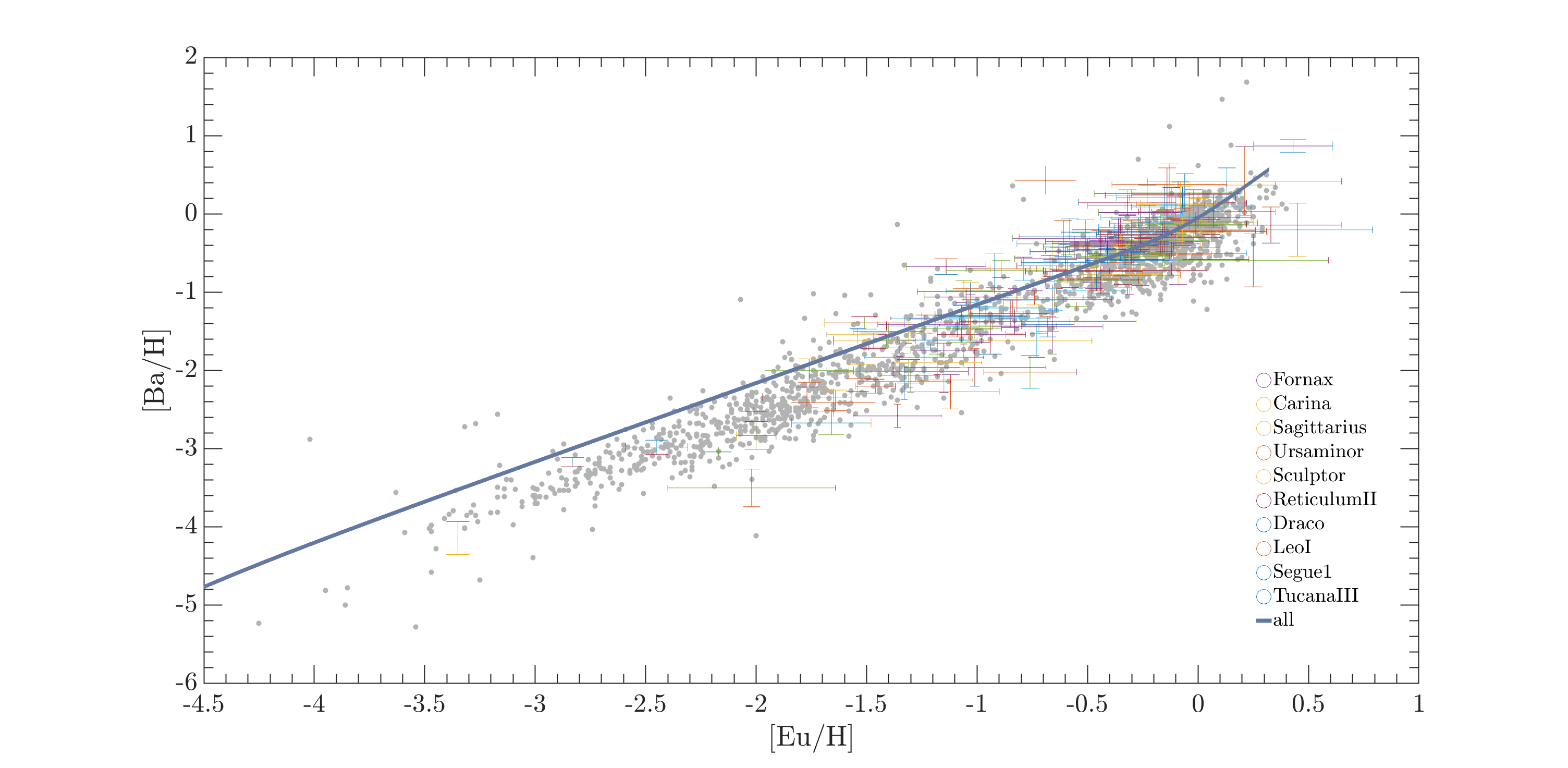}
    
    \caption{SAGA database \cite{SudaT+08_SAGA,SudaT+11_SAGA,SudaT+17_SAGA} stellar data (MW stars in grey and dwarf galaxies in color) of [Y/Eu],[Ba/H] vs [Eu/H] (left) and [Y/H] and [Ba/H] vs [Eu/H] (right). We over-plot (blue lines) the fiducial three-channel model presented in this work with r, s process contributions as given in Tab.\ref{tab:Values}.}
    \label{fig:XEu_Euh_X_Eu}
\end{figure*}

\section{Basic Arguments}
\label{sec:highlight}

Many studies in the literature consider the $r$-process enrichment history by treating an $ r$-process-dominated element $X$ (typically Eu) and examining [X/Fe] as a function of [Fe/H]. However, this approach represents only one projection of the multi-dimensional abundance data and hinges on accurate modeling of Fe production from both Core-collapse supernovae (CCSN) and Type Ia SNe, which introduces additional uncertainties. While our model (\S\ref{sec:onezone}) accounts for Fe production to ensure consistency with MW chemical evolution, we emphasize a different approach, examining first-peak and second-peak $r$-process element abundances as functions of one another. This directly isolates the relative contributions and delay times of different heavy-element channels while minimizing sensitivity to uncertainties in Fe nucleosynthesis.

We focus on Y \& Sr (first-peak elements with extensive stellar data), Ba (second-peak with significant $s$-process contribution), and Eu (second-peak with negligible $s$-process contribution). If Y (or Sr) \& Eu were produced by the same channel throughout MW history, then [Y/Eu] \& [Sr/Eu] would be constant as functions of [Eu/H] in the absence of metallicity-dependent yields or mixing of channels with different DTDs. 
Instead, Figures~\ref{fig:XEu_Euh_X_Eu} and \ref{fig:Obs_SrYEu_D_MW} reveal three distinct phases, a decline of [Y/Eu] ([Sr/Eu]) at [Eu/H] $\lesssim -2.5$, a plateau or very shallow decline extending to [Eu/H] $\approx -0.2$, and then a subsequent increase at [Eu/H] $\gtrsim -0.2$. The decline at [Eu/H] $\lesssim -2.5$ indicates that [Y/Eu] ([Sr/Eu]) decrease as [Eu/H] increases from the earliest enrichment phases. This suggests a prompt channel that preferentially produces first-peak elements dominates at very low metallicity. The plateau phase from [Eu/H] $\approx -2.5$ to $-0.2$ shows relatively constant [Y/Eu] and [Sr/Eu] ratios, indicating that a delayed main $r$-process channel producing robust second-peak elements (Ba, Eu) has become the dominant source, establishing a quasi-steady-state ratio between first- and second-peak production. The subsequent rise at [Eu/H] $\gtrsim -0.2$ requires a further delayed channel (compared to the primary $r$-process channel) that produces high Y/Eu \& Sr/Eu ratios with negligible Eu contribution. The timescale and abundance pattern are consistent with AGB $s$-process enrichment operating on Gyr timescales, which becomes the dominant source of first-peak elements at late times while contributing $\lesssim 5\%$ of total Eu (Table~\ref{tab:Values}).

The two-dimensional [Y/Eu]--[Y/Ba] plane provides complementary diagnostics (Fig.~\ref{fig:YEu_YBa}). Robust $r$-process sources like BNS mergers populate the low [Y/Eu], low [Y/Ba] region \citep[e.g.,][]{RosswogS+14,FernandezR+16,RadiceD+18}. AGB $s$-process sources populate the moderate [Y/Ba], very large [Y/Eu] region. If these were the only dominant producers of heavy elements, stellar data would be restricted to the area connecting these two regions. Observationally, this is not the case. A significant population exhibits large [Y/Ba] and moderate [Y/Eu], preferentially at low metallicity. This suggests at least one additional prompt channel represented by elevated [Y/Ba] and moderate [Y/Eu].

With these observational indications of a prompt channel producing mostly first-peak elements, we turn in \S\ref{sec:onezone} to a detailed description of our GCE model, including this additional channel, and attempt to constrain its main characteristics (yields and DTD) through a detailed comparison with stellar data.

\section{One-Zone Model}
\label{sec:onezone}

Our framework is a one-zone GCE model treating the interstellar medium as a well-mixed reservoir with prescribed star formation. CCSN supplies $\alpha$-elements and Fe. Type Ia SNe contribute Fe through a DTD. BNS mergers contribute $r$-process material through distinct DTDs and yields. AGB stars contribute $s$-process material on Gyr timescales. The $s$-process contribution is calibrated to reproduce present-day MW abundances. An additional prompt channel, motivated by the arguments in \S \ref{sec:highlight}, contributes mostly first peak elements.
The relative contributions of prompt and delayed heavy-element production channels, along with their characteristic delay times, are free parameters determined by fitting the observed evolutionary trends. The synthesized mass of a given heavy element $j$ may have contributions from channel A (prompt $r$ or $s$ process), channel B (delayed $r$-process, taken as BNS mergers), and channel C (delayed $s$ process from AGB stars). The enrichment is a convolution between the SFR $\Psi(t)$ and the effective DTD of channel $X$, $D_X(t')$. The result is (see \citealt{Hotokezaka+18} for a version of this model without channels A \& C)

\begin{equation}
\begin{split}
\frac{dM_j}{dt} = &\int_0^t m_{A,j}\,D_A(t')\,\Psi(t-t')\,dt' \\
&+ \int_0^t m_{B,j}\,D_B(t')\,\Psi(t-t')\,dt' \\
&+ \int_0^t m_{C,j}\,D_{C}(t')\,\Psi(t-t')\,dt'\, - F(t)\,M_j(t).
\end{split}
\label{eq:DTD_sr}
\end{equation}

where $m_{X,j}$ is the mass of element $j$ synthesized by channel $X$ (if channel $X$ does not produce any amount of element $j$, it can also be zero, see Table \ref{tab:Values}), and $F(t) = (1+O)\Psi(t)/M_{\rm ISM}(t)$ represents the mass-loss rate from the ISM due to star formation and outflows, with $O$ the mass-loading factor. We adopt power-law DTDs of the form

\begin{equation}
  D_l(t) = G_l \left(\frac{t}{t_{\min,l}}\right)^{-b_l}
  \Theta(t - t_{\min,l}),
  \label{eq:DTD}
\end{equation}

for $l \in \{A, B, C\}$, where $G_l$ is a normalization, $t_{\min,l}$ is the minimum delay time, $b_l$ is the slope, and $\Theta$ is the Heaviside step function. We construct a cosmic time grid by numerically integrating $dt/dz = -[(1+z)H(z)]^{-1}$ for a flat $\Lambda$CDM cosmology and assign a SFR history using the parametrization of \citep[e.g.,][]{Madau_Dickinson14}.

\begin{figure}[t]
    \centering
    \includegraphics[width=\linewidth]{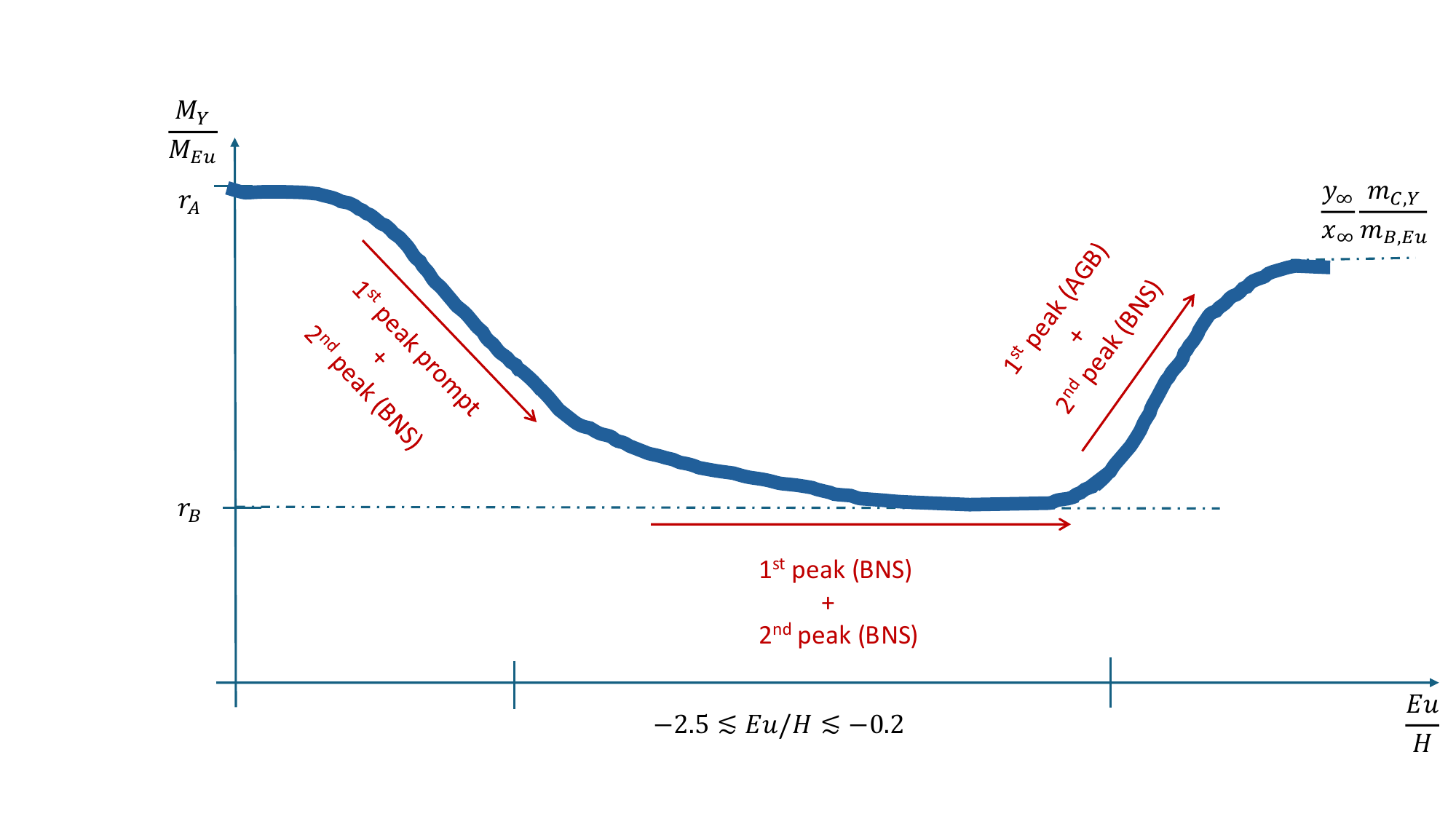}
    \caption{Schematic overview of the enrichment channels included in our fiducial one-zone model and how they map onto the abundance trends analyzed in this work. We include a prompt, primarily first-peak producing, channel (A), and a delayed BNS-merger channel (B). First-peak enrichment additionally receives a delayed AGB $s$-process (channel C) component with long delays compared to channel B. The prompt channel sets the low-$\rm[Eu/H]$ first-to-second peak ratios, while the AGB enrichment sets the rise of [Y/Eu] at large [Eu/H]. The BNS mergers dominate Eu production throughout the observable range and first peak production at $-2.5\lesssim \mbox{[Eu/H]}\lesssim -0.2$.}
    \label{fig:Cartoon}
\end{figure}

\subsection{$s$-process Calibration}
\label{sec:s-process}

The slow neutron-capture process $s$-process synthesizes heavy nuclei by successive neutron captures that occur on timescales longer than typical $\beta$-decay lifetimes, so the reaction flow proceeds close to the valley of $\beta$-stability. The dominant neutron sources in stellar interiors are typically ${}^{13}\mathrm{C}(\alpha,n){}^{16}\mathrm{O}$, operating at relatively low temperatures, and ${}^{22}\mathrm{Ne}(\alpha,n){}^{25}\mathrm{Mg}$ activated at higher temperatures. In GCE, the main $s$-process component is associated with thermally pulsing AGB stars of low and intermediate mass, which can account for a large fraction of the solar abundances of first- and second-peak elements ($\rm Sr, Y, Ba$). A secondary contribution (optional in our model), the \emph{weak $s$-process} component operating in massive stars, contributes primarily to lighter heavy elements (see \citealt{PignatariM+23, ArconesA+23} for updated solar $s$/$r$ decompositions and yields). This secondary contribution, if required by the data, would be captured by our prompt channel A.
We decompose the solar abundance of each element $X$ into $s$- \& $r$-process contributions using an adopted solar $s$-fraction $f_s(X)$,

\begin{equation}
X_{\odot,s}(X)=f_s(X)\,X_\odot(X); \ X_{\odot,r}(X)=\left[1-f_s(X)\right]X_\odot(X).
\end{equation}

We adopt representative solar $s$-fractions $f_s({\rm Sr})=0.76$, $f_s({\rm Y})=0.72$, $f_s({\rm Ba})=0.81$, $f_s({\rm Eu})=0.03$, consistent with the standard picture that Sr--Ba have substantial solar $s$-process contributions, whereas Eu is overwhelmingly $r$-dominated \citep[e.g.,][]{KappelerF+11,BisterzoS+14,BisterzoS+15,Sales-SilvaV+22,RoedererI+23}.

\begin{figure*}[t]
    \includegraphics[width=\linewidth]{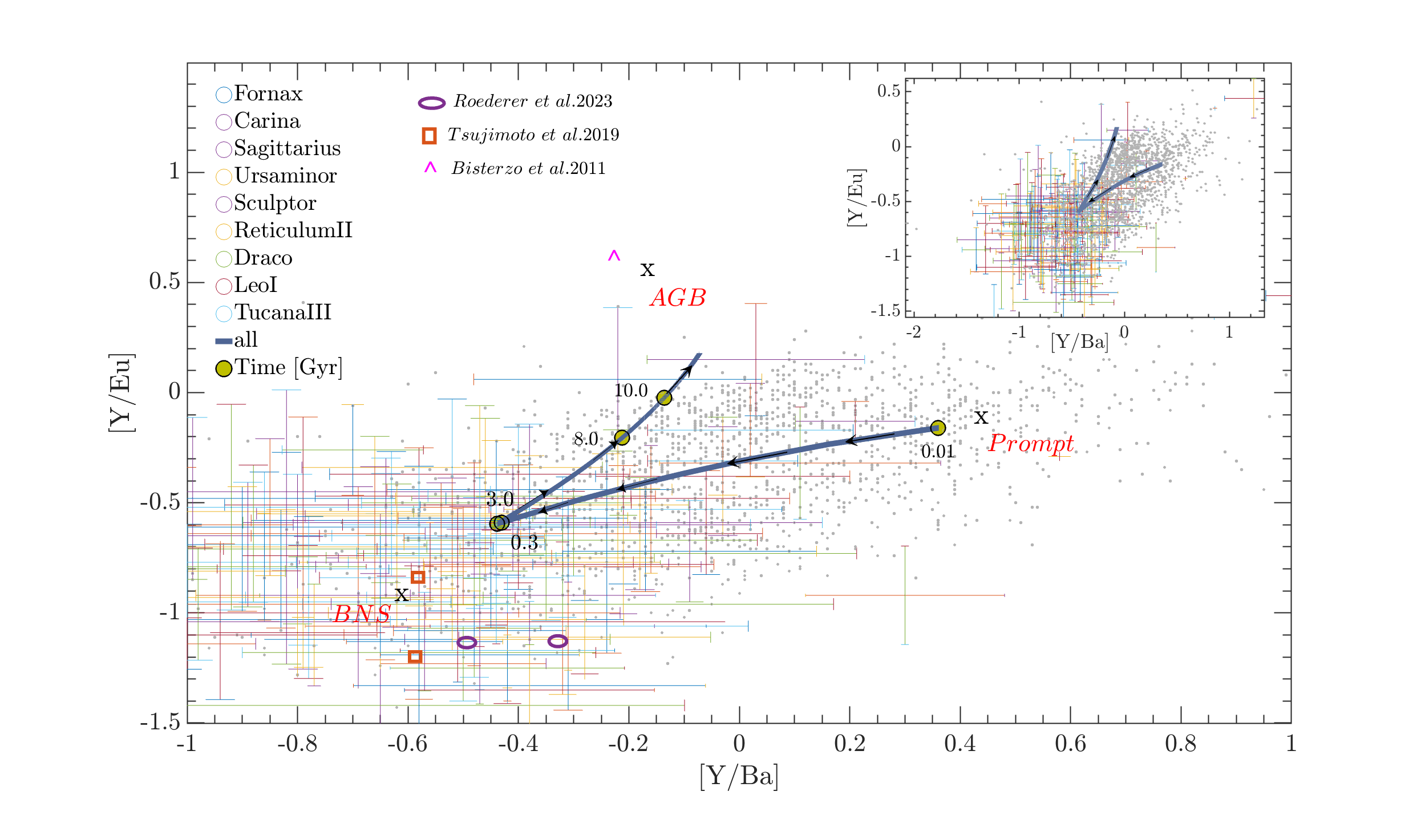}
    
    \caption{$\rm [Y/Eu]$ versus $\rm [Y/Ba]$ for MW (gray) and dwarf galaxies (color), overlaid with the fiducial one-zone chemical-evolution track. Markers indicate cosmic time along the model trajectory. Early times reflect increased prompt contribution (channel~A) to first peak production, intermediate times move toward a more Eu-rich mixture as the delayed BNS channel~B becomes important, and late times deviate as AGB $s$-process enrichment (channel~C) modifies Y and Ba relative to Eu. Reference yield points from noteworthy stars as discussed in \S ~\ref{sec:intro} is marked with special symbols \citep{RoedererI+23,TsujimotoT_BabaJ19,BisterzoS+14}. The model track transitions through the pure BNS $r$-process yield point (lower left) at intermediate times when channel~B dominates, then evolves toward the AGB $s$-process regime (upper right) at late times, successfully reproducing the observed progression from $r$-process-dominated to $s$-process-enhanced composition.}
    \label{fig:YEu_YBa}
\end{figure*}

We calibrate the effective $s$-process normalization by requiring that the $s$-process ISM mass of each element matches its adopted present-day $s$-component target, $M^{\rm targ}_{X,s}(z{=}0) = f_s(X)\,M^{\rm targ}_{X}(z{=}0)$, where $M^{\rm targ}_X(z=0)$ is the total present-day target mass derived from the solar mass fraction converted consistently with the fiducial present-day ISM mass in the model. This ensures that the $s+r$ model reproduces the chosen present-day decomposition by construction.

\subsection{Analytic three-channel mixing}
\label{sec:Analyticmodel}

We derive a simple analytic expression for the instantaneous ISM abundance ratio $\mathrm{[Y/Eu]}$ produced by a mixture of three enrichment channels. This relation is useful for interpreting the shape and limiting behavior of the numerical tracks in Fig.~\ref{fig:XEu_Euh_X_Eu}, and for clarifying which combinations of yield ratios are constrained by the data. We assume the per-event yields of Y, Sr, and Eu from each channel are time-independent constants. The relative contributions of the channels vary over time through their cumulative event counts. 

We define $N_A(t)$, $N_B(t)$, and $N_C(t)$ as the cumulative number of events of type A, B, \& C that have enriched the ISM up to time $t$. The time dependence $N_B(t)/N_A(t)$ \& $N_C(t)/N_A(t)$ reflects the adopted DTDs and SFH. The per-event yield ratios $m_{X,{\rm Y}}/m_{X,{\rm Eu}}$ for each channel $X \in \{A, B, C\}$ are constrained below by the requirement that the model reproduce the observed [Y/Eu] versus [Eu/H] ([Sr/Eu] versus [Eu/H]) trend.
We do not apriori impose the yield ratio, $m_{A,{\rm Y}}/m_{A,{\rm Eu}}$, produced by channel~A but, motivated by the argument in \S~\ref{sec:highlight} we search for solutions with first peak dominance, $m_{A,{\rm Y}}/m_{A,{\rm Eu}}\gtrsim 12.7$. Channel~B satisfies $m_{B,{\rm Y}}/m_{B,{\rm Eu}} \simeq 4.4$ consistent with robust $r$-process simulations \citep{RosswogS+14,RadiceD+18}; and channel~C lies in the range $m_{C,{\rm Y}}/m_{C,{\rm Eu}} \gg 1$ consistent with $s$-process yields \citep{KarakasA_LattanzioJ14}. The total accumulated Y and Eu masses are $M_{\rm Y}(t) = N_A(t)\, m_{A,{\rm Y}} + N_B(t)\, m_{B,{\rm Y}} + N_C(t)\, m_{C,{\rm Y}}$ and $M_{\rm Eu}(t) = N_A(t)\, m_{A,{\rm Eu}} + N_B(t)\, m_{B,{\rm Eu}} + N_C(t)\, m_{C,{\rm Eu}}$, yielding the instantaneous abundance ratio. For the general three-channel case, taking [Y/Eu]\footnote{throughout, we use the standard bracket notation $[{\rm X/Y}] \equiv \log_{10}[(M_{\rm X}/M_{\rm Y})/(M_{\rm X}/M_{\rm Y})_\odot]$ 
with solar normalization \citep[e.g.,][]{AsplundM+09,AsplundM+20}, example [Y/Eu] $\equiv \log_{10}[(M_{\rm Y}/M_{\rm Eu})/(X_{\rm Y,\odot}/X_{\rm Eu,\odot})] \simeq \log_{10}(3/49) \sim -1.2$.} as an illustrative example, the abundance ratio is

\begin{equation}
\frac{\rm{M_Y}}{\rm{M_{Eu}}}(t)\!
=\!
\frac{N_A(t)\,m_{A,{\rm Y}} + N_B(t)\,m_{B,{\rm Y}} + N_C(t)\,m_{C,{\rm Y}}}
     {N_A(t)\,m_{A,{\rm Eu}} + N_B(t)\,m_{B,{\rm Eu}} + N_C(t)\,m_{C,{\rm Eu}}}.
\end{equation}

A convenient minimal set of dimensionless yield ratios required to explain the [Y/Eu] evolution is

\begin{equation}
\begin{split}
r_A &\equiv \frac{m_{A,{\rm Y}}}{m_{A,{\rm Eu}}}, \quad
r_B \equiv \frac{m_{B,{\rm Y}}}{m_{B,{\rm Eu}}}, \quad
r_C \equiv \frac{m_{C,{\rm Y}}}{m_{C,{\rm Eu}}}, \\
k_B &\equiv \frac{m_{A,{\rm Y}}}{m_{B,{\rm Y}}}, \quad
k_C \equiv \frac{m_{A,{\rm Y}}}{m_{C,{\rm Y}}}.
\end{split}
\end{equation}

Here, $r_A$, $r_B$, \& $r_C$ are the intrinsic first-to-second peak yield ratios of channels~A, B, \& C, while $k_B$ \& $k_C$ control the relative Y normalization between the channels. Defining $x(t) \equiv N_B(t)/N_A(t)$ \& $y(t) \equiv N_C(t)/N_A(t)$, the mixture may be written as

\begin{equation}
\frac{\rm{M_Y}}{\rm{M_{Eu}}}(t) = \frac{1 + x(t)/k_B + y(t)/k_C}{1/r_A + x(t)/(k_B r_B) + y(t)/(k_C r_C)}.
\label{eq:threechannel_general}
\end{equation}

Four analytical limits of the three-channel mixing relation (Eq.~\ref{eq:threechannel_general}) provide physical insight into the model constraints. The very early-time limit ($x, y \to 0$) yields $\frac{\rm{M_Y}}{\rm{M_{Eu}}} = r_A$\footnote{As shown in appendix \ref{app:limits}, this plateau is obtained at extremely early values of $\rm{[Eu/H]}\lesssim -5$, where there is little to no observed stellar data. As such the lowest Eu enriched stars provide a lower limit on $r_A$.}, while at intermediate times BNS dominance ($x \to \infty$, $y/x \to 0$) gives $\frac{\rm{M_Y}}{\rm{M_{Eu}}} = r_B$. The prompt-to-delayed amplitude of [Y/Eu] variation, $\Delta [\mathrm{Y/Eu}]_{\rm A \to B} = \log(r_A/r_B)$, is independent of DTD normalizations and constrains directly the ratio of intrinsic yield ratios. The late-time limit ($y \to \infty$) is dominated by AGB $s$-process with $r_C \gg r_A, r_B$, which drives [Y/Eu] upward once AGB stars become the dominant Sr, Y production channel. 
The fourth diagnostic, the [Eu/H] at which [Y/Eu] reaches its minimum, depends on the point at which the first peak contributions of channel B and C are comparable, and therefore constrains the DTD shapes and minimum delay times. These limits and their evaluation are derived in Appendix~\ref{app:limits} \& Fig.~\ref{fig:x_y}. In the appropriate limit in which channel C contributes negligibly to Eu production ($m_{C,{\rm Eu}} \simeq 0$, or equivalently $r_C \to \infty$), Eq.~(\ref{eq:threechannel_general}) reduces to,

\begin{equation}
\frac{\rm{M_Y}}{\rm{M_{Eu}}}(t)\simeq \frac{1 + x(t)/k_B + y(t)/k_C}{1/r_A + x(t)/(k_B r_B)}.
\label{eq:threechannel_limit}
\end{equation}

This approximation clarifies that late-time [Y/Eu] evolution is controlled by the ratio $y(t)/[1 + x(t)/k_B]$: as $y(t)$ increases, the numerator grows while the denominator remains approximately constant, driving [Y/Eu] upward, see Fig. \ref{fig:x_y} for the $x(t),y(t)$ behavior.

\begin{equation}
\frac{\rm{M_Y}}{\rm{M_{Eu}}}(t)
\simeq
\frac{1 + 12.504\,x(t) + 2310.43\,y(t)}
     {0.4001 + 18.754\,x(t)}.
\label{eq:threechannel_CnoEu_numeric}
\end{equation}

This approximation is excellent for our fiducial parameters, where channel~C contributes only $\sim 5\%$ of total Eu (Table~\ref{tab:Values}), and clarifies that late-time [Y/Eu] evolution is controlled by the interplay between the enrichment provided by the different channels. The explicit time evolution of $x(t)$ and $y(t)$ and their asymptotic behavior are derived in §\ref{app:rates}. For additional useful limiting cases and constraints, see Appendix~\ref{app:limits}. 

\section{Results} \label{sec:results}

We analyze SAGA abundances \citep{SudaT+08_SAGA,SudaT+11_SAGA,SudaT+17_SAGA} over a broad metallicity range, focusing on first-peak tracers (Sr, Y) and second-peak tracers (Ba, Eu; Figs.~\ref{fig:XEu_Euh_X_Eu} \& \ref{fig:SrEu_Euh_Sr_Eu}).
In absolute-abundance space, $\rm [Ba/H]$ and $\rm [Eu/H]$ follow a tight, nearly linear relation with dispersion $\lesssim 0.2$ dex over four decades, indicating nearly constant [Ba/Eu] production. By contrast, $\rm [Sr/H]$ and $\rm [Y/H]$ versus $\rm [Eu/H]$ show larger dispersion ($\sim 0.4$--$0.6$ dex) and clear curvature. In ratio space, $\rm [Ba/Eu]$ remains nearly flat ($\lesssim 0.3$ dex). Instead, $\rm [Y/Eu]$ declines by $\sim 1$ dex while $\rm [Eu/H]$ changes from $-3.5$ to $-1.5$. The ratio then increases at large $\rm [Eu/H]$. This non-monotonic behavior rules out a single enrichment channel. A similar trend is seen in [Sr/H] (Fig. \ref{fig:SrEu_Euh_Sr_Eu}).
BNS merger models typically produce [Y/Eu]$\sim 0.08$--$0.2$ \& [Sr/Eu] $\sim 0.1$--$0.3$, whereas collapsar-like models reach $\sim 1$--$10$ \citep{Siegel+19_Nat, DeanC_FernandezR24}. The data are naturally explained if the very early enrichment is dominated by a prompt high-[Sr/Eu] source and later enrichment by delayed low-[Sr/Eu], [Y/Eu] BNS mergers. The reversal at $\rm [Eu/H]\gtrsim -1$ requires a third component, delayed AGB $s$-process enrichment boosting Sr \& Y with minimal Eu.

We apply the GCE model from \S\ref{sec:onezone} to the SAGA data. The preferred AGB parameters are $t_{\min,\rm agb}\simeq 0.45~\rm{Gyr}$ and $b_{\rm C}\approx 1.17$, reproducing the observed turnover without overproducing first-peak elements at high $\rm [Eu/H]$. Channel A is consistent with having little to no delay.

The median curves in Figs.~\ref{fig:XEu_Euh_X_Eu} \& \ref{fig:SrEu_Euh_Sr_Eu} are well described by the analytic three-channel mixing relation (\S\ref{sec:Analyticmodel}). The fiducial model evolves through four phases. At early times (up to $t \sim 15~\rm{Myr}$, see figure \ref{fig:M_t_ABC}.), $x(t) \ll 1$ and $y(t) \approx 0$, so $[\rm Y/Eu] \equiv \log{10}\left(\frac{n_{\rm Y,A}/n_{\rm Eu,A}}{n_{\rm Y,\odot}/n_{\rm Eu,\odot}}\right) \equiv \log_{10}\left(\frac{r_A\times(A_{Eu}/A_{Y})}{n_{\rm Y,\odot}/n_{\rm Eu,\odot}}\right) \approx -0.15$. This phase only persists up to $[\rm Eu/H] \lesssim -4.8$ and as such is not observationally accessible.
After this point, Eu starts building up rapidly due to contribution from BNS mergers and $[\rm Eu/H] \approx -2.5$ is reached by $t\sim 0.08$\,Gyr. During this stage [Y/Eu] drops to $\approx -0.55$, approaching the value expected from pure channel B enrichment, $\log{10}\left(\frac{n_{\rm Y,B}/n_{\rm Eu,B}}{n_{\rm Y,\odot}/n_{\rm Eu,\odot}}\right) \approx -0.8$. At later times ($0.8~\rm{Gyr} \lesssim t \lesssim 4~\rm{Gyr}$), the evolution resembles a quasi-plateau (or shallow declining phase), in which channel B continues to accumulates and drives $[\rm{Y/Eu}]$ further toward $\log{10}\left(\frac{n_{\rm Y,B}/n_{\rm Eu,B}}{n_{\rm Y,\odot}/n_{\rm Eu,\odot}}\right) \approx -0.80$ ($r_B \simeq 4.8$), reaching a minimum at $[\rm{Eu/H}] \sim -0.3$ (with [Y/Eu] $\approx -0.61$).
Finally, at $t \gtrsim 4~\rm{Gyr}$, AGB $s$-process enrichment becomes a dominant producer of first peak elements, resulting in the observed upturn in $[\rm{Y/Eu}]$. This sequence reproduces the observed non-monotonic $[\rm{Y/Eu}]$ trend across $\sim 5$ dex in [Eu/H]. Early enrichment from channel A sets initial first-to-second peak ratios, while delayed BNS mergers (channel B) dominate the Eu budget at practically all metallicities ($\sim 95\%$ of total Eu production) and first peak production in the range $-2.5\lesssim [\rm{Eu/H}]\lesssim -0.2$.

The $\rm [Y/Eu]$--$\rm [Y/Ba]$ plane (Fig.~\ref{fig:YEu_YBa}) provides an additional diagnostic. The vertical axis traces first-peak production relative to Eu (predominantly $r$-process), while the horizontal axis traces the light-to-heavy peak ratio relative to Ba (predominantly $s$-process). A single source would occupy a narrow locus; mixtures generate curved, time-ordered tracks. The thick blue curve shows the fiducial model's $s{+}r$ totals, with markers denoting cosmic time. Early times reflect the prompt channel; intermediate times become Eu-rich as BNS mergers accumulate; late times move upward and rightward as AGB enrichment raises Y and Ba with negligible Eu. Because the trajectory represents total $s{+}r$ abundances, it does not converge to a pure BNS yield even when BNS mergers dominate Eu—Y and Ba continue receiving AGB contributions, so the late-time locus reflects BNS-driven Eu and AGB-driven Y and Ba.

\section{Conclusions} \label{sec:conclusions}

We find that neither a single enrichment channel nor any two-channel combination of $r$-process, $s$-process, or mixed sources can reproduce the observed Sr--Y--Ba--Eu patterns in dwarf galaxies and the MW without conflicting with independent constraints on the BNS population. A three-channel model is preferred: a prompt channel (A) with $t_{\rm min,A} \simeq 0 - 5~\mathrm{Myr}$ producing mostly first peak elements, a delayed BNS channel~B with minimal delay times $\sim 12~\mathrm{Myr}$, and a further delayed AGB $s$-process channel~C with onset $t_{\rm min,C} \sim 0.3$--$0.6~\mathrm{Gyr}$ contributing negligible Eu. This framework reproduces the Eu abundance evolution (Fig.~\ref{fig:XEu_Euh_X_Eu}) and the distinct behaviors of first- and second-peak elements in both dwarfs and the MW. The preferred solutions require a channel~A rate $\sim 0.4$ times the BNS rate (corresponding to $\sim 36$--$50~{\rm Gpc}^{-3}\,{\rm yr}^{-1}$ versus $\sim 90~{\rm Gpc}^{-3}\,{\rm yr}^{-1}$), consistent with channel~A arising from a rare subset of CCSN-like events. Observationally, $\Delta[\rm{Y/Eu}] \gtrsim 0.6$ dex\, over a range of [Eu/H] from $\sim -4$ to $\sim -0.3$, establishes a lower limit $r_A/r_B \gtrsim 10^{0.6}\sim 4$. Our model achieves $\Delta[\rm{Y/Eu}]_{\rm model} \approx 0.3\rm{-}0.8~\rm{dex}$ for $r_A/r_B \sim 2.8\rm{-}9.2$. This constraint is independent of DTD normalizations and rules out models in which both channels have similar [Y/Eu] ratios ($r_A \sim r_B$).

Our model is a proof of concept of the need for three distinct channels - however we have not fully explored the model parameter space and other solutions might exist with markedly different values of $\mathcal{C}_A$ and the abundances per event of channel A. This will be explored in a follow-up work. Future progress will come from extending the analysis to additional first- and heavy-peak tracers, including metallicity-dependent weak-$s$ yields, and testing whether the preferred parameter combinations remain consistent across multiple galactic environments.

\begin{acknowledgments}
The authors thank Elias Most, Shivani Shah, Todd Thompson for helpful discussions. YZ acknowledges support from the MAOF grant 12641898 and visitor support from the Observatories of the Carnegie Institution for Science, Pasadena, CA, where part of this work was completed.
PB's work was funded by grants (no. 2020747, 2024788) from the United States-Israel Binational Science Foundation (BSF), Jerusalem, Israel, by a grant (no. 1649/23) from the Israel Science Foundation and by a NASA grant (80NSSC24K0770).
KH's work  was supported by the JST FOREST Program (JPMJFR2136) and the JSPS Grant-in-Aid for Scientific Research (20H05639, 20H00158, 23H01169, 23H04900).

\end{acknowledgments}

\facilities{\texttt{SAGA database}\citep{SudaT+08_SAGA,SudaT+11_SAGA,SudaT+17_SAGA}}

\software{\texttt{AstroColour} \citep{Lane_2025_colour}, \texttt{astropy} \citep{Astropy_2013, Astropy_2018, Astropy_2022}, \texttt{matplotlib} \citep{Hunter_2007}, \texttt{numpy} \citep{Harris_2020}, \texttt{pandas} \citep{McKinney_2010}, \texttt{scipy} \citep{Virtanen_2020}}


\appendix
\restartappendixnumbering
\onecolumngrid

\section{Mass Ratio of events} 
\label{app:rates}

The three-channel mixing relation (Eq.~\ref{eq:threechannel_general}) depends on the time-dependent cumulative event ratios $x(t) \equiv N_B(t)/N_A(t)$ and $y(t) \equiv N_C(t)/N_A(t)$, which encode the relative buildup of delayed enrichment channels B and C compared to the prompt channel A. Here $N_i(t)$ denotes the cumulative number of enrichment events from channel $i$ that have contributed to the ISM up to cosmic time $t$. These ratios evolve from zero at early times (when only the prompt channel A has contributed) to asymptotic values $x_\infty$ and $y_\infty$ at late times (when all channels have fully activated). The transition timescales and functional forms are determined by the convolution of the star formation history $\Psi(t)$ with the channel-specific delay-time distributions $D_A(t)$, $D_B(t)$, and $D_C(t)$ (Eq.~\ref{eq:DTD}). Thus, the event ratios become

\begin{equation}
x(t) = \frac{\int_0^t \Psi(t') \, D_B(t - t') \, dt'}{\int_0^t \Psi(t') \, D_A(t - t') \, dt'}, \quad
y(t) = \frac{\int_0^t \Psi(t') \, D_C(t - t') \, dt'}{\int_0^t \Psi(t') \, D_A(t - t') \, dt'}.
\label{eq:xy_exact}
\end{equation}

For the power-law DTDs adopted in this work (Eq.~\ref{eq:DTD}), we evaluate these convolution integrals numerically using our adopted MW SFR. As shown in Fig.~\ref{fig:x_y}, the resulting time evolution is well approximated by saturating power laws of the form

\begin{equation}
x(t) \simeq x_\infty \left[1 - \left(\frac{\tau_B}{t}\right)^p\right] \quad \text{for } t \geq \tau_B,
\label{eq:x_powerlaw}
\end{equation}

\begin{equation}
y(t) \simeq y_\infty \left[1 - \left(\frac{\tau_C}{t}\right)^q\right] \quad \text{for } t \geq \tau_C,
\label{eq:y_powerlaw}
\end{equation}

Crucially, as clear from Eq. \ref{eq:xy_exact}, the asymptotic values, $x_{\infty},y_{\infty}$ are independent of the SFH and DTDs, and they directly reflect the ratio of intrinsic rates between the respective channels.

For our fiducial model parameters (Table~\ref{tab:Values}), we obtain $x_\infty \equiv \lim_{t \to \infty} N_B(t)/N_A(t)\approx 2.33$, indicating that BNS mergers (channel~B) produce 2.33 times events per unit stellar mass as the prompt channel~A over Galactic history. Similarly, $y_\infty \equiv \lim_{t \to \infty} N_C(t)/N_A(t)$ is the asymptotic AGB-to-prompt event ratio, with $y_\infty \approx 2100$ for our fiducial model, reflecting the substantial cumulative contribution of AGB stars to first-peak element production.

The characteristic timescale $\tau_B$, approximately satisfies $x(\tau_B) \simeq 0.1\cdot x_\infty \approx 0.23$ or $\tau_B \approx 25$ Myr. $\tau_B$ represents the transition time at which the cumulative contribution from channel~B becomes comparable to that from channel~A. This is approximately $\sim 1.5-2 \times t_{\rm min,B}$, where $t_{\rm min,B} = 12$ Myr is the minimum delay for BNS mergers (Table~\ref{tab:Values}). The characteristic timescale $\tau_C \approx 450\mbox{ Myr}\approx 1.5 t_{\rm min,C}$ represents the onset time for significant AGB enrichment, reflecting the evolutionary timescale required for intermediate-mass stars to reach the AGB phase. 

The saturation exponents $p \approx 1.2$ and $q \approx 0.92$ are determined numerically by fitting Eqs.~(\ref{eq:x_powerlaw})--(\ref{eq:y_powerlaw}) to the convolution integrals. These exponents depend on the DTD power-law slopes $b_A$, $b_B$, $b_C$ (Eq.~\ref{eq:DTD}) and the SFH, but the relationship is non-trivial due to the convolution structure. We therefore determine $p$ and $q$ empirically rather than deriving analytic expressions.

Figure~\ref{fig:x_y} shows the numerically computed $x(t)$ and $y(t)$ (solid curves). The corresponding cumulative mass ratios are related to the event ratios by $M_{\rm Y,B}(t)/M_{\rm Y,A}(t) = x(t) \cdot [m_{B,\rm Y}/m_{A,\rm Y}]$ and $M_{\rm Y,C}(t)/M_{\rm Y,A}(t) = y(t) \cdot [m_{C,\rm Y}/m_{A,\rm Y}]$, where the yield ratios are determined from the per-event Y masses (Table~\ref{tab:Values}):

\begin{equation}
\frac{m_{B,\rm Y}}{m_{A,\rm Y}} = \frac{9.71 \times 10^{-4}}{8.58 \times 10^{-5}} \approx 11.3.
\end{equation}

For channel C (AGB stars), the per-event yield $m_{C,\rm Y}$ is not listed directly in Table~\ref{tab:Values}, but can be derived from the total Y mass and event number. Using $y_\infty \equiv N_C(t\to \infty)/N_A(t \to \infty)$ and the total masses $M_{\rm Y,C} = 4.72 M_\odot$ and $M_{\rm Y,A} = 3.94\times 10^{-2} M_\odot$ (Table~\ref{tab:Values}), we have,

\begin{equation}
\frac{m_{C,\rm Y}}{m_{A,\rm Y}} \approx 0.06.
\end{equation}

The asymptotic cumulative mass ratios are therefore, $\frac{M_{\rm Y,B}(\infty)}{M_{\rm Y,A}(\infty)} = x_\infty \cdot \frac{m_{B,\rm Y}}{m_{A,\rm Y}} \approx 2.33 \times 11.3 \approx 27.2$, and, $\frac{M_{\rm Y,C}(\infty)}{M_{\rm Y,A}(\infty)} = y_\infty \cdot \frac{m_{C,\rm Y}}{m_{A,\rm Y}} \approx =2100 \times 0.06 \approx 119$.

At early times ($t \ll \tau_B \sim 12$ Myr), $x(t) \ll 1$ and the [Y/Eu] ratio is set by the prompt channel yield $r_A$ (Eq.~\ref{eq:limit_A}). At intermediate times ($\tau_B \lesssim t \lesssim \tau_C$, corresponding to $\sim 0.15$--$0.5$ Gyr), $x(t)$ saturates toward $x_\infty$ while $y(t) \approx 0$, driving [Y/Eu] toward the BNS minimum $r_B$ (Eq.~\ref{eq:limit_B}). At late times ($t \gtrsim \tau_C \sim 0.5$ Gyr), $y(t)$ rises and the AGB contribution lifts [Y/Eu] upward.

The asymptotic values are constrained by observed abundances. BNS mergers dominate Eu production at late times (contributing $\gtrsim 95\%$ of total Eu; Table~\ref{tab:Values}) implying that channel B must be the primary $r-$process source, with $M_{\rm Eu,B}(t\to \infty) / M_{\rm Eu,tot}(t\to \infty) \gtrsim 0.96$.

From Table~\ref{tab:Values}, we find that AGB stars (channel~C) contribute $\sim 84\%$ of the total Y mass at late times, with BNS mergers providing only $\lesssim 14\%$. This is consistent with the fact that Y is predominantly an s-process element ($\sim 95\%$ s-process in solar system material; \citealt{TravaglioC+04,SnedenC+08,CowanJ+21}), and validates our adopted AGB yields and DTD normalization.

Substituting Eqs.~(\ref{eq:x_powerlaw})--(\ref{eq:y_powerlaw}) into the three-channel mixing relation (Eq.~\ref{eq:threechannel_general}) yields an explicit time-dependent prediction for [Y/Eu] as a function of cosmic time, which can be mapped to [Eu/H] via the time-integrated Eu enrichment history. The location of the [Y/Eu] minimum in [Eu/H] space is determined by the condition $d[{\rm Y/Eu}]/d[{\rm Eu/H}] = 0$, which occurs when the rate of AGB Y enrichment begins to exceed the rate of BNS Y enrichment. This roughly corresponds to the time $t_{\rm min}$, where approximately,

\begin{equation}
\frac{M_{\rm Y,C}(t_{\rm min})}{M_{\rm Y,B}(t_{\rm min})} = \frac{N_C(t_{\rm min}) \cdot m_{C,\rm Y}}{N_B(t_{\rm min}) \cdot m_{B,\rm Y}} \sim 1,
\label{eq:minimum_condition}
\end{equation}

or equivalently, $y(t_{\rm min})/x(t_{\rm min}) \sim k_C/k_B \approx 127$.
This corresponds to [Eu/H] $\sim -0.2$ in our fiducial model (Fig.~\ref{fig:M_t_ABC}), in excellent agreement with the observed location of the [Y/Eu] minimum in MW disk stars at [Eu/H] $\sim -0.3$ to $0.0$ (Fig.~\ref{fig:XEu_Euh_X_Eu}).

\begin{figure*}[t]
    \centering
    \includegraphics[width=0.49\linewidth]{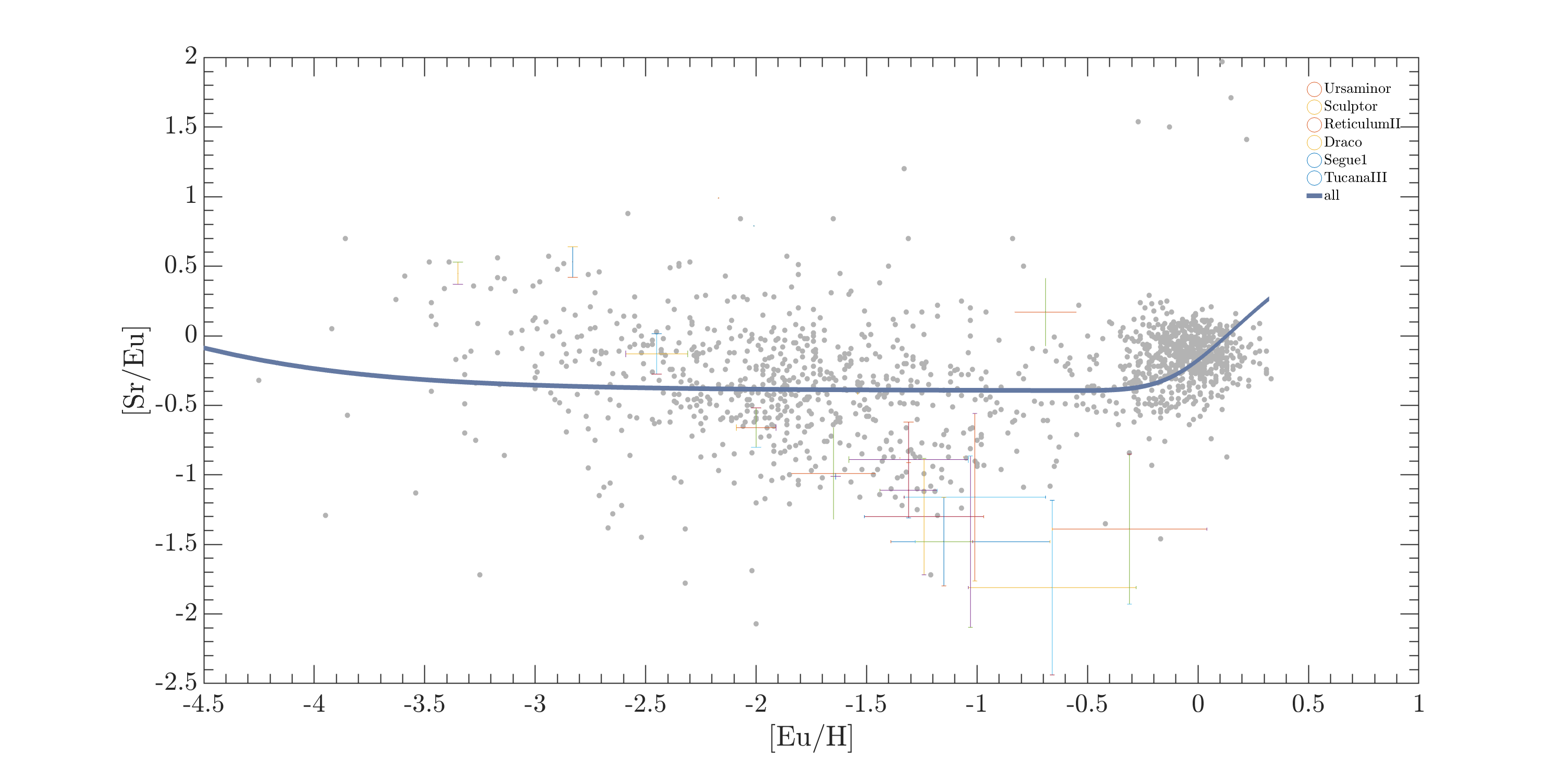}
    \includegraphics[width=0.49\linewidth]{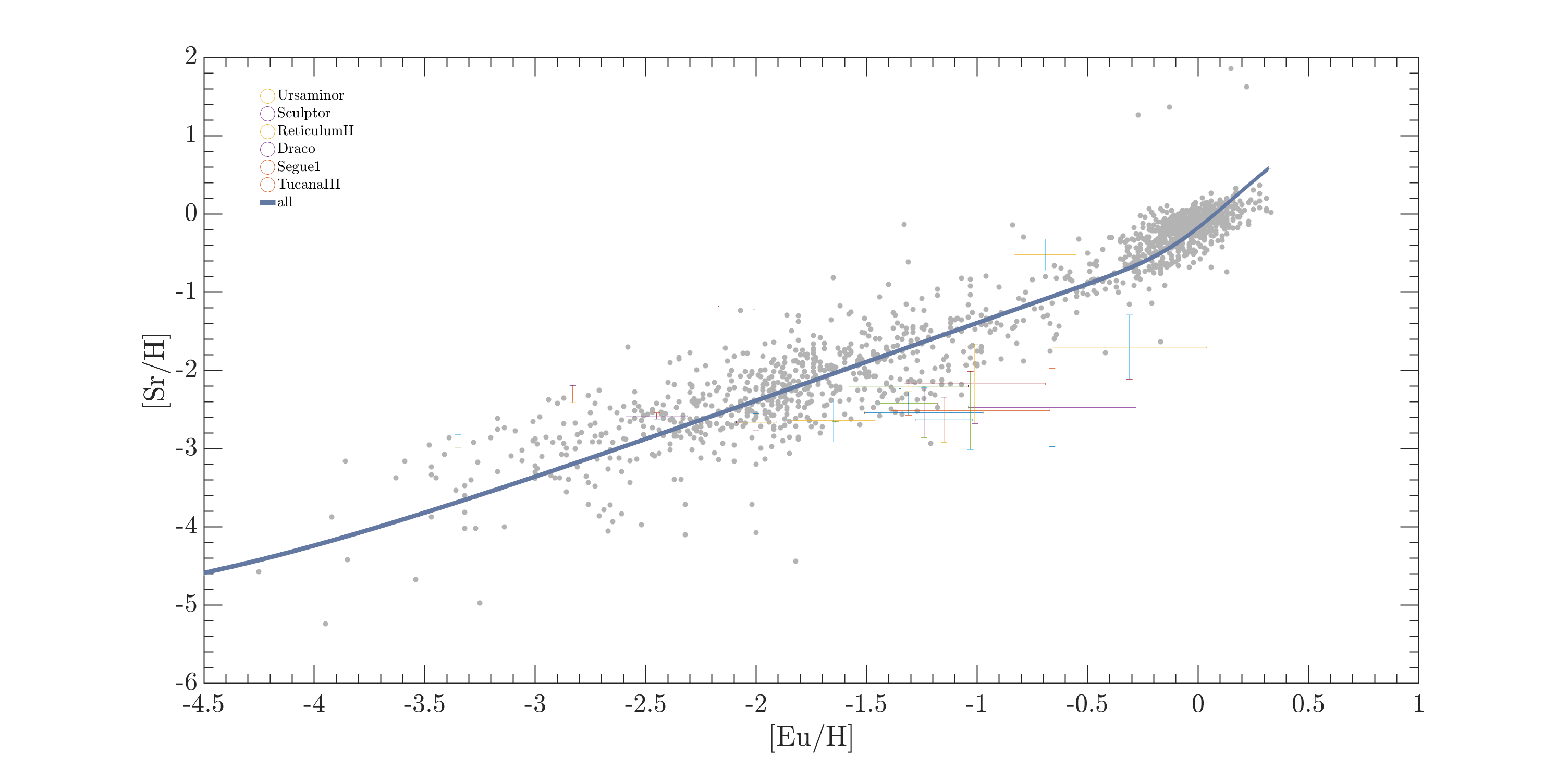}
    
    \caption{The same as Fig.~\ref{fig:XEu_Euh_X_Eu} for Sr (strontium) instead of Y (Yttrium). The evolution of both of these first peak elements is comparable (while specific ratios are slightly different).}
    \label{fig:SrEu_Euh_Sr_Eu}
\end{figure*}

\begin{figure}[h]
\centering
    \includegraphics[width=0.7\linewidth]{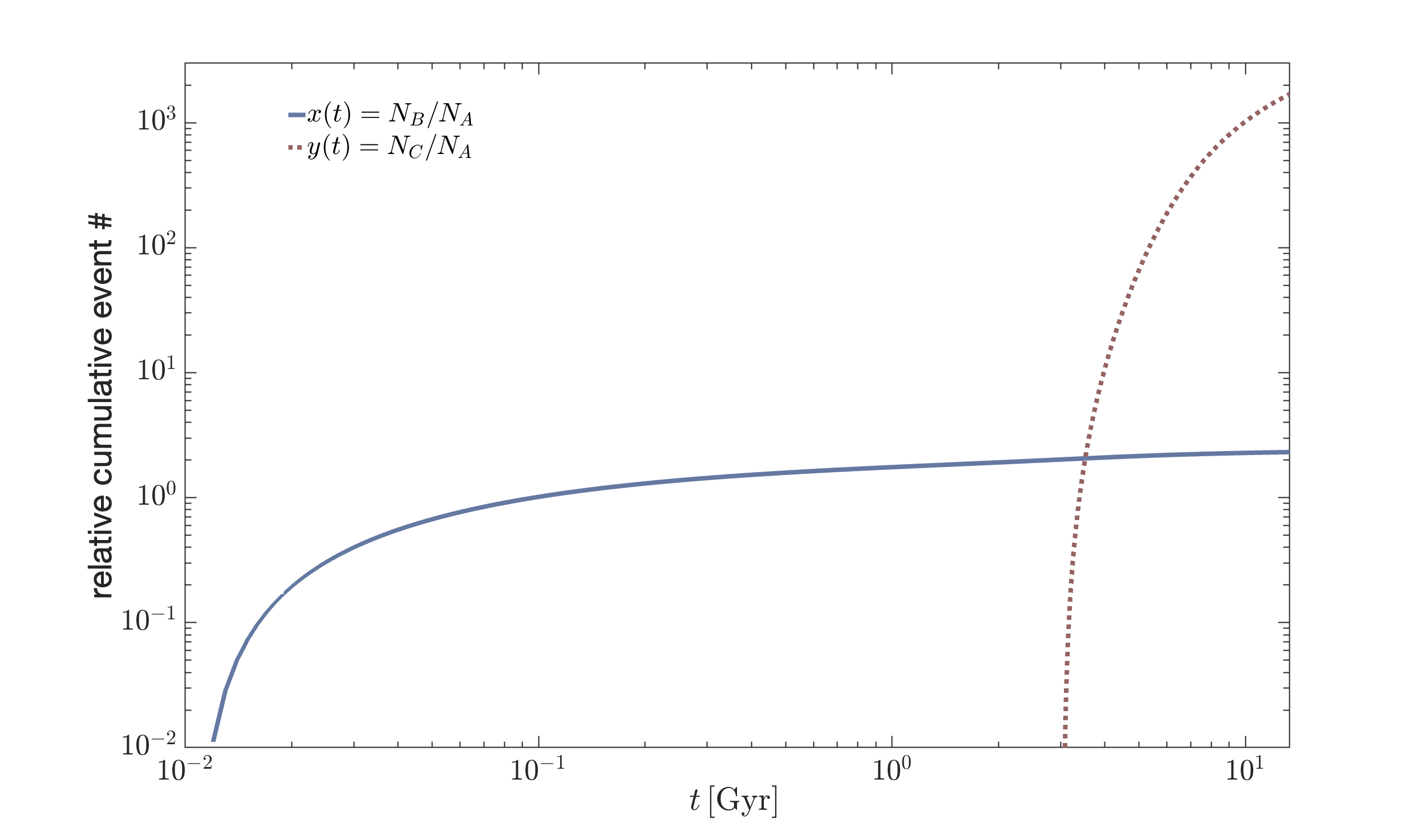}
   \caption{Time evolution of the cumulative enrichment-event number ratios, $x(t) \equiv N_B(t)/N_A(t)$ (blue solid curve) and $y(t) \equiv N_C(t)/N_A(t)$ (purple solid curve). The event ratio $x(t)$ saturates to $x_\infty \approx 2.3$ by $\sim 1$ Gyr, indicating that BNS mergers produce $\sim 2.3$ times as many events as the prompt channel per unit stellar mass. In contrast, $y(t)$ saturates to $y_\infty \approx 2100$ by $\approx 16$ Gyr, reflecting the fact that essentially all intermediate-mass stars contribute to AGB enrichment. Despite the large event number ratio, the cumulative Y mass from AGB stars is only $\sim 15$ times that from BNS mergers at late times, due to the lower per-event Y yield from AGB stars ($m_{C,\rm Y} \approx 0.11 \, m_{A,\rm Y}$ vs. $m_{B,\rm Y} \approx 11.3 \, m_{A,\rm Y}$; Table~\ref{tab:Values}). At late times, the cumulative Y masses satisfy $M_{\rm Y,C}(t\to\infty) : M_{\rm Y,B}(t\to\infty) : M_{\rm Y,A}(t\to\infty) \approx 185 : 34 : 1$, with AGB stars (channel C) contributing $\sim 86\%$ of the total Y production, consistent with Y being predominantly an s-process element.}
\label{fig:x_y}
    \label{fig:x_y}
\end{figure}

\section{Analysis of useful limits, boundaries, and rates.} \label{app:limits}

The three-channel mixing relation (Eq.~\ref{eq:threechannel_general}) can be understood through four key observational diagnostics, each constraining distinct combinations of model parameters: (i) the [Y/Eu] ratio at very low [Eu/H], controlled by the prompt channel yield ratio $r_A$; (ii) the [Y/Eu] ratio at very high [Eu/H], governed by the interplay of delayed BNS and AGB contributions; (iii) the [Eu/H] at which the minimum of [Y/Eu] occurs, set by the relative timing of channels A and B; and (iv) the value of the minimum [Y/Eu], determined primarily by the delayed channel C and the yield ratio $r_B$. We first derive these limits analytically in the general case, then evaluate them numerically using the fiducial parameters from Table~\ref{tab:Values} using the three-channel mixing relation derived in \S\ref{sec:Analyticmodel}, 
\paragraph{Low-[Eu/H] Plateau (Early-Time, Prompt Dominance)}

At early times, when channels B and C have not yet contributed significantly, $x(t) \to 0$ and $y(t) \to 0$. The $r_A$ scaled by solar abundances, $r_A = \frac{m_{Y,A}}{m_{Eu,A}}= \frac{8.58\times 10^{-5}}{6.75\times 10^{-6}} = 12.75$. In this limit, Eq.~(\ref{eq:threechannel_general}) reduces to

\begin{equation}
\left[\frac{\rm{Y}}{\rm Eu}\right]_A = \log{10}\left(\frac{n_{\rm Y,A}/n_{\rm Eu,A}}{n_{\rm Y,\odot}/n_{\rm Eu,\odot}}\right) =  \log{10}\left(\frac{12.75\times \left(152/88.9\right)}{48.9}\right)  \simeq -0.35;
\label{eq:limit_A}
\end{equation}


This sets an upper limit on the [Y/Eu] ratio observed at the lowest metallicities and provides a direct constraint on the first-to-second peak yield ratio of the prompt channel.

\begin{equation}
[\rm{Y/Eu}]_{\rm{low}} \lesssim \log{10}\left(\frac{n_{\rm Y,A}/n_{\rm Eu,A}}{n_{\rm Y,\odot}/n_{\rm Eu,\odot}}\right).
\end{equation}

This regime corresponds to the very low-$[\rm{Eu}/\rm{H}]$ tail of the evolutionary track, before the delayed BNS channel and the $s$-process channel have accumulated sufficient enrichment. In our fiducial model, this limit applies at $[\rm{Eu}/\rm{H}] \lesssim -4.8$, although the exact boundary depends on $\mathcal{C}_A$, SFH, and DTD. Observationally, the lowest-$[\rm{Eu}/\rm{H}]$ stars in ultra-faint dwarfs, regular dwarf galaxies, and the MW halo exhibit $[\rm{Y/Eu}] \sim -0.65$ to $+0.05$ at $[\rm{Eu}/\rm{H}] \lesssim -3.9$ (Fig.~\ref{fig:XEu_Euh_X_Eu}), with substantial star-to-star scatter reflecting stochastic enrichment. The upper envelope of this distribution constrains the prompt-source yield ratio, requiring $[\rm{Y/Eu}]_{\rm{low}} \gtrsim -0.65$ to $0.05$. This early-time behavior confirms that channel~A produces relatively weak first-peak enrichment compared to second-peak elements.

\paragraph{Intermediate-[Eu/H] Plateau (BNS Dominance)}

At intermediate $[\rm Eu/H]$, channel B dominates over A while C remains negligible. Here $x(t) \gg 1$ and $y(t)/x(t) \to 0$. Dividing numerator and denominator of Eq.~(\ref{eq:threechannel_general}) by $x(t)$ yields

\begin{equation}
\left[\frac{\rm{Y}}{\rm{Eu}}\right]_B = \log{10}\left(\frac{n_{\rm Y,B}/n_{\rm Eu,B}}{n_{\rm Y,\odot}/n_{\rm Eu,\odot}}\right) = \log{10}\left(\frac{4.8\times (152/89.8)}{48.9}\right) \simeq -0.8,
\label{eq:limit_B}
\end{equation}

which governs the [Y/Eu] plateau (minimum), $[\rm{Y/Eu}]_{\rm{min}} = \log{10}\left(\frac{n_{\rm Y,B}/n_{\rm Eu,B}}{n_{\rm Y,\odot}/n_{\rm Eu,\odot}}\right)$. Observationally, $[\rm{Y/Eu}] \sim -0.3$ to $-1.0$ at $[\rm Eu/H] \sim -1.5$ to $-0.25$ requires $r_B \sim 2.8$--$9.2$, consistent with lanthanide-rich BNS merger ejecta. 

Consequentiality, $\Delta[\rm{Y}/\rm{Eu}]$ provides us with the difference in first/second peak production by channel~A vs B. The location of the plateau gives us the relative delay between channel B and A. Observational excursions of order $\Delta[\rm{Y/Eu}] \sim 0.5$--$0.88$ dex, requiring

\begin{equation}
\frac{r_A}{r_B} \gtrsim 10^{0.6}\sim 4.
\label{eq:ratio_constraint}
\end{equation}

This constraint is independent of DTD normalizations and depends only on the intrinsic yield ratios. Models in which both channels have similar [Y/Eu] ratios ($r_A \sim r_B$) cannot reproduce the observed [Y/Eu] variation.

While the mean depth of the plateau is set by $r_B$, its location in terms of $[\rm Eu/H]$ depends on the relative timing of channel activation. The decline from the early plateau begins at $[\rm Eu/H] \lesssim -5$ when channel~B becomes comparable to A,

\begin{equation}
\frac{x(t)}{k_B}\sim 1,
\label{eq:decline_condition}
\end{equation}

This transition is controlled by the BNS DTD and the ratio $N_B/N_A$. Steeper BNS DTDs shift the decline onset to lower $[\rm Eu/H]$, while higher $N_B/N_A$ shifts it to higher $[\rm Eu/H]$ (Fig.~\ref{fig:XEu_Euh_X_Eu}, upper right panel). Beyond this point, channel B dominates both Y and Eu production ($x(t) \gg 1$).
The late rise at $[\rm Eu/H] \gtrsim -0.2$ begins when channel~C becomes comparable to B in first-peak production,

\begin{equation}
 \frac{N_C(t_{\rm min}) \, m_{C,{\rm Y}}}{N_B(t_{\rm min}) \, m_{B,{\rm Y}}} \approx 1,
\label{eq:rise_condition_alt}
\end{equation}

The $[\rm Eu/H]$ at which this occurs constrains the AGB onset time and normalization $N_C/N_B$ (Table~\ref{tab:Values} and \S \ref{app:rates}).




\paragraph{High-$[\rm{Eu}/\rm{H}]$ behavior late-time rise from $s$-process Y,\& Sr.}

At the late-time limit channel C can dominate the Y (Sr) budget while channel B still dominates the Eu budget. The relevant inequalities are therefore

\begin{equation}
\frac{y(t)}{k_C} \gg max \left(1,\ \frac{x(t)}{k_B}\right),
\qquad\text{and}\qquad
\frac{x(t)}{k_B r_B} \gg max \left(\frac{1}{r_A},\ \frac{y(t)}{k_C r_C}\right).
\label{eq:late_time_conditions}
\end{equation}

In this regime, Equation~(\ref{eq:threechannel_general}) together using the definitions of $k_B$, $k_C$, and $r_B$, this be written as,

\begin{equation}
\frac{\rm{M_Y}}{\rm{M_{Eu}}}|_{\rm late}
\simeq \frac{y(t)/k_C}{x(t)/(k_B r_B)} \simeq\frac{y(t)}{x(t)}\,\frac{k_B r_B}{k_C} \simeq
\frac{N_C(t)}{N_B(t)}\,\frac{m_{C,\rm Y}}{m_{B,\rm Eu}} \approx \frac{y_{\infty}}{x_{\infty}}\,\frac{m_{C,\rm Y}}{m_{B,\rm Eu}},
\label{eq:late_time_limit_alt}
\end{equation}

i.e.\ the late-time abundance ratio is controlled by the cumulative buildup of channel~C events relative to channel~B events, multiplied by the Y yield of channel~C relative to the Eu yield of channel~B. At the highest metallicities in our sample ($[\rm{Eu/H}] \gtrsim -0.3$, corresponding to $t \gtrsim 3.2$ Gyr; Figs.~\ref{fig:M_t_ABC} \& \ref{fig:XEu_Euh_X_Eu}), the observed [Y/Eu] ratios span approximately $-0.2$ to $+0.4$ (Fig.~\ref{fig:Obs_SrYEu_D_MW} \& \ref{fig:XEu_Euh_X_Eu}), with our fiducial model predicting $[{\rm Y}/{\rm Eu}]_{\rm late} \approx +0.15$ at solar [Eu/H]. This asymptotic value reflects the dominance of AGB stars (channel~C) in Y production and BNS mergers (channel~B) in Eu production at late times.

\begin{equation}
\frac{M_{\rm Y}}{M_{\rm Eu}}|_{\rm late} 
\approx \frac{y_{\infty}}{x_{\infty}}\,\frac{m_{B,\rm Y}}{m_{C,\rm Y}} \cdot r_B
= \frac{y_{\infty}}{x_{\infty}}\,\frac{k_C}{k_B}, \longrightarrow \left[\frac{\rm Y}{\rm Eu}\right]_{\rm late} \approx \log{10}\left(\frac{36\times(152/88.9)}{48.9}\right) \simeq +0.11
\end{equation}

where this represents the late-time Y/Eu mass ratio in the ISM, thus, the Converting to bracket notation yields basically, $[{\rm Y}/{\rm Eu}]_{\rm late} \approx +0.15$ (Fig.~\ref{fig:Obs_SrYEu_D_MW}), in good agreement with observations at solar [Eu/H]. 

The late-time rise is not determined solely by $r_C$, but rather by the interplay between the cumulative AGB event rate, the BNS event rate, the relative yield normalizations ($k_C/k_B$), and the Y/Eu yield ratio from BNS mergers ($r_B$). Only if $N_C/N_B$ asymptotes to a constant does one recover an approximately constant late-time plateau. In the more general case, the high-$[\rm{Eu}/\rm{H}]$ evolution remains time-dependent and reflects the gradual onset of $s$-process material. Thus, the observed high-$[\rm{Eu}/\rm{H}]$ upturn in $[\rm{Y}/\rm{Eu}]$ should be interpreted as a consequence of increasing channel~C Y, Sr enrichment superposed on a Eu budget that is still largely controlled by channel~B. At late times, cumulative AGB enrichment contributes $\gtrsim 86\%$ of the total Y production (Table~\ref{tab:Values}).


\begin{table*}
    \centering
    \begin{tabular}{lc}
        \hline
        \hline
        Component & Value \\
        \hline
        $m_{A,\rm Sr}$ (1st peak) & $3.93\times 10^{-5}M_\odot$ \\
        $m_{A, \rm Y}$ (1st peak) & $8.58\times 10^{-5}M_\odot $ \\
        $m_{A,\rm Ba}$ (2nd peak) & $2.32\times 10^{-5}M_\odot$ \\
        $m_{A,\rm Eu}$ (2nd peak) & $6.75\times 10^{-6}M_\odot$ \\
        $m_{B,\rm Sr}$ (1st peak) & $2.84\times 10^{-4}M_\odot$ \\
        $m_{B,\rm Y}$ (1st peak) & $9.71\times 10^{-4}M_\odot$ \\
        $m_{B,\rm Ba}$ (2nd peak) & $8.09\times 10^{-4}M_\odot$ \\
        $m_{B,\rm Eu}$ (2nd peak) & $2.17\times 10^{-4}M_\odot $ \\
        $\mathcal{C}_{\rm B}=\mathcal{C}_{\rm BNS}$ & $7\times 10^{-6}$ per $M_\odot$; ($90$ Gpc$^{-3}$ yr$^{-1}$) \\
        $\mathcal{C}_{\rm A}$ & $3\times 10^{-6}$ per $M_\odot$; ($42$ Gpc$^{-3}$ yr$^{-1}$) \\
        $M_{\rm Sr,X}(X=A,B,C)$ & ($3.12\times 10^{-1}$, 6.11, $2.41\times 10^1$)$M_\odot$ \\
        $M_{\rm Y,X}(X=A, B, C)$ & ($3.94\times 10^{-2}$, $8.72\times 10^{-1}$, $4.72$)$M_\odot$ \\
        $M_{\rm Ba,X}(X=A, B, C)$ & ($1.69\times 10^{-2}$, $2.79$, $4.54$)$M_\odot$ \\
        $M_{\rm Eu,X}(X=A, B, C)$ & ($1.03\times 10^{-3}$, $6.66\times 10^{-2}$, $2.01\times 10^{-3}$) $M_\odot$\\
        $\frac{\chi}{(r+s)}|_{Sr} (A, B, C)$ & ($1.02\times10^{-2}$, $1.99\times10^{-1}$, 0.81)\\
        $\frac{\chi}{(r+s)}|_{Y} (A, B, C$ & ($4.54\times10^{-3}$, $1.55\times10^{-1}$, 0.86)\\
        $\frac{\chi}{(r+s)}|_{Ba} (A, B, C)$ & ($2.21\times10^{-3}$, $3.78\times10^{-1}$, 0.63)\\
        $\frac{\chi}{(r+s)}|_{Eu} (A, B, C)$ & ($1.31\times10^{-2}$, $9.56\times10^{-1}$, 0.030)\\
        $b_{\rm B}$ & $-1.5$ \\
        $t_{\rm min, B}$ & 12 Myr \\
        $t_{\rm min, C}$--$t_{\rm max, C}$ & 0.3--12 Gyr \\
        $A_{\rm Sr,Y, Ba, Eu}$ & $88, 89, 137, 152 $ \\ 
        \hline
    \end{tabular}
    \caption{Fiducial model parameters. Channel A is a prompt collapsar-like source; channel B represents delayed BNS mergers.}
    \label{tab:Values}
\end{table*}

\begin{figure}[h]
    \centering
    \includegraphics[width=\linewidth]{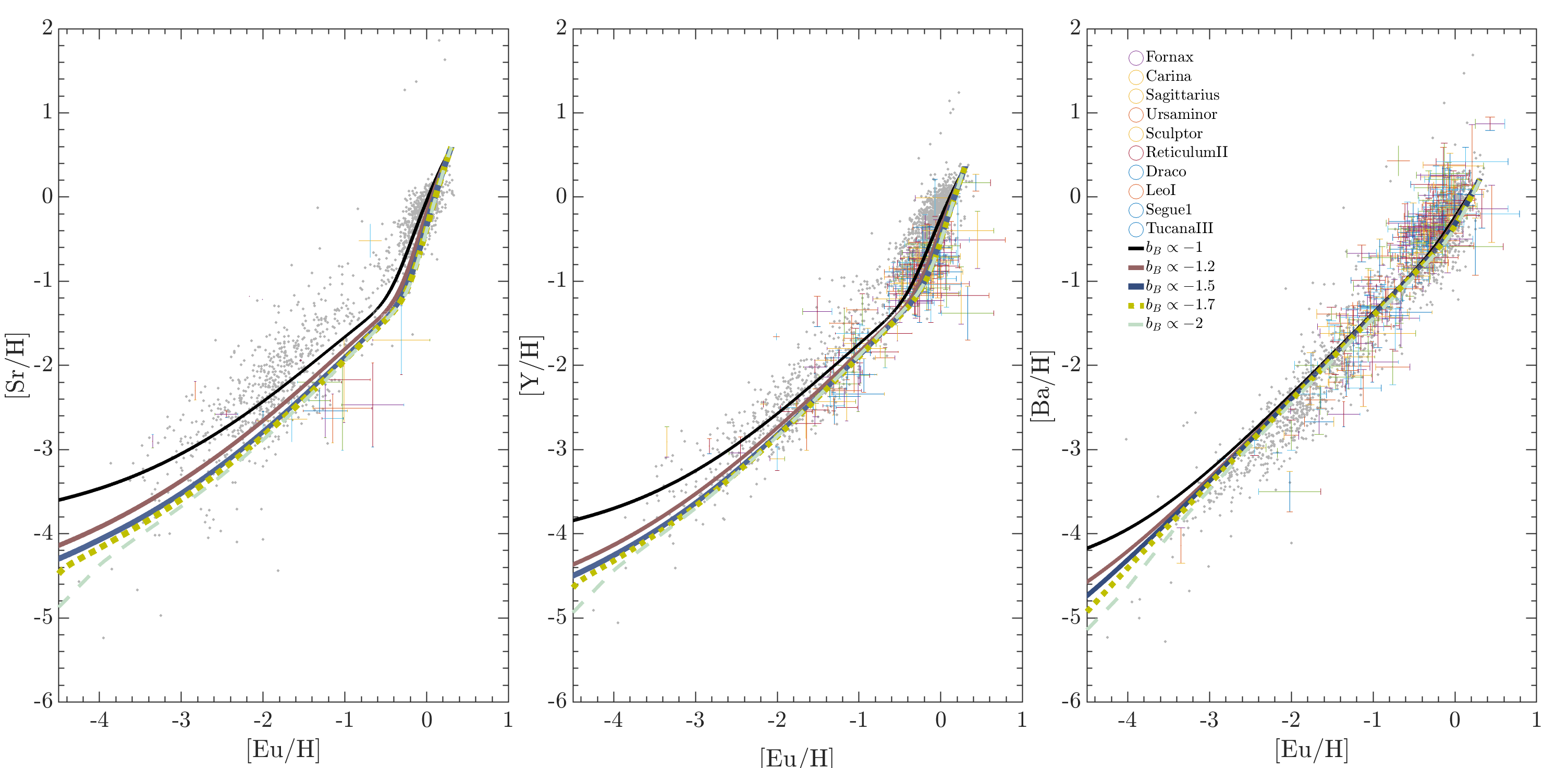}
    \caption{[Sr/H], [Y/H], [Ba/H] versus [Eu/H] for BNS DTD power-law slopes $b_{\rm B} = -1.0, -1.2, -1.5, -1.7, -2.0$ with fixed channel normalization ratio $\mathcal{C}_{\rm B}/\mathcal{C}_{\rm A} = 2.33$.}
    \label{fig:SrYBa_H_3dtd}
\end{figure}


\begin{figure*}[h]
    \centering
    \includegraphics[width=0.49\linewidth]{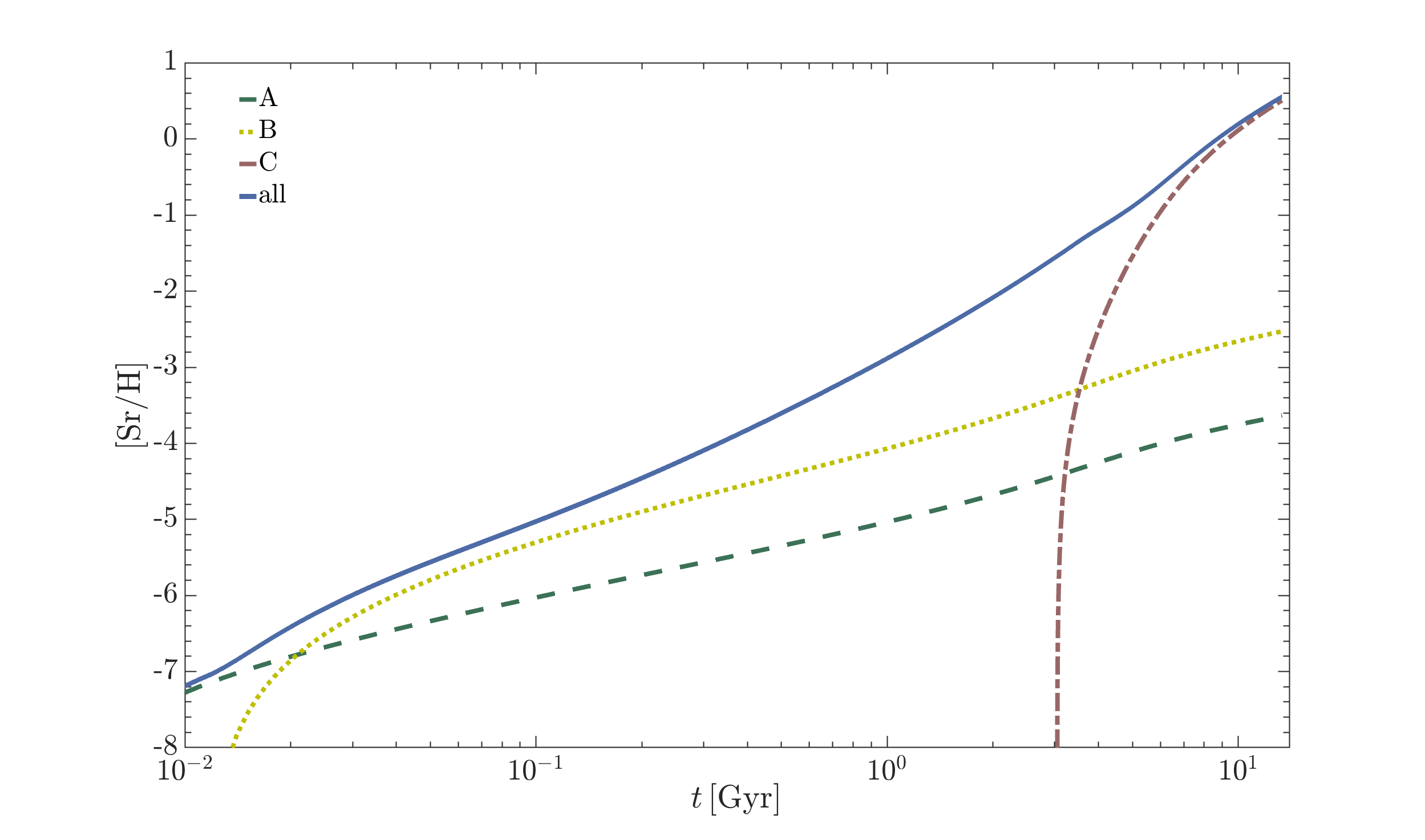}
    \includegraphics[width=0.49\linewidth]{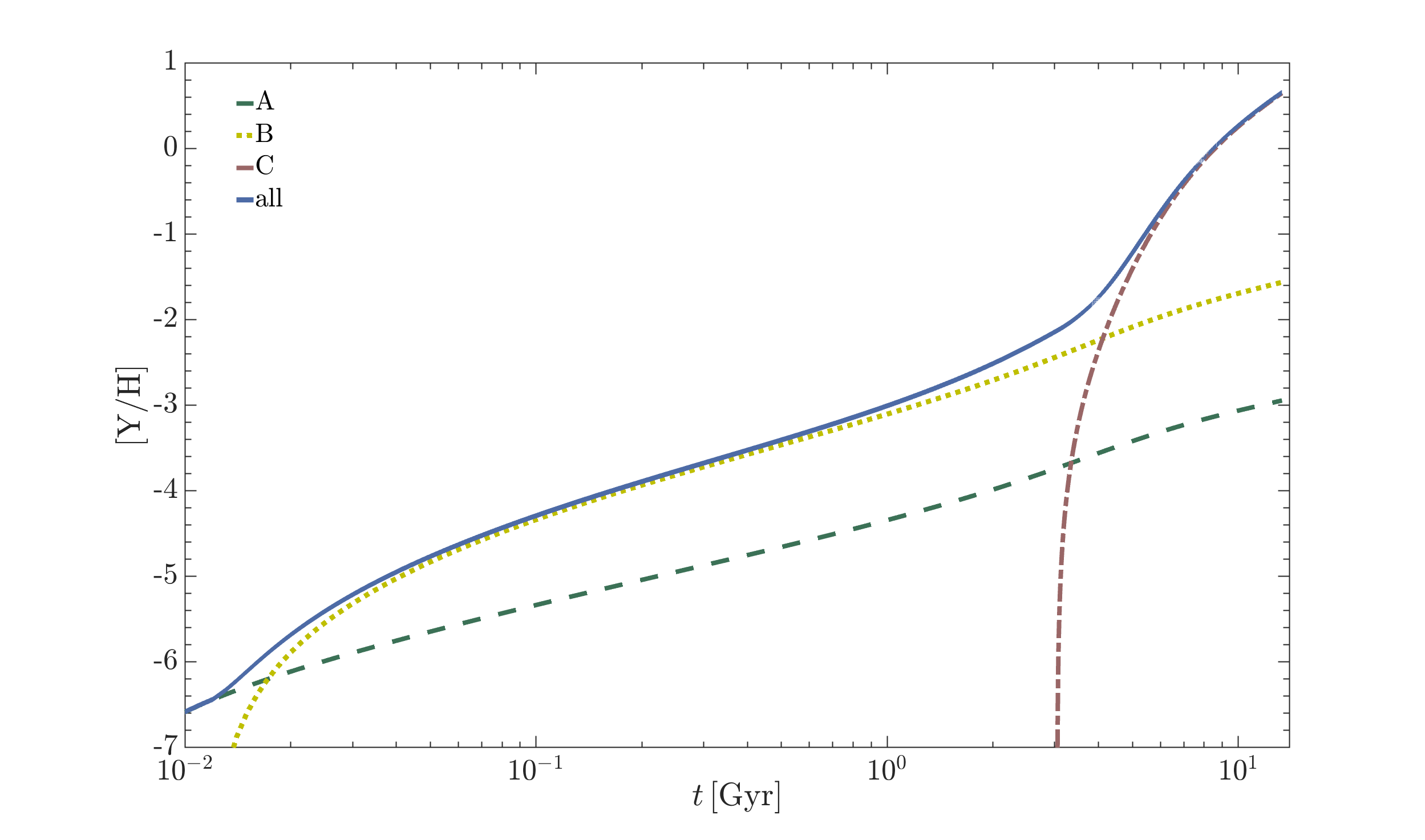}
    \includegraphics[width=0.49\linewidth]{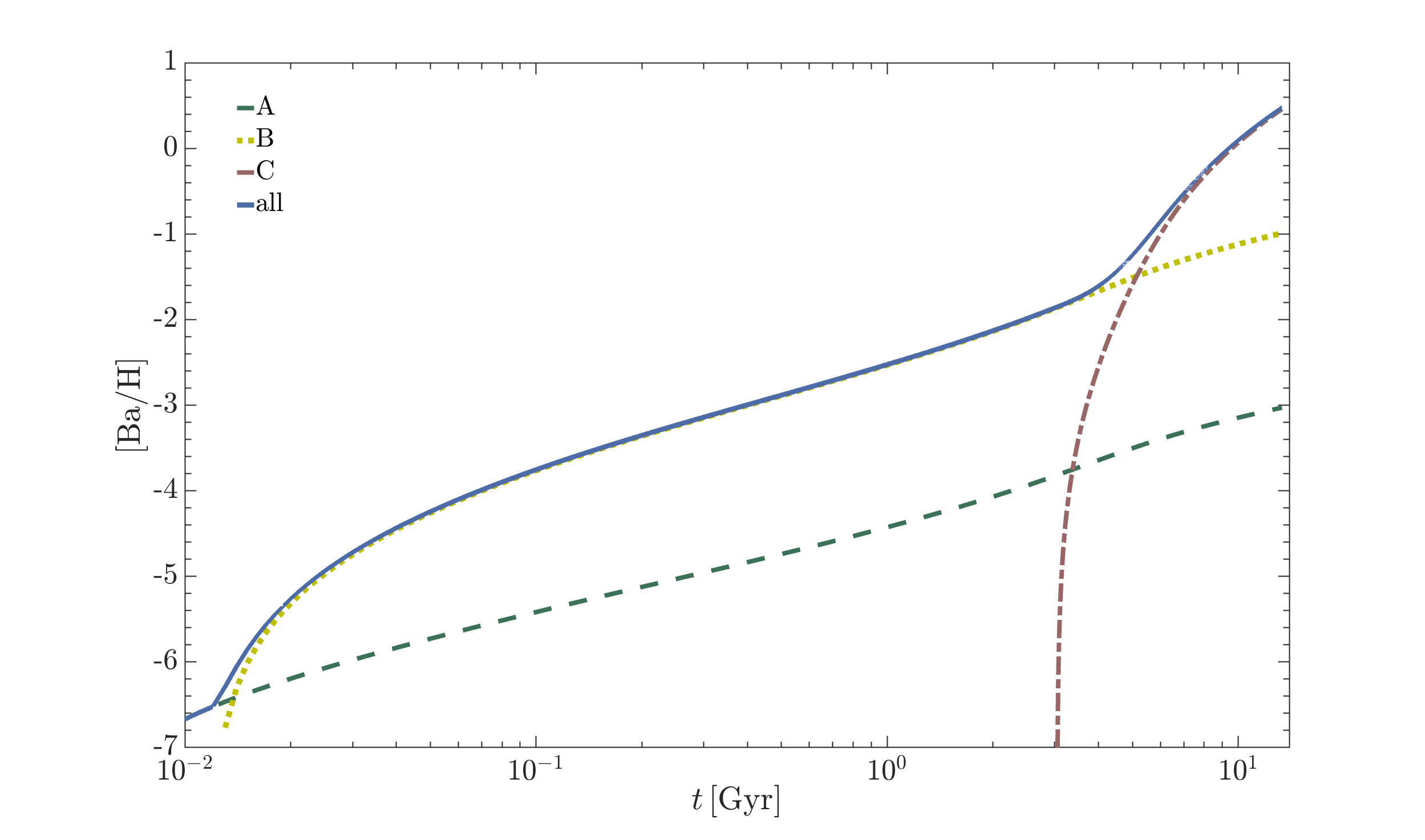}
    \includegraphics[width=0.49\linewidth]{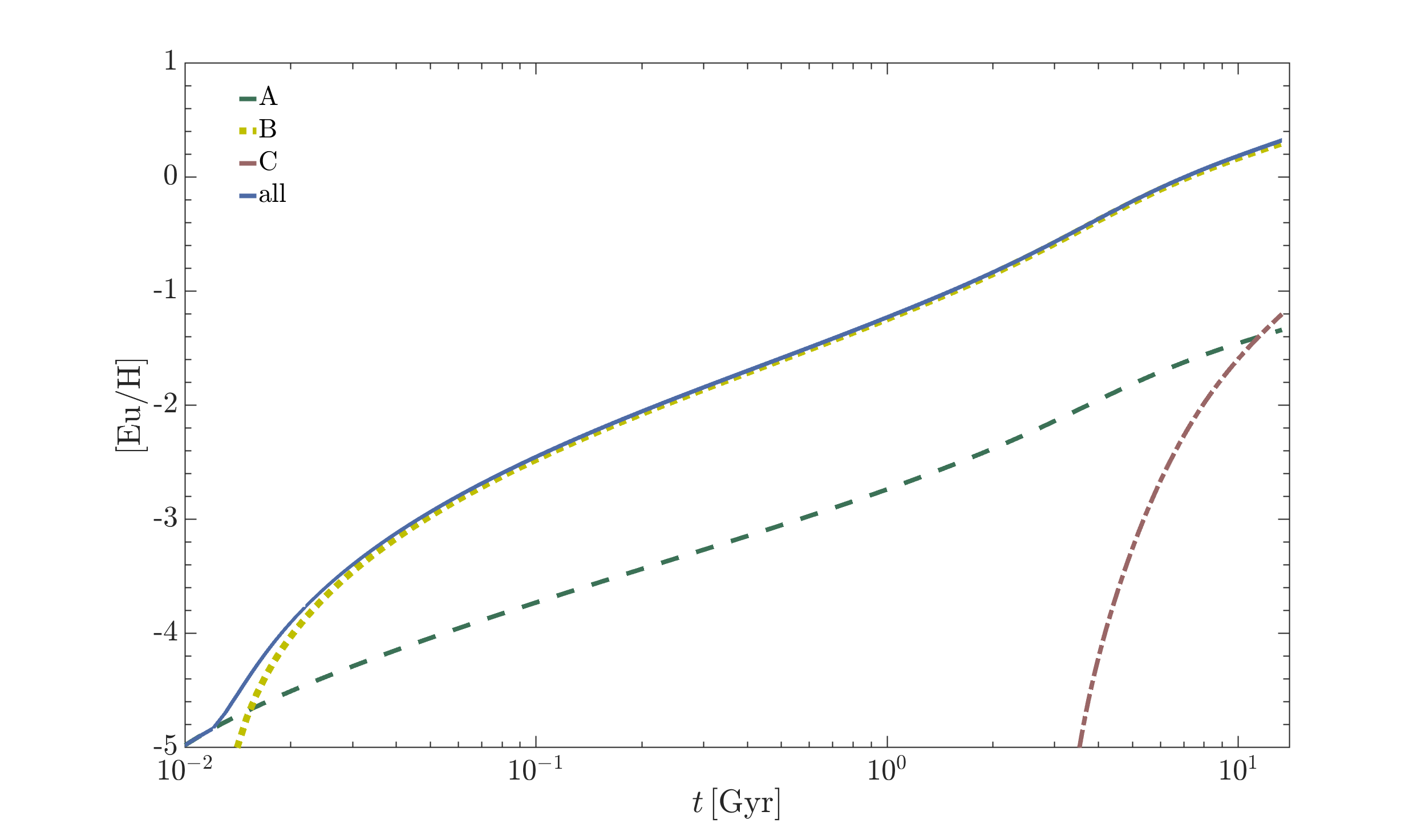}
    \caption{Time evolution of the modeled $\rm{[X/H]}$ abundance as a function of time, where $\mathrm{X}=\mathrm{Sr}, \mathrm{Y}, \mathrm{Ba}, \mathrm{Eu}$. The individual curves show the contributions from channel A, channel B, and channel C, as well as the total.}
    \label{fig:M_t_ABC}
\end{figure*}

\section{Sensitivity to BNS delay-time distribution}
\label{app:dtd_sensitivity}

Figure~\ref{fig:SrYBa_H_3dtd} explores how varying the BNS DTD power-law slope affects predicted abundance evolution. We compare slopes $b_{\rm B} = -1.0, -1.2, -1.5, -1.7, -2.0$ while holding the channel normalization ratio fixed at $\mathcal{C}_{\rm B}/\mathcal{C}_{\rm A} = 2.33$. For all three elements, the [X/H]--[Eu/H] tracks remain nearly parallel with offsets $\lesssim 0.2$ dex. Ba correlates most tightly with Eu, consistent with both being second-peak $r$-process tracers, while Sr and Y there is a single value of [Sr/H] at a given [Eu/H] reflecting additional $s$-process contributions. Steepening the BNS DTD accelerates evolution along a fixed abundance locus rather than altering its shape, demonstrating that Sr--Y--Ba versus Eu trends constrain channel yield ratios.

\section{Prompt versus delayed $r$-process sources}
\label{app:theory}

Multiple chemical evolution studies \citep{Hotokezaka+18, CoteB+19_NotBNS, Beniamini_Tsvi19, KobayashiC+23, Maoz_Nakar25, Molero2025} confirm that reproducing the [Eu/Fe] knee near [Fe/H] $\sim -1$ requires a BNS DTD steeper than $t^{-1.5}$. This has motivated proposals for additional prompt $r$-process channels including collapsars \citep{Siegel+19_Nat, Gottlieb2025}, magnetorotational supernovae \citep{MostaP+18}, and magnetar-powered explosions \citep{MetzgerB+08, ThompsonT_udDoula18}, as well as environmental modifications such as natal kicks \citep{BanerjeeP+20, vande_Voort+22} or inhomogeneous ISM mixing \citep{2015NatPh..11.1042H, Beniamini_Hotokezaka20}.

Binary evolution models predict merger delay times $\tau_{\rm GW} \propto a^4 (1-e^2)^{7/2}$ \citep{Peters1964}. If initial separations follow $dN/da \propto a^{-\alpha}$ then $dN/dt\propto t^{-(\alpha+3)/4}$, so that $t^{-1.5}$ means $\alpha=3$. Obtaining $b_{\rm BNS} > -1.5$ requires $\alpha \gtrsim 6$, significantly steeper than the canonical Öpik distribution ($\alpha \simeq 1$) and inconsistent with observed Galactic BNS demographics \citep{Beniamini_Tsvi19}. 
Natal kicks produce a tail $dN/dt \propto t^{-5/7}$ at short delays, below $t_{\rm min}$ \citep{Beniamini_Piran24} but cannot steepen the DTD at $t \gtrsim 100$ Myr, as needed in order to form a [Eu/Fe] knee rather than a flattening at large [Fe/H].

\begin{figure*}[t]
    \centering
    \includegraphics[width=0.49\linewidth]{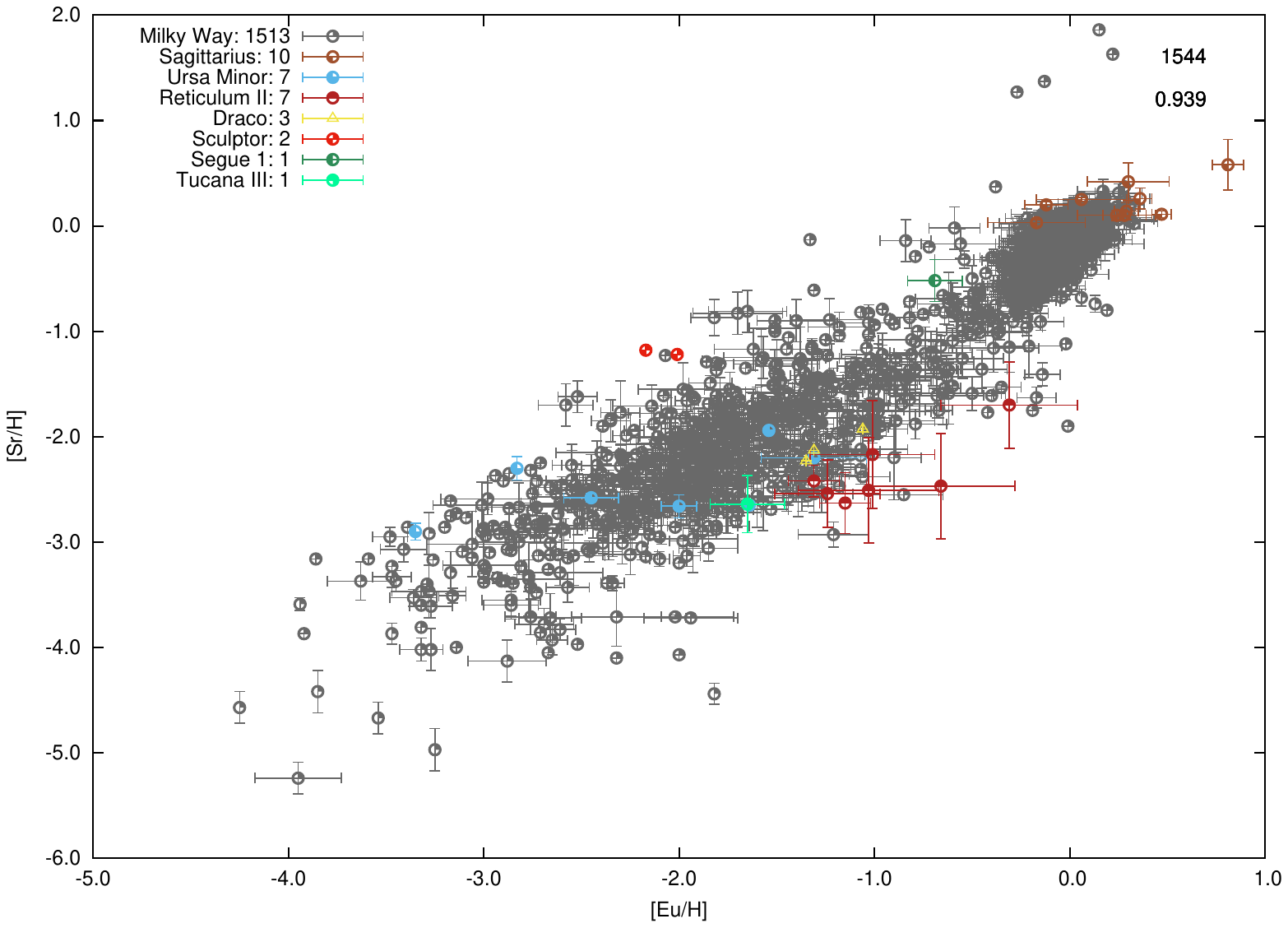}
    \includegraphics[width=0.49\linewidth]{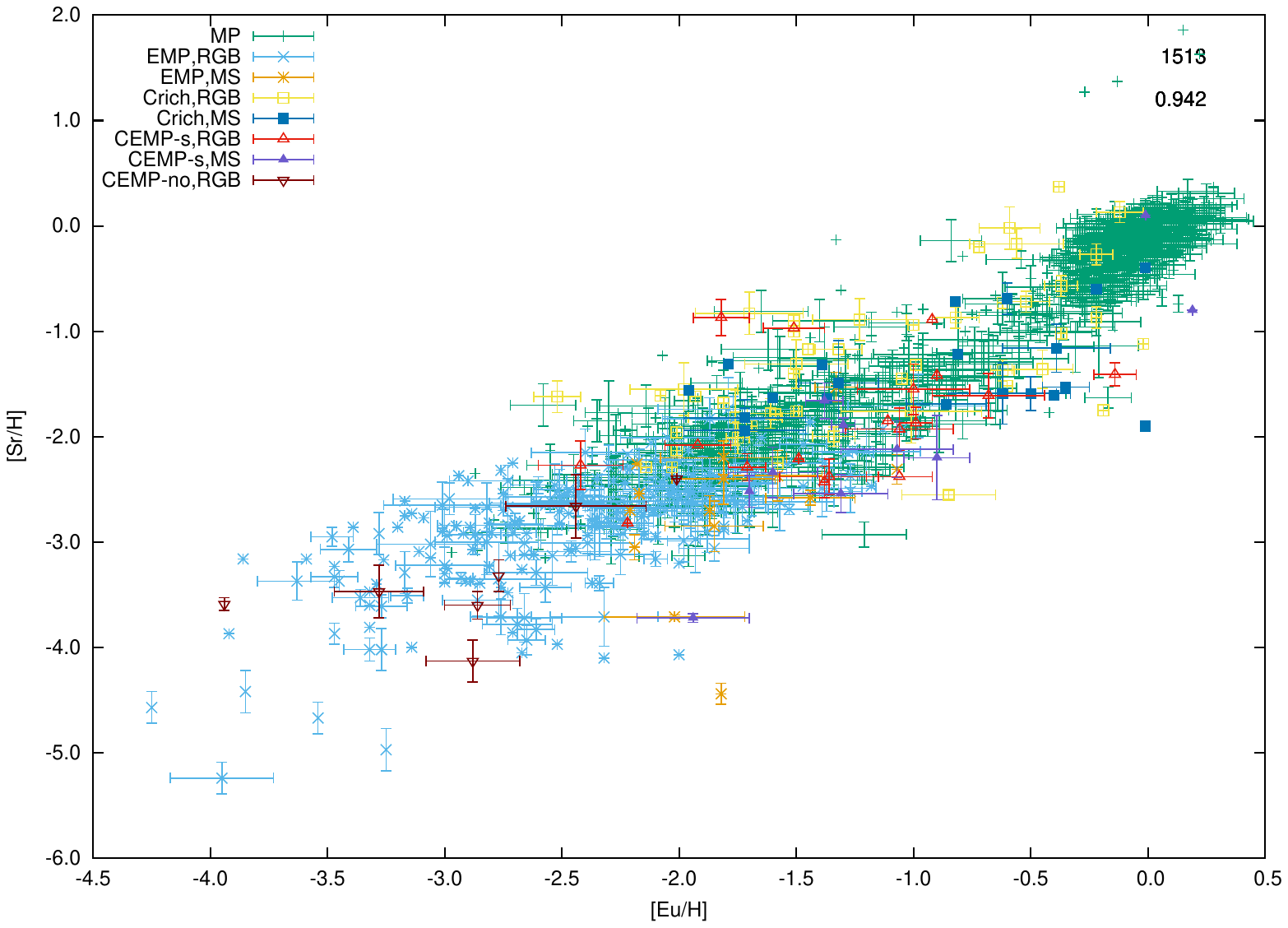}
    \includegraphics[width=0.49\linewidth]{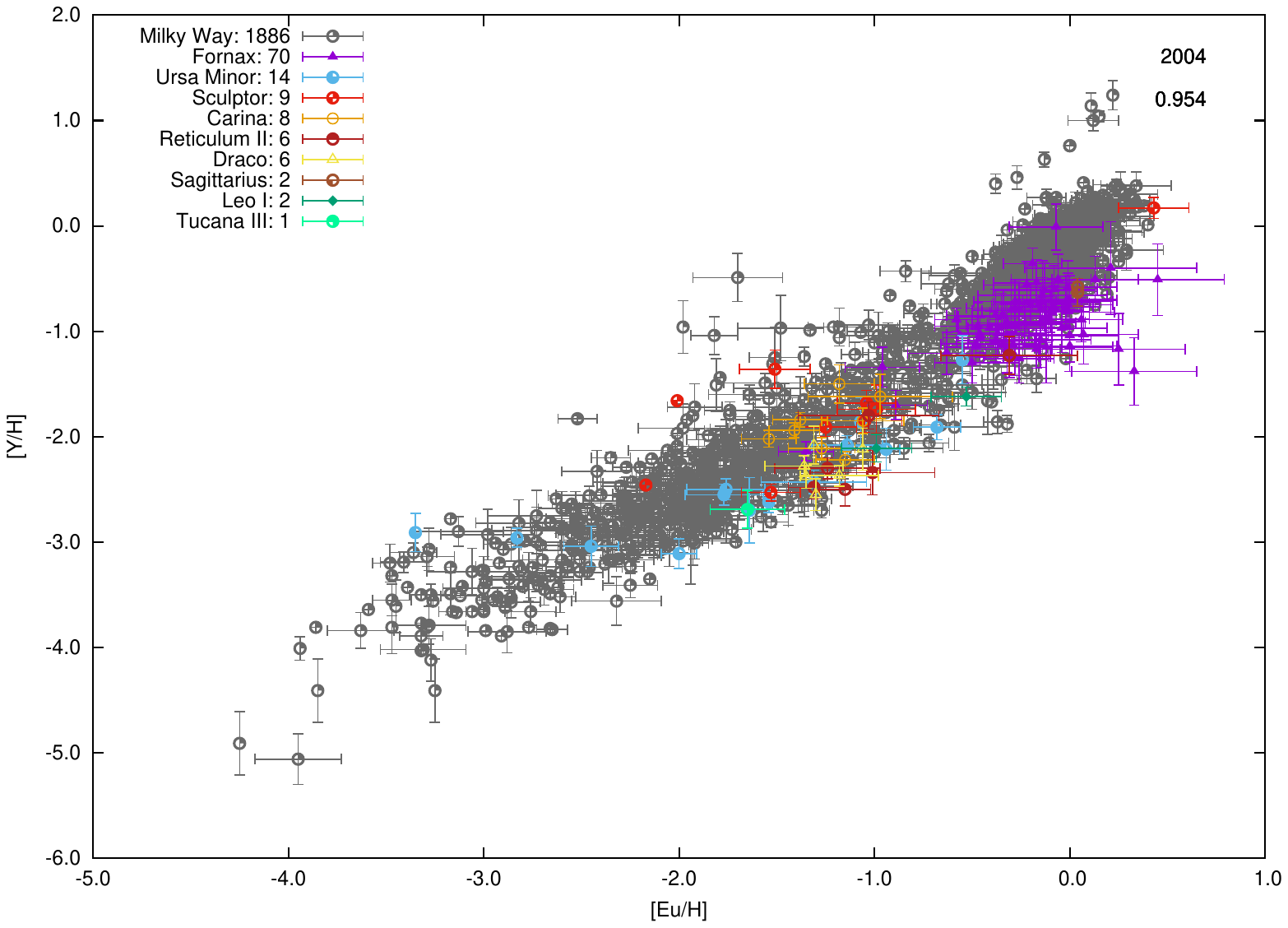}
    \includegraphics[width=0.49\linewidth]{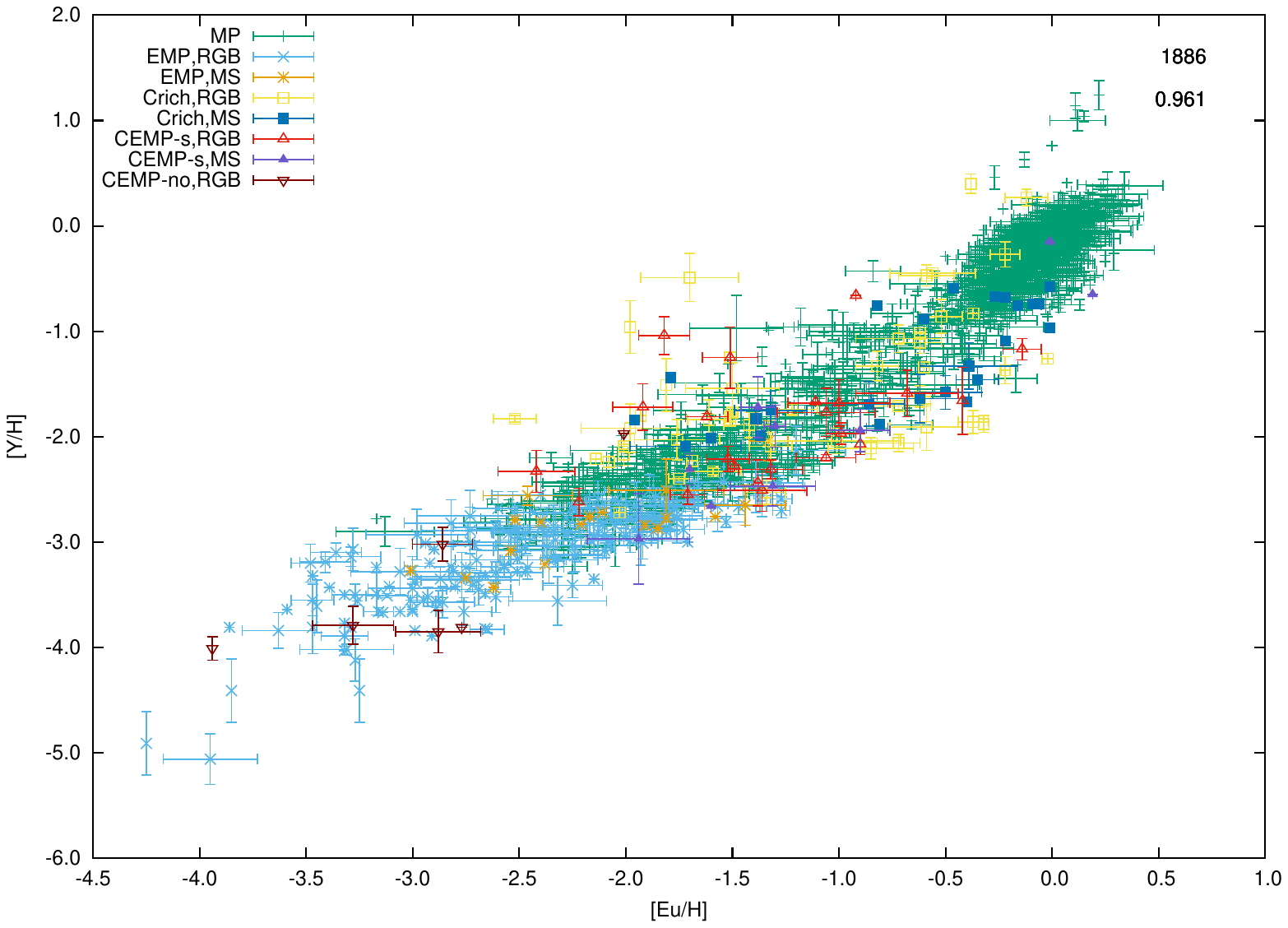}
    \includegraphics[width=0.49\linewidth]{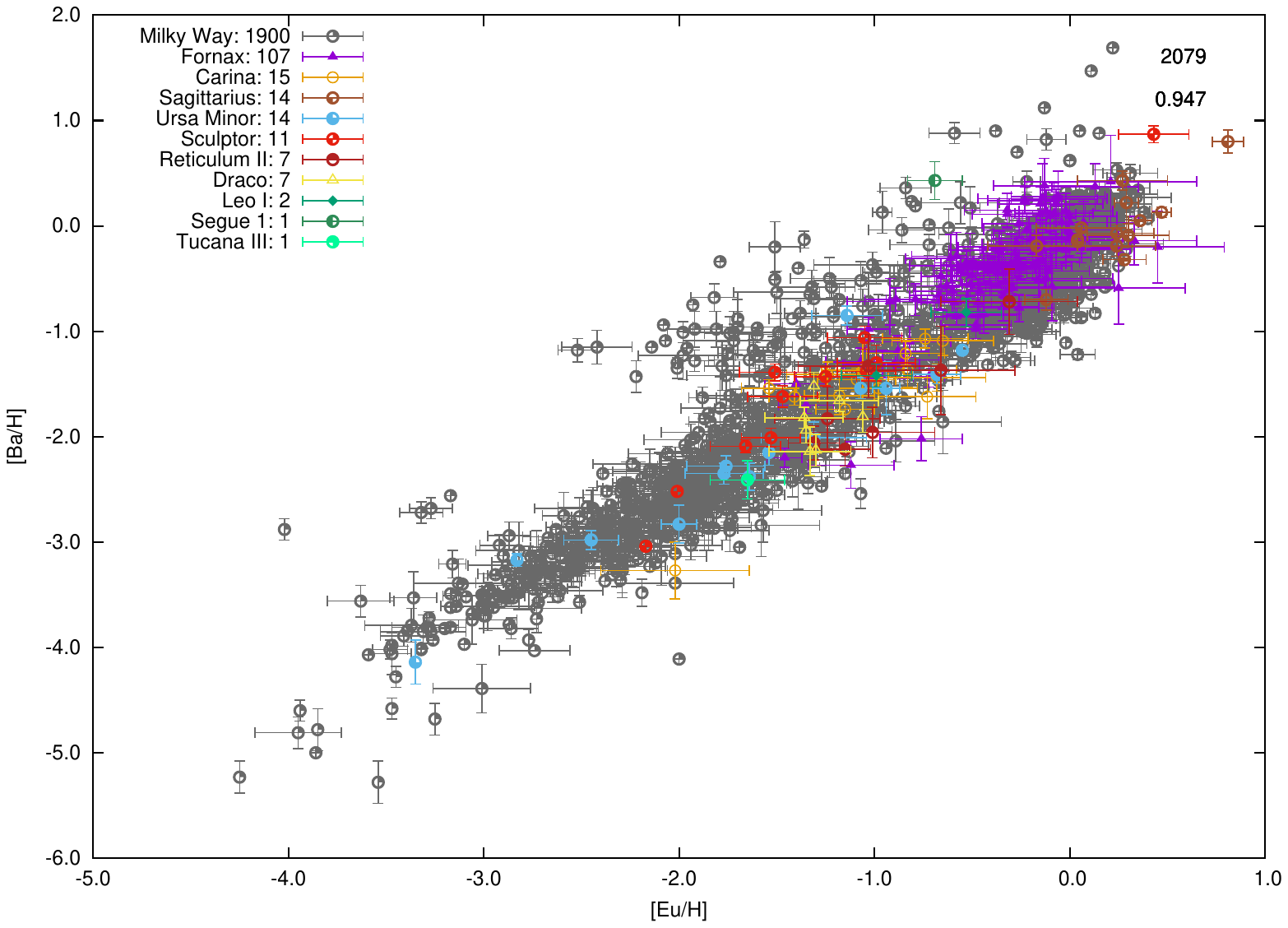}
    \includegraphics[width=0.49\linewidth]{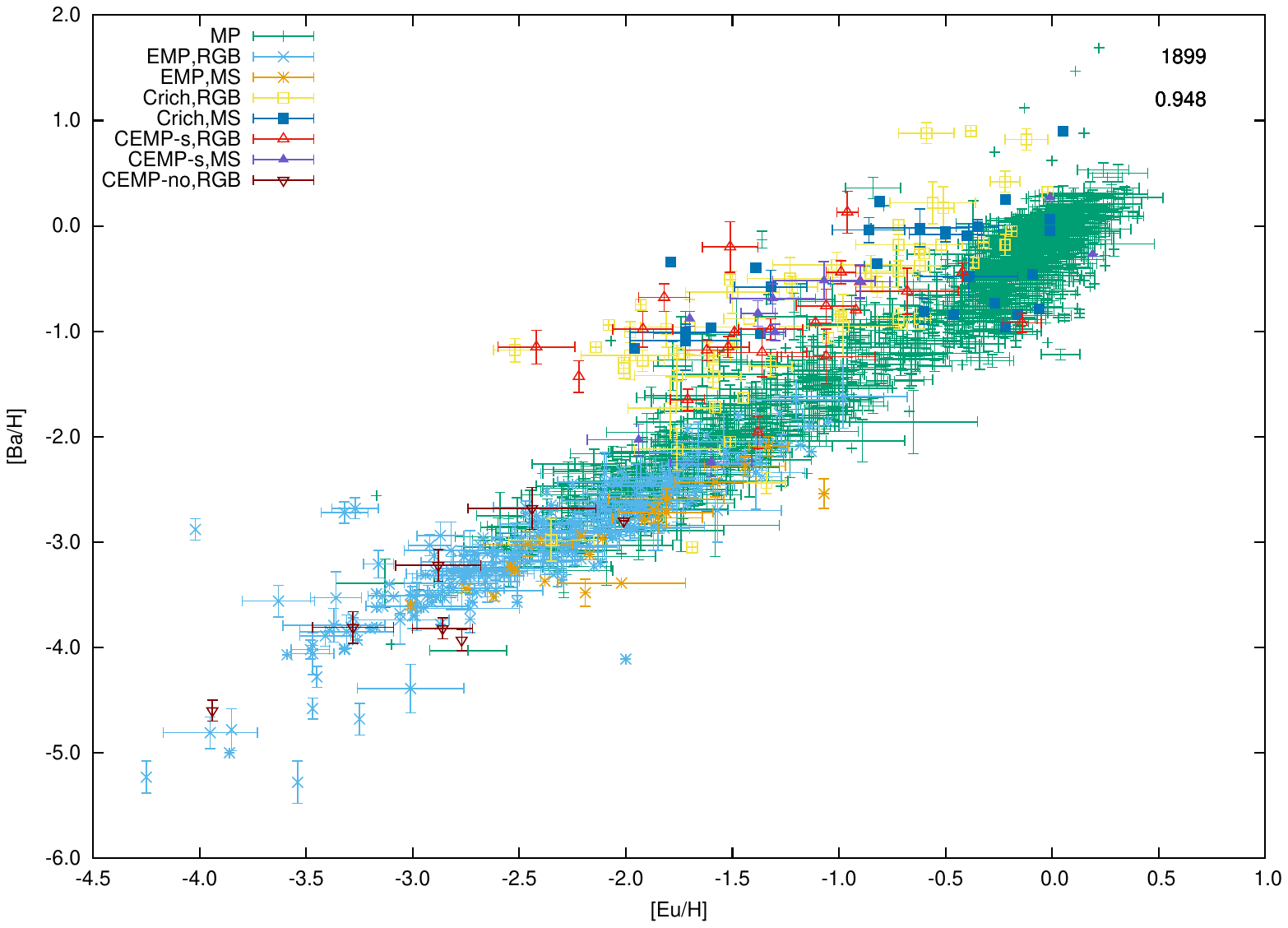}
    \caption{Observed abundance ratios from SAGA database \citep{SudaT+08_SAGA,SudaT+11_SAGA,SudaT+17_SAGA}. Left: dwarf galaxies (colored points); middle: MW (gray points); right: [X/Eu] versus [Eu/H] for all data. Top row: [Sr/Eu]; middle row: [Y/Eu]; bottom row: [Ba/Eu].}
    \label{fig:Obs_SrYEu_D_MW}
\end{figure*}

\bibliography{rprocess}

@ARTICLE{Surman2008ApJ,
       author = {{Surman}, R. and {McLaughlin}, G.~C. and {Ruffert}, M. and {Janka}, H. -Th. and {Hix}, W.~R.},
        title = "{r-Process Nucleosynthesis in Hot Accretion Disk Flows from Black Hole-Neutron Star Mergers}",
      journal = {\apjl},
         year = 2008,
        month = jun,
       volume = {679},
       number = {2},
        pages = {L117},
          doi = {10.1086/589507},
archivePrefix = {arXiv},
       eprint = {0803.1785},
 primaryClass = {astro-ph},
       adsurl = {https://ui.adsabs.harvard.edu/abs/2008ApJ...679L.117S}
}

@ARTICLE{Bartos2019Natur,
       author = {{Bartos}, Imre and {Marka}, Szabolcs},
        title = "{A nearby neutron-star merger explains the actinide abundances in the early Solar System}",
      journal = {\nat},
         year = 2019,
        month = may,
       volume = {569},
       number = {7754},
        pages = {85-88},
          doi = {10.1038/s41586-019-1113-7},
       adsurl = {https://ui.adsabs.harvard.edu/abs/2019Natur.569...85B}
}

@ARTICLE{Cote2019ApJ,
       author = {{C{\^o}t{\'e}}, Benoit and {Lugaro}, Maria and {Reifarth}, Rene and {Pignatari}, Marco and {Vil{\'a}gos}, Blanka and {Yag{\"u}e}, Andr{\'e}s and {Gibson}, Brad K.},
        title = "{Galactic Chemical Evolution of Radioactive Isotopes}",
      journal = {\apj},
         year = 2019,
        month = jun,
       volume = {878},
       number = {2},
          eid = {156},
        pages = {156},
          doi = {10.3847/1538-4357/ab21d1},
archivePrefix = {arXiv},
       eprint = {1905.07828},
 primaryClass = {astro-ph.GA},
       adsurl = {https://ui.adsabs.harvard.edu/abs/2019ApJ...878..156C}
}

@ARTICLE{Thompson2003ApJ,
       author = {{Thompson}, Todd A.},
        title = "{Magnetic Protoneutron Star Winds and r-Process Nucleosynthesis}",
      journal = {\apjl},
         year = 2003,
        month = mar,
       volume = {585},
       number = {1},
        pages = {L33-L36},
          doi = {10.1086/374261},
archivePrefix = {arXiv},
       eprint = {astro-ph/0302132},
 primaryClass = {astro-ph},
       adsurl = {https://ui.adsabs.harvard.edu/abs/2003ApJ...585L..33T}
}

@ARTICLE{Nishimura2015ApJ,
       author = {{Nishimura}, Nobuya and {Takiwaki}, Tomoya and {Thielemann}, Friedrich-Karl},
        title = "{The r-process Nucleosynthesis in the Various Jet-like Explosions of Magnetorotational Core-collapse Supernovae}",
      journal = {\apj},
         year = 2015,
        month = sep,
       volume = {810},
       number = {2},
          eid = {109},
        pages = {109},
          doi = {10.1088/0004-637X/810/2/109},
archivePrefix = {arXiv},
       eprint = {1501.06567},
 primaryClass = {astro-ph.SR},
       adsurl = {https://ui.adsabs.harvard.edu/abs/2015ApJ...810..109N}
}

@ARTICLE{Shibata2025PhRvD,
       author = {{Shibata}, Masaru and {Fujibayashi}, Sho and {Wanajo}, Shinya and {Ioka}, Kunihito and {Lam}, Alan Tsz-Lok and {Sekiguchi}, Yuichiro},
        title = "{Self-consistent scenario for jet and stellar explosions in collapsar: General relativistic magnetohydrodynamics simulation with a dynamo}",
      journal = {\prd},
         year = 2025,
        month = jun,
       volume = {111},
       number = {12},
          eid = {123017},
        pages = {123017},
          doi = {10.1103/msy2-fwhx},
archivePrefix = {arXiv},
       eprint = {2502.02077},
 primaryClass = {astro-ph.HE},
       adsurl = {https://ui.adsabs.harvard.edu/abs/2025PhRvD.111l3017S}
}

@ARTICLE{Molero2023MNRAS,
       author = {{Molero}, Marta and {Magrini}, Laura and {Matteucci}, Francesca and {Romano}, Donatella and {Palla}, Marco and {Cescutti}, Gabriele and {Viscasillas V{\'a}zquez}, Carlos and {Spitoni}, Emanuele},
        title = "{Origin of neutron-capture elements with the Gaia-ESO survey: the evolution of s- and r-process elements across the Milky Way}",
      journal = {\mnras},
         year = 2023,
        month = aug,
       volume = {523},
       number = {2},
        pages = {2974-2989},
          doi = {10.1093/mnras/stad1577},
archivePrefix = {arXiv},
       eprint = {2304.06452},
 primaryClass = {astro-ph.GA},
       adsurl = {https://ui.adsabs.harvard.edu/abs/2023MNRAS.523.2974M}
}

@ARTICLE{Vincenzo2021MNRAS,
       author = {{Vincenzo}, Fiorenzo and {Thompson}, Todd A. and {Weinberg}, David H. and {Griffith}, Emily J. and {Johnson}, James W. and {Johnson}, Jennifer A.},
        title = "{Nucleosynthesis signatures of neutrino-driven winds from proto-neutron stars: a perspective from chemical evolution models}",
      journal = {\mnras},
         year = 2021,
        month = dec,
       volume = {508},
       number = {3},
        pages = {3499-3507},
          doi = {10.1093/mnras/stab2828},
archivePrefix = {arXiv},
       eprint = {2102.04920},
 primaryClass = {astro-ph.GA},
       adsurl = {https://ui.adsabs.harvard.edu/abs/2021MNRAS.508.3499V}
}

@ARTICLE{Ishimaru2005,
       author = {{Ishimaru}, Yuhri and {Wanajo}, Shinya and {Aoki}, Wako and {Ryan}, Sean G. and {Prantzos}, Nikos},
        title = "{First enrichment of r-process elements in our Galaxy}",
      journal = {\nphysa},
         year = 2005,
        month = jul,
       volume = {758},
        pages = {603-606},
          doi = {10.1016/j.nuclphysa.2005.05.109},
       adsurl = {https://ui.adsabs.harvard.edu/abs/2005NuPhA.758..603I}
}

@ARTICLE{Montes2007ApJ,
       author = {{Montes}, F. and {Beers}, T.~C. and {Cowan}, J. and {Elliot}, T. and {Farouqi}, K. and {Gallino}, R. and {Heil}, M. and {Kratz}, K.-L. and {Pfeiffer}, B. and {Pignatari}, M. and {Schatz}, H.},
        title = "{Nucleosynthesis in the Early Galaxy}",
      journal = {\apj},
         year = 2007,
        month = dec,
       volume = {671},
       number = {2},
        pages = {1685-1695},
          doi = {10.1086/523084},
archivePrefix = {arXiv},
       eprint = {0709.0417},
 primaryClass = {astro-ph},
       adsurl = {https://ui.adsabs.harvard.edu/abs/2007ApJ...671.1685M}
}

@ARTICLE{Abbott+17,
       author = {{Abbott}, B.~P. and {Abbott}, R. and {Abbott}, T.~D. and {Acernese}, F. and {Ackley}, K. and {Adams}, C. and {Adams}, T. and {Addesso}, P. and {Adhikari}, R.~X. and {Adya}, V.~B. and {Affeldt}, C. and {Afrough}, M. and {Agarwal}, B. and {Agathos}, M. and {Agatsuma}, K. and {Aggarwal}, N. and {Aguiar}, O.~D. and {Aiello}, L. and {Ain}, A. and {Ajith}, P. and {Allen}, B. and {Allen}, G. and {Allocca}, A. and {Altin}, P.~A. and {Amato}, A. and {Ananyeva}, A. and {Anderson}, S.~B. and {Anderson}, W.~G. and {Angelova}, S.~V. and {Antier}, S. and {Appert}, S. and {Arai}, K. and {Araya}, M.~C. and {Areeda}, J.~S. and {Arnaud}, N. and {Arun}, K.~G. and {Ascenzi}, S. and {Ashton}, G. and {Ast}, M. and {Aston}, S.~M. and {Astone}, P. and {Atallah}, D.~V. and {Aufmuth}, P. and {Aulbert}, C. and {AultONeal}, K. and {Austin}, C. and {Avila-Alvarez}, A. and {Babak}, S. and {Bacon}, P. and {Bader}, M.~K.~M. and {Bae}, S. and {Baker}, P.~T. and {Baldaccini}, F. and {Ballardin}, G. and {Ballmer}, S.~W. and {Banagiri}, S. and {Barayoga}, J.~C. and {Barclay}, S.~E. and {Barish}, B.~C. and {Barker}, D. and {Barkett}, K. and {Barone}, F. and {Barr}, B. and {Barsotti}, L. and {Barsuglia}, M. and {Barta}, D. and {Barthelmy}, S.~D. and {Bartlett}, J. and {Bartos}, I. and {Bassiri}, R. and {Basti}, A. and {Batch}, J.~C. and {Bawaj}, M. and {Bayley}, J.~C. and {Bazzan}, M. and {B{\'e}csy}, B. and {Beer}, C. and {Bejger}, M. and {Belahcene}, I. and {Bell}, A.~S. and {Berger}, B.~K. and {Bergmann}, G. and {Bero}, J.~J. and {Berry}, C.~P.~L. and {Bersanetti}, D. and {Bertolini}, A. and {Betzwieser}, J. and {Bhagwat}, S. and {Bhandare}, R. and {Bilenko}, I.~A. and {Billingsley}, G. and {Billman}, C.~R. and {Birch}, J. and {Birney}, R. and {Birnholtz}, O. and {Biscans}, S. and {Biscoveanu}, S. and {Bisht}, A. and {Bitossi}, M. and {Biwer}, C. and {Bizouard}, M.~A. and {Blackburn}, J.~K. and {Blackman}, J. and {Blair}, C.~D. and {Blair}, D.~G. and {Blair}, R.~M. and {Bloemen}, S. and {Bock}, O. and {Bode}, N. and {Boer}, M. and {Bogaert}, G. and {Bohe}, A. and {Bondu}, F. and {Bonilla}, E. and {Bonnand}, R. and {Boom}, B.~A. and {Bork}, R. and {Boschi}, V. and {Bose}, S. and {Bossie}, K. and {Bouffanais}, Y. and {Bozzi}, A. and {Bradaschia}, C. and {Brady}, P.~R. and {Branchesi}, M. and {Brau}, J.~E. and {Briant}, T. and {Brillet}, A. and {Brinkmann}, M. and {Brisson}, V. and {Brockill}, P. and {Broida}, J.~E. and {Brooks}, A.~F. and {Brown}, D.~A. and {Brown}, D.~D. and {Brunett}, S. and {Buchanan}, C.~C. and {Buikema}, A. and {Bulik}, T. and {Bulten}, H.~J. and {Buonanno}, A. and {Buskulic}, D. and {Buy}, C. and {Byer}, R.~L. and {Cabero}, M. and {Cadonati}, L. and {Cagnoli}, G. and {Cahillane}, C. and {Calder{\'o}n Bustillo}, J. and {Callister}, T.~A. and {Calloni}, E. and {Camp}, J.~B. and {Canepa}, M. and {Canizares}, P. and {Cannon}, K.~C. and {Cao}, H. and {Cao}, J. and {Capano}, C.~D. and {Capocasa}, E. and {Carbognani}, F. and {Caride}, S. and {Carney}, M.~F. and {Casanueva Diaz}, J. and {Casentini}, C. and {Caudill}, S. and {Cavagli{\`a}}, M. and {Cavalier}, F. and {Cavalieri}, R. and {Cella}, G. and {Cepeda}, C.~B. and {Cerd{\'a}-Dur{\'a}n}, P. and {Cerretani}, G. and {Cesarini}, E. and {Chamberlin}, S.~J. and {Chan}, M. and {Chao}, S. and {Charlton}, P. and {Chase}, E. and {Chassande-Mottin}, E. and {Chatterjee}, D. and {Chatziioannou}, K. and {Cheeseboro}, B.~D. and {Chen}, H.~Y. and {Chen}, X. and {Chen}, Y. and {Cheng}, H.-P. and {Chia}, H. and {Chincarini}, A. and {Chiummo}, A. and {Chmiel}, T. and {Cho}, H.~S. and {Cho}, M. and {Chow}, J.~H. and {Christensen}, N. and {Chu}, Q. and {Chua}, A.~J.~K. and {Chua}, S. and {Chung}, A.~K.~W. and {Chung}, S. and {Ciani}, G.},
        title = "{Multi-messenger Observations of a Binary Neutron Star Merger}",
      journal = {\apjl},
         year = 2017,
        month = oct,
       volume = {848},
       number = {2},
          eid = {L12},
        pages = {L12},
          doi = {10.3847/2041-8213/aa91c9},
archivePrefix = {arXiv},
       eprint = {1710.05833},
 primaryClass = {astro-ph.HE},
       adsurl = {https://ui.adsabs.harvard.edu/abs/2017ApJ...848L..12A}
}

@ARTICLE{VillarV+17,
       author = {{Villar}, V.~A. and {Guillochon}, J. and {Berger}, E. and {Metzger}, B.~D. and {Cowperthwaite}, P.~S. and {Nicholl}, M. and {Alexander}, K.~D. and {Blanchard}, P.~K. and {Chornock}, R. and {Eftekhari}, T. and {Fong}, W. and {Margutti}, R. and {Williams}, P.~K.~G.},
        title = "{The Combined Ultraviolet, Optical, and Near-infrared Light Curves of the Kilonova Associated with the Binary Neutron Star Merger GW170817: Unified Data Set, Analytic Models, and Physical Implications}",
      journal = {\apjl},
         year = 2017,
        month = dec,
       volume = {851},
       number = {1},
          eid = {L21},
        pages = {L21},
          doi = {10.3847/2041-8213/aa9c84},
archivePrefix = {arXiv},
       eprint = {1710.11576},
 primaryClass = {astro-ph.HE},
       adsurl = {https://ui.adsabs.harvard.edu/abs/2017ApJ...851L..21V}
}

@ARTICLE{CoulterD+17,
       author = {{Coulter}, D.~A. and {Foley}, R.~J. and {Kilpatrick}, C.~D. and {Drout}, M.~R. and {Piro}, A.~L. and {Shappee}, B.~J. and {Siebert}, M.~R. and {Simon}, J.~D. and {Ulloa}, N. and {Kasen}, D. and {Madore}, B.~F. and {Murguia-Berthier}, A. and {Pan}, Y.-C. and {Prochaska}, J.~X. and {Ramirez-Ruiz}, E. and {Rest}, A. and {Rojas-Bravo}, C.},
        title = "{Swope Supernova Survey 2017a (SSS17a), the optical counterpart to a gravitational wave source}",
      journal = {Science},
         year = 2017,
        month = dec,
       volume = {358},
       number = {6370},
        pages = {1556-1558},
          doi = {10.1126/science.aap9811},
archivePrefix = {arXiv},
       eprint = {1710.05452},
 primaryClass = {astro-ph.HE},
       adsurl = {https://ui.adsabs.harvard.edu/abs/2017Sci...358.1556C}
}

@ARTICLE{LIGO+23_Rates,
       author = {{Abbott}, R. and {Abbott}, T.~D. and {Acernese}, F. and {Ackley}, K. and {Adams}, C. and {Adhikari}, N. and {Adhikari}, R.~X. and {Adya}, V.~B. and {Affeldt}, C. and {Agarwal}, D. and {Agathos}, M. and {Agatsuma}, K. and {Aggarwal}, N. and {Aguiar}, O.~D. and {Aiello}, L. and {Ain}, A. and {Ajith}, P. and {Akutsu}, T. and {de Alarc{\'o}n}, P.~F. and {Akcay}, S. and {Albanesi}, S. and {Allocca}, A. and {Altin}, P.~A. and {Amato}, A. and {Anand}, C. and {Anand}, S. and {Ananyeva}, A. and {Anderson}, S.~B. and {Anderson}, W.~G. and {Ando}, M. and {Andrade}, T. and {Andres}, N. and {Andri{\'c}}, T. and {Angelova}, S.~V. and {Ansoldi}, S. and {Antelis}, J.~M. and {Antier}, S. and {Antonini}, F. and {Appert}, S. and {Arai}, Koji and {Arai}, Koya and {Arai}, Y. and {Araki}, S. and {Araya}, A. and {Araya}, M.~C. and {Areeda}, J.~S. and {Ar{\`e}ne}, M. and {Aritomi}, N. and {Arnaud}, N. and {Arogeti}, M. and {Aronson}, S.~M. and {Arun}, K.~G. and {Asada}, H. and {Asali}, Y. and {Ashton}, G. and {Aso}, Y. and {Assiduo}, M. and {Aston}, S.~M. and {Astone}, P. and {Aubin}, F. and {Austin}, C. and {Babak}, S. and {Badaracco}, F. and {Bader}, M.~K.~M. and {Badger}, C. and {Bae}, S. and {Bae}, Y. and {Baer}, A.~M. and {Bagnasco}, S. and {Bai}, Y. and {Baiotti}, L. and {Baird}, J. and {Bajpai}, R. and {Ball}, M. and {Ballardin}, G. and {Ballmer}, S.~W. and {Balsamo}, A. and {Baltus}, G. and {Banagiri}, S. and {Bankar}, D. and {Barayoga}, J.~C. and {Barbieri}, C. and {Barish}, B.~C. and {Barker}, D. and {Barneo}, P. and {Barone}, F. and {Barr}, B. and {Barsotti}, L. and {Barsuglia}, M. and {Barta}, D. and {Bartlett}, J. and {Barton}, M.~A. and {Bartos}, I. and {Bassiri}, R. and {Basti}, A. and {Bawaj}, M. and {Bayley}, J.~C. and {Baylor}, A.~C. and {Bazzan}, M. and {B{\'e}csy}, B. and {Bedakihale}, V.~M. and {Bejger}, M. and {Belahcene}, I. and {Benedetto}, V. and {Beniwal}, D. and {Bennett}, T.~F. and {Bentley}, J.~D. and {Benyaala}, M. and {Bergamin}, F. and {Berger}, B.~K. and {Bernuzzi}, S. and {Berry}, C.~P.~L. and {Bersanetti}, D. and {Bertolini}, A. and {Betzwieser}, J. and {Beveridge}, D. and {Bhandare}, R. and {Bhardwaj}, U. and {Bhattacharjee}, D. and {Bhaumik}, S. and {Bilenko}, I.~A. and {Billingsley}, G. and {Bini}, S. and {Birney}, R. and {Birnholtz}, O. and {Biscans}, S. and {Bischi}, M. and {Biscoveanu}, S. and {Bisht}, A. and {Biswas}, B. and {Bitossi}, M. and {Bizouard}, M.-A. and {Blackburn}, J.~K. and {Blair}, C.~D. and {Blair}, D.~G. and {Blair}, R.~M. and {Bobba}, F. and {Bode}, N. and {Boer}, M. and {Bogaert}, G. and {Boldrini}, M. and {Bonavena}, L.~D. and {Bondu}, F. and {Bonilla}, E. and {Bonnand}, R. and {Booker}, P. and {Boom}, B.~A. and {Bork}, R. and {Boschi}, V. and {Bose}, N. and {Bose}, S. and {Bossilkov}, V. and {Boudart}, V. and {Bouffanais}, Y. and {Bozzi}, A. and {Bradaschia}, C. and {Brady}, P.~R. and {Bramley}, A. and {Branch}, A. and {Branchesi}, M. and {Brandt}, J. and {Brau}, J.~E. and {Breschi}, M. and {Briant}, T. and {Briggs}, J.~H. and {Brillet}, A. and {Brinkmann}, M. and {Brockill}, P. and {Brooks}, A.~F. and {Brooks}, J. and {Brown}, D.~D. and {Brunett}, S. and {Bruno}, G. and {Bruntz}, R. and {Bryant}, J. and {Bulik}, T. and {Bulten}, H.~J. and {Buonanno}, A. and {Buscicchio}, R. and {Buskulic}, D. and {Buy}, C. and {Byer}, R.~L. and {Cadonati}, L. and {Cagnoli}, G. and {Cahillane}, C. and {Bustillo}, J. Calder{\'o}n and {Callaghan}, J.~D. and {Callister}, T.~A. and {Calloni}, E. and {Cameron}, J. and {Camp}, J.~B. and {Canepa}, M. and {Canevarolo}, S. and {Cannavacciuolo}, M. and {Cannon}, K.~C. and {Cao}, H. and {Cao}, Z. and {Capocasa}, E. and {Capote}, E. and {Carapella}, G.},
        title = "{Population of Merging Compact Binaries Inferred Using Gravitational Waves through GWTC-3}",
      journal = {Physical Review X},
         year = 2023,
        month = jan,
       volume = {13},
       number = {1},
          eid = {011048},
        pages = {011048},
          doi = {10.1103/PhysRevX.13.011048},
archivePrefix = {arXiv},
       eprint = {2111.03634},
 primaryClass = {astro-ph.HE},
       adsurl = {https://ui.adsabs.harvard.edu/abs/2023PhRvX..13a1048A}
}

@INPROCEEDINGS{FernandezR+16,
       author = {{Fernandez}, Rodrigo and {Lippuner}, Jonas and {Roberts}, Luke and {Tchekhovskoy}, Alexander and {Foucart}, Francois and {Metzger}, Brian and {Kasen}, Daniel and {Quataert}, Eliot},
        title = "{Outflows from neutron star merger remnant disks: nucleosynthesis and kilonovae}",
    booktitle = {APS April Meeting Abstracts},
         year = 2016,
       series = {APS Meeting Abstracts},
       volume = {2016},
        month = mar,
          eid = {K9.008},
        pages = {K9.008},
       adsurl = {https://ui.adsabs.harvard.edu/abs/2016APS..APR.K9008F}
}

@ARTICLE{JiA+16_RetII,
       author = {{Ji}, Alexander P. and {Frebel}, Anna and {Simon}, Joshua D. and {Chiti}, Anirudh},
        title = "{Complete Element Abundances of Nine Stars in the r-process Galaxy Reticulum II}",
      journal = {\apj},
         year = 2016,
        month = oct,
       volume = {830},
       number = {2},
          eid = {93},
        pages = {93},
          doi = {10.3847/0004-637X/830/2/93},
archivePrefix = {arXiv},
       eprint = {1607.07447},
 primaryClass = {astro-ph.GA},
       adsurl = {https://ui.adsabs.harvard.edu/abs/2016ApJ...830...93J}
}

@ARTICLE{SudaT+08_SAGA,
       author = {{Suda}, Takuma and {Katsuta}, Yutaka and {Yamada}, Shimako and {Suwa}, Tamon and {Ishizuka}, Chikako and {Komiya}, Yutaka and {Sorai}, Kazuo and {Aikawa}, Masayuki and {Fujimoto}, Masayuki Y.},
        title = "{Stellar Abundances for the Galactic Archeology (SAGA) Database --- Compilation of the Characteristics of Known Extremely Metal-Poor Stars}",
      journal = {\pasj},
         year = 2008,
        month = oct,
       volume = {60},
        pages = {1159},
          doi = {10.1093/pasj/60.5.1159},
archivePrefix = {arXiv},
       eprint = {0806.3697},
 primaryClass = {astro-ph},
       adsurl = {https://ui.adsabs.harvard.edu/abs/2008PASJ...60.1159S}
}

@ARTICLE{SudaT+11_SAGA,
       author = {{Suda}, Takuma and {Yamada}, Shimako and {Katsuta}, Yutaka and {Komiya}, Yutaka and {Ishizuka}, Chikako and {Aoki}, Wako and {Fujimoto}, Masayuki Y.},
        title = "{The Stellar Abundances for Galactic Archaeology (SAGA) data base - II. Implications for mixing and nucleosynthesis in extremely metal-poor stars and chemical enrichment of the Galaxy}",
      journal = {\mnras},
         year = 2011,
        month = apr,
       volume = {412},
       number = {2},
        pages = {843-874},
          doi = {10.1111/j.1365-2966.2011.17943.x},
archivePrefix = {arXiv},
       eprint = {1010.6272},
 primaryClass = {astro-ph.GA},
       adsurl = {https://ui.adsabs.harvard.edu/abs/2011MNRAS.412..843S}
}

@ARTICLE{SudaT+17_SAGA,
       author = {{Suda}, Takuma and {Hidaka}, Jun and {Aoki}, Wako and {Katsuta}, Yutaka and {Yamada}, Shimako and {Fujimoto}, Masayuki Y. and {Ohtani}, Yukari and {Masuyama}, Miyu and {Noda}, Kazuhiro and {Wada}, Kentaro},
        title = "{Stellar Abundances for Galactic Archaeology Database. IV. Compilation of stars in dwarf galaxies}",
      journal = {\pasj},
         year = 2017,
        month = oct,
       volume = {69},
       number = {5},
          eid = {76},
        pages = {76},
          doi = {10.1093/pasj/psx059},
archivePrefix = {arXiv},
       eprint = {1703.10009},
 primaryClass = {astro-ph.GA},
       adsurl = {https://ui.adsabs.harvard.edu/abs/2017PASJ...69...76S}
}

@ARTICLE{TravaglioC+04,
       author = {{Travaglio}, Claudia and {Gallino}, Roberto and {Arnone}, Enrico and {Cowan}, John and {Jordan}, Faith and {Sneden}, Christopher},
        title = "{Galactic Evolution of Sr, Y, And Zr: A Multiplicity of Nucleosynthetic Processes}",
      journal = {\apj},
         year = 2004,
        month = feb,
       volume = {601},
       number = {2},
        pages = {864-884},
          doi = {10.1086/380507},
archivePrefix = {arXiv},
       eprint = {astro-ph/0310189},
 primaryClass = {astro-ph},
       adsurl = {https://ui.adsabs.harvard.edu/abs/2004ApJ...601..864T}
}

@ARTICLE{KarakasA_LattanzioJ14,
       author = {{Karakas}, Amanda I. and {Lattanzio}, John C.},
        title = "{The Dawes Review 2: Nucleosynthesis and Stellar Yields of Low- and Intermediate-Mass Single Stars}",
      journal = {\pasa},
         year = 2014,
        month = jul,
       volume = {31},
          eid = {e030},
        pages = {e030},
          doi = {10.1017/pasa.2014.21},
archivePrefix = {arXiv},
       eprint = {1405.0062},
 primaryClass = {astro-ph.SR},
       adsurl = {https://ui.adsabs.harvard.edu/abs/2014PASA...31...30K}
}

@ARTICLE{PalmeriniS+21,
       author = {{Palmerini}, Sara and {Busso}, Maurizio and {Vescovi}, Diego and {Naselli}, Eugenia and {Pidatella}, Angelo and {Mucciola}, Riccardo and {Cristallo}, Sergio and {Mascali}, David and {Mengoni}, Alberto and {Simonucci}, Stefano and {Taioli}, Simone},
        title = "{Presolar Grain Isotopic Ratios as Constraints to Nuclear and Stellar Parameters of Asymptotic Giant Branch Star Nucleosynthesis}",
      journal = {\apj},
         year = 2021,
        month = nov,
       volume = {921},
       number = {1},
          eid = {7},
        pages = {7},
          doi = {10.3847/1538-4357/ac1786},
archivePrefix = {arXiv},
       eprint = {2107.12037},
 primaryClass = {astro-ph.SR},
       adsurl = {https://ui.adsabs.harvard.edu/abs/2021ApJ...921....7P}
}

@ARTICLE{SnedenC+08,
       author = {{Sneden}, C. and {Cowan}, J.~J. and {Gallino}, R.},
        title = "{Neutron-capture elements in the early galaxy.}",
      journal = {\araa},
         year = 2008,
        month = sep,
       volume = {46},
        pages = {241-288},
          doi = {10.1146/annurev.astro.46.060407.145207},
       adsurl = {https://ui.adsabs.harvard.edu/abs/2008ARA&A..46..241S}
}

@ARTICLE{BussoM+99,
       author = {{Busso}, M. and {Gallino}, R. and {Wasserburg}, G.~J.},
        title = "{Nucleosynthesis in Asymptotic Giant Branch Stars: Relevance for Galactic Enrichment and Solar System Formation}",
      journal = {\araa},
         year = 1999,
        month = jan,
       volume = {37},
        pages = {239-309},
          doi = {10.1146/annurev.astro.37.1.239},
       adsurl = {https://ui.adsabs.harvard.edu/abs/1999ARA&A..37..239B}
}

@ARTICLE{BussoM+01,
       author = {{Busso}, Maurizio and {Gallino}, Roberto and {Lambert}, David L. and {Travaglio}, Claudia and {Smith}, Verne V.},
        title = "{Nucleosynthesis and Mixing on the Asymptotic Giant Branch. III. Predicted and Observed s-Process Abundances}",
      journal = {\apj},
         year = 2001,
        month = aug,
       volume = {557},
       number = {2},
        pages = {802-821},
          doi = {10.1086/322258},
archivePrefix = {arXiv},
       eprint = {astro-ph/0104424},
 primaryClass = {astro-ph},
       adsurl = {https://ui.adsabs.harvard.edu/abs/2001ApJ...557..802B}
}

@ARTICLE{RosswogS+14,
       author = {{Rosswog}, S. and {Korobkin}, O. and {Arcones}, A. and {Thielemann}, F.-K. and {Piran}, T.},
        title = "{The long-term evolution of neutron star merger remnants - I. The impact of r-process nucleosynthesis}",
      journal = {\mnras},
         year = 2014,
        month = mar,
       volume = {439},
       number = {1},
        pages = {744-756},
          doi = {10.1093/mnras/stt2502},
archivePrefix = {arXiv},
       eprint = {1307.2939},
 primaryClass = {astro-ph.HE},
       adsurl = {https://ui.adsabs.harvard.edu/abs/2014MNRAS.439..744R}
}

@ARTICLE{RadiceD+18,
       author = {{Radice}, David and {Perego}, Albino and {Hotokezaka}, Kenta and {Fromm}, Steven A. and {Bernuzzi}, Sebastiano and {Roberts}, Luke F.},
        title = "{Binary Neutron Star Mergers: Mass Ejection, Electromagnetic Counterparts, and Nucleosynthesis}",
      journal = {\apj},
         year = 2018,
        month = dec,
       volume = {869},
       number = {2},
          eid = {130},
        pages = {130},
          doi = {10.3847/1538-4357/aaf054},
archivePrefix = {arXiv},
       eprint = {1809.11161},
 primaryClass = {astro-ph.HE},
       adsurl = {https://ui.adsabs.harvard.edu/abs/2018ApJ...869..130R}
}

@ARTICLE{CowanJ+21,
       author = {{Cowan}, John J. and {Sneden}, Christopher and {Lawler}, James E. and {Aprahamian}, Ani and {Wiescher}, Michael and {Langanke}, Karlheinz and {Mart{\'\i}nez-Pinedo}, Gabriel and {Thielemann}, Friedrich-Karl},
        title = "{Origin of the heaviest elements: The rapid neutron-capture process}",
      journal = {Reviews of Modern Physics},
         year = 2021,
        month = jan,
       volume = {93},
       number = {1},
          eid = {015002},
        pages = {015002},
          doi = {10.1103/RevModPhys.93.015002},
archivePrefix = {arXiv},
       eprint = {1901.01410},
 primaryClass = {astro-ph.HE},
       adsurl = {https://ui.adsabs.harvard.edu/abs/2021RvMP...93a5002C}
}

@ARTICLE{HotokezakaH+16,
       author = {{Hotokezaka}, K. and {Wanajo}, S. and {Tanaka}, M. and {Bamba}, A. and {Terada}, Y. and {Piran}, T.},
        title = "{Radioactive decay products in neutron star merger ejecta: heating efficiency and {\ensuremath{\gamma}}-ray emission}",
      journal = {\mnras},
         year = 2016,
        month = jun,
       volume = {459},
       number = {1},
        pages = {35-43},
          doi = {10.1093/mnras/stw404},
archivePrefix = {arXiv},
       eprint = {1511.05580},
 primaryClass = {astro-ph.HE},
       adsurl = {https://ui.adsabs.harvard.edu/abs/2016MNRAS.459...35H}
}

@ARTICLE{AsplundM+09,
       author = {{Asplund}, Martin and {Grevesse}, Nicolas and {Sauval}, A. Jacques and {Scott}, Pat},
        title = "{The Chemical Composition of the Sun}",
      journal = {\araa},
         year = 2009,
        month = sep,
       volume = {47},
       number = {1},
        pages = {481-522},
          doi = {10.1146/annurev.astro.46.060407.145222},
archivePrefix = {arXiv},
       eprint = {0909.0948},
 primaryClass = {astro-ph.SR},
       adsurl = {https://ui.adsabs.harvard.edu/abs/2009ARA&A..47..481A}
}

@ARTICLE{DomotoN+21,
       author = {{Domoto}, Nanae and {Tanaka}, Masaomi and {Wanajo}, Shinya and {Kawaguchi}, Kyohei},
        title = "{Signatures of r-process Elements in Kilonova Spectra}",
      journal = {\apj},
         year = 2021,
        month = may,
       volume = {913},
       number = {1},
          eid = {26},
        pages = {26},
          doi = {10.3847/1538-4357/abf358},
archivePrefix = {arXiv},
       eprint = {2103.15284},
 primaryClass = {astro-ph.HE},
       adsurl = {https://ui.adsabs.harvard.edu/abs/2021ApJ...913...26D}
}

@ARTICLE{LiX_PaczynskiB98,
       author = {{Li}, Li-Xin and {Paczy{\'n}ski}, Bohdan},
        title = "{Transient Events from Neutron Star Mergers}",
      journal = {\apjl},
         year = 1998,
        month = nov,
       volume = {507},
       number = {1},
        pages = {L59-L62},
          doi = {10.1086/311680},
archivePrefix = {arXiv},
       eprint = {astro-ph/9807272},
 primaryClass = {astro-ph},
       adsurl = {https://ui.adsabs.harvard.edu/abs/1998ApJ...507L..59L}
}

@ARTICLE{Beniamini_Piran24,
       author = {{Beniamini}, Paz and {Piran}, Tsvi},
        title = "{Ultrafast Compact Binary Mergers}",
      journal = {\apj},
         year = 2024,
        month = may,
       volume = {966},
       number = {1},
          eid = {17},
        pages = {17},
          doi = {10.3847/1538-4357/ad32cd},
archivePrefix = {arXiv},
       eprint = {2312.02269},
 primaryClass = {astro-ph.HE},
       adsurl = {https://ui.adsabs.harvard.edu/abs/2024ApJ...966...17B}
}

@ARTICLE{Tarumi_Hotokezaka_Beniamini21,
       author = {{Tarumi}, Yuta and {Hotokezaka}, Kenta and {Beniamini}, Paz},
        title = "{Evidence for r-process Delay in Very Metal-poor Stars}",
      journal = {\apjl},
         year = 2021,
        month = jun,
       volume = {913},
       number = {2},
          eid = {L30},
        pages = {L30},
          doi = {10.3847/2041-8213/abfe13},
archivePrefix = {arXiv},
       eprint = {2102.03368},
 primaryClass = {astro-ph.GA},
       adsurl = {https://ui.adsabs.harvard.edu/abs/2021ApJ...913L..30T}
}

@ARTICLE{MostaP+18,
       author = {{M{\"o}sta}, Philipp and {Roberts}, Luke F. and {Halevi}, Goni and {Ott}, Christian D. and {Lippuner}, Jonas and {Haas}, Roland and {Schnetter}, Erik},
        title = "{r-process Nucleosynthesis from Three-dimensional Magnetorotational Core-collapse Supernovae}",
      journal = {\apj},
         year = 2018,
        month = sep,
       volume = {864},
       number = {2},
          eid = {171},
        pages = {171},
          doi = {10.3847/1538-4357/aad6ec},
archivePrefix = {arXiv},
       eprint = {1712.09370},
 primaryClass = {astro-ph.HE},
       adsurl = {https://ui.adsabs.harvard.edu/abs/2018ApJ...864..171M}
}

@ARTICLE{ThompsonT_udDoula18,
       author = {{Thompson}, Todd A. and {ud-Doula}, Asif},
        title = "{High-entropy ejections from magnetized proto-neutron star winds: implications for heavy element nucleosynthesis}",
      journal = {\mnras},
         year = 2018,
        month = jun,
       volume = {476},
       number = {4},
        pages = {5502-5515},
          doi = {10.1093/mnras/sty480},
archivePrefix = {arXiv},
       eprint = {1709.03997},
 primaryClass = {astro-ph.HE},
       adsurl = {https://ui.adsabs.harvard.edu/abs/2018MNRAS.476.5502T}
}

@ARTICLE{CoteB+19_NotBNS,
       author = {{C{\^o}t{\'e}}, Benoit and {Eichler}, Marius and {Arcones}, Almudena and {Hansen}, Camilla J. and {Simonetti}, Paolo and {Frebel}, Anna and {Fryer}, Chris L. and {Pignatari}, Marco and {Reichert}, Moritz and {Belczynski}, Krzysztof and {Matteucci}, Francesca},
        title = "{Neutron Star Mergers Might Not Be the Only Source of r-process Elements in the Milky Way}",
      journal = {\apj},
         year = 2019,
        month = apr,
       volume = {875},
       number = {2},
          eid = {106},
        pages = {106},
          doi = {10.3847/1538-4357/ab10db},
archivePrefix = {arXiv},
       eprint = {1809.03525},
 primaryClass = {astro-ph.HE},
       adsurl = {https://ui.adsabs.harvard.edu/abs/2019ApJ...875..106C}
}

@ARTICLE{BanerjeeP+20,
       author = {{Banerjee}, Projjwal and {Wu}, Meng-Ru and {Yuan}, Zhen},
        title = "{Neutron Star Mergers as the Main Source of r-process: Natal Kicks and Inside-out Evolution to the Rescue}",
      journal = {\apjl},
         year = 2020,
        month = oct,
       volume = {902},
       number = {2},
          eid = {L34},
        pages = {L34},
          doi = {10.3847/2041-8213/abbc0d},
archivePrefix = {arXiv},
       eprint = {2007.04442},
 primaryClass = {astro-ph.GA},
       adsurl = {https://ui.adsabs.harvard.edu/abs/2020ApJ...902L..34B}
}

@ARTICLE{vande_Voort+22,
       author = {{van de Voort}, Freeke and {Pakmor}, R{\"u}diger and {Bieri}, Rebekka and {Grand}, Robert J.~J.},
        title = "{The impact of natal kicks on galactic r-process enrichment by neutron star mergers}",
      journal = {\mnras},
         year = 2022,
        month = jun,
       volume = {512},
       number = {4},
        pages = {5258-5268},
          doi = {10.1093/mnras/stac710},
archivePrefix = {arXiv},
       eprint = {2110.11963},
 primaryClass = {astro-ph.GA},
       adsurl = {https://ui.adsabs.harvard.edu/abs/2022MNRAS.512.5258V}
}

@ARTICLE{Peters1964,
       author = {{Peters}, P.~C.},
        title = "{Gravitational Radiation and the Motion of Two Point Masses}",
      journal = {Physical Review},
         year = 1964,
        month = nov,
       volume = {136},
       number = {4B},
        pages = {1224-1232},
          doi = {10.1103/PhysRev.136.B1224},
       adsurl = {https://ui.adsabs.harvard.edu/abs/1964PhRv..136.1224P}
}

@ARTICLE{MetzgerB+08,
       author = {{Metzger}, B.~D. and {Quataert}, E. and {Thompson}, T.~A.},
        title = "{Short-duration gamma-ray bursts with extended emission from protomagnetar spin-down}",
      journal = {\mnras},
         year = 2008,
        month = apr,
       volume = {385},
       number = {3},
        pages = {1455-1460},
          doi = {10.1111/j.1365-2966.2008.12923.x},
archivePrefix = {arXiv},
       eprint = {0712.1233},
 primaryClass = {astro-ph},
       adsurl = {https://ui.adsabs.harvard.edu/abs/2008MNRAS.385.1455M}
}

@ARTICLE{Maoz_Nakar25,
       author = {{Maoz}, Dan and {Nakar}, Ehud},
        title = "{The Neutron Star Merger Delay-time Distribution, R-process ``Knees,'' and the Metal Budget of the Galaxy}",
      journal = {\apj},
         year = 2025,
        month = apr,
       volume = {982},
       number = {2},
          eid = {179},
        pages = {179},
          doi = {10.3847/1538-4357/ada3bd},
archivePrefix = {arXiv},
       eprint = {2406.08630},
 primaryClass = {astro-ph.HE},
       adsurl = {https://ui.adsabs.harvard.edu/abs/2025ApJ...982..179M}
}

@ARTICLE{ThielemannF+17,
       author = {{Thielemann}, F.-K. and {Eichler}, M. and {Panov}, I.~V. and {Wehmeyer}, B.},
        title = "{Neutron Star Mergers and Nucleosynthesis of Heavy Elements}",
      journal = {Annual Review of Nuclear and Particle Science},
         year = 2017,
        month = oct,
       volume = {67},
        pages = {253-274},
          doi = {10.1146/annurev-nucl-101916-123246},
archivePrefix = {arXiv},
       eprint = {1710.02142},
 primaryClass = {astro-ph.HE},
       adsurl = {https://ui.adsabs.harvard.edu/abs/2017ARNPS..67..253T}
}

@ARTICLE{Lattimer_Schramm74,
       author = {{Lattimer}, J.~M. and {Schramm}, D.~N.},
        title = "{Black-Hole-Neutron-Star Collisions}",
      journal = {\apjl},
         year = 1974,
        month = sep,
       volume = {192},
        pages = {L145},
          doi = {10.1086/181612},
       adsurl = {https://ui.adsabs.harvard.edu/abs/1974ApJ...192L.145L}
}

@ARTICLE{Siegel+19_Nat,
       author = {{Siegel}, Daniel M. and {Barnes}, Jennifer and {Metzger}, Brian D.},
        title = "{Collapsars as a major source of r-process elements}",
      journal = {\nat},
         year = 2019,
        month = may,
       volume = {569},
       number = {7755},
        pages = {241-244},
          doi = {10.1038/s41586-019-1136-0},
archivePrefix = {arXiv},
       eprint = {1810.00098},
 primaryClass = {astro-ph.HE},
       adsurl = {https://ui.adsabs.harvard.edu/abs/2019Natur.569..241S}
}

@ARTICLE{DeanC_FernandezR24,
       author = {{Dean}, Coleman and {Fern{\'a}ndez}, Rodrigo},
        title = "{Collapsar disk outflows: Heavy element production}",
      journal = {\prd},
         year = 2024,
        month = oct,
       volume = {110},
       number = {8},
          eid = {083024},
        pages = {083024},
          doi = {10.1103/PhysRevD.110.083024},
archivePrefix = {arXiv},
       eprint = {2408.15338},
 primaryClass = {astro-ph.HE},
       adsurl = {https://ui.adsabs.harvard.edu/abs/2024PhRvD.110h3024D}
}

@ARTICLE{Gottlieb2025,
       author = {{Gottlieb}, Ore},
        title = "{The Landscape of Collapsar Outflows: Structure, Signatures, and Origins of Einstein Probe Relativistic Supernova Transients}",
      journal = {\apjl},
         year = 2025,
        month = oct,
       volume = {992},
       number = {1},
          eid = {L3},
        pages = {L3},
          doi = {10.3847/2041-8213/ae09af},
archivePrefix = {arXiv},
       eprint = {2509.04551},
 primaryClass = {astro-ph.HE},
       adsurl = {https://ui.adsabs.harvard.edu/abs/2025ApJ...992L...3G}
}

@ARTICLE{MacFadyen_Woosley99,
       author = {{MacFadyen}, A.~I. and {Woosley}, S.~E.},
        title = "{Collapsars: Gamma-Ray Bursts and Explosions in ``Failed Supernovae''}",
      journal = {\apj},
         year = 1999,
        month = oct,
       volume = {524},
       number = {1},
        pages = {262-289},
          doi = {10.1086/307790},
archivePrefix = {arXiv},
       eprint = {astro-ph/9810274},
 primaryClass = {astro-ph},
       adsurl = {https://ui.adsabs.harvard.edu/abs/1999ApJ...524..262M}
}

@ARTICLE{Beniamini_Hotokezaka20,
       author = {{Beniamini}, Paz and {Hotokezaka}, Kenta},
        title = "{Turbulent mixing of r-process elements in the Milky Way}",
      journal = {\mnras},
         year = 2020,
        month = aug,
       volume = {496},
       number = {2},
        pages = {1891-1901},
          doi = {10.1093/mnras/staa1690},
archivePrefix = {arXiv},
       eprint = {2003.01129},
 primaryClass = {astro-ph.HE},
       adsurl = {https://ui.adsabs.harvard.edu/abs/2020MNRAS.496.1891B}
}

@ARTICLE{Hotokezaka+18,
       author = {{Hotokezaka}, Kenta and {Beniamini}, Paz and {Piran}, Tsvi},
        title = "{Neutron star mergers as sites of r-process nucleosynthesis and short gamma-ray bursts}",
      journal = {International Journal of Modern Physics D},
         year = 2018,
        month = jan,
       volume = {27},
       number = {13},
          eid = {1842005},
        pages = {1842005},
          doi = {10.1142/S0218271818420051},
archivePrefix = {arXiv},
       eprint = {1801.01141},
 primaryClass = {astro-ph.HE},
       adsurl = {https://ui.adsabs.harvard.edu/abs/2018IJMPD..2742005H}
}

@ARTICLE{Madau_Dickinson14,
       author = {{Madau}, Piero and {Dickinson}, Mark},
        title = "{Cosmic Star-Formation History}",
      journal = {\araa},
         year = 2014,
        month = aug,
       volume = {52},
        pages = {415-486},
          doi = {10.1146/annurev-astro-081811-125615},
archivePrefix = {arXiv},
       eprint = {1403.0007},
 primaryClass = {astro-ph.CO},
       adsurl = {https://ui.adsabs.harvard.edu/abs/2014ARA&A..52..415M}
}

@misc{Lane_2025_colour,
    author = {Zachary G. Lane},
    title = {AstroColour},
    year = 2025,
    url = {https://github.com/ZacharyLane1204/AstroColour.git}
}

@ARTICLE{KappelerF+11,
       author = {{K{\"a}ppeler}, F. and {Gallino}, R. and {Bisterzo}, S. and {Aoki}, Wako},
        title = "{The s process: Nuclear physics, stellar models, and observations}",
      journal = {Reviews of Modern Physics},
         year = 2011,
        month = jan,
       volume = {83},
       number = {1},
        pages = {157-194},
          doi = {10.1103/RevModPhys.83.157},
archivePrefix = {arXiv},
       eprint = {1012.5218},
 primaryClass = {astro-ph.SR},
       adsurl = {https://ui.adsabs.harvard.edu/abs/2011RvMP...83..157K}
}

@ARTICLE{BisterzoS+15,
       author = {{Bisterzo}, S. and {Gallino}, R. and {K{\"a}ppeler}, F. and {Wiescher}, M. and {Imbriani}, G. and {Straniero}, O. and {Cristallo}, S. and {G{\"o}rres}, J. and {deBoer}, R.~J.},
        title = "{The branchings of the main s-process: their sensitivity to {\ensuremath{\alpha}}-induced reactions on $^{13}$C and $^{22}$Ne and to the uncertainties of the nuclear network}",
      journal = {\mnras},
         year = 2015,
        month = may,
       volume = {449},
       number = {1},
        pages = {506-527},
          doi = {10.1093/mnras/stv271},
archivePrefix = {arXiv},
       eprint = {1507.06798},
 primaryClass = {astro-ph.SR},
       adsurl = {https://ui.adsabs.harvard.edu/abs/2015MNRAS.449..506B}
}

@ARTICLE{ArconesA+23,
       author = {{Arcones}, Almudena and {Thielemann}, Friedrich-Karl},
        title = "{Origin of the elements}",
      journal = {\aapr},
         year = 2023,
        month = dec,
       volume = {31},
       number = {1},
          eid = {1},
        pages = {1},
          doi = {10.1007/s00159-022-00146-x},
       adsurl = {https://ui.adsabs.harvard.edu/abs/2023A&ARv..31....1A}
}

@ARTICLE{BurbidgeM+57,
       author = {{Burbidge}, E. Margaret and {Burbidge}, G.~R. and {Fowler}, William A. and {Hoyle}, F.},
        title = "{Synthesis of the Elements in Stars}",
      journal = {Reviews of Modern Physics},
         year = 1957,
        month = oct,
       volume = {29},
       number = {4},
        pages = {547-650},
          doi = {10.1103/RevModPhys.29.547},
       adsurl = {https://ui.adsabs.harvard.edu/abs/1957RvMP...29..547B}
}

@ARTICLE{PignatariM+23,
       author = {{Pignatari}, Marco and {Gallino}, Roberto and {Reifarth}, Rene},
        title = "{The s process in massive stars, a benchmark for neutron capture reaction rates}",
      journal = {European Physical Journal A},
         year = 2023,
        month = dec,
       volume = {59},
       number = {12},
          eid = {302},
        pages = {302},
          doi = {10.1140/epja/s10050-023-01206-1},
archivePrefix = {arXiv},
       eprint = {2510.18124},
 primaryClass = {astro-ph.SR},
       adsurl = {https://ui.adsabs.harvard.edu/abs/2023EPJA...59..302P}
}

@ARTICLE{Sales-SilvaV+22,
       author = {{Sales-Silva}, J.~V. and {Daflon}, S. and {Cunha}, K. and {Souto}, D. and {Smith}, V.~V. and {Chiappini}, C. and {Donor}, J. and {Frinchaboy}, P.~M. and {Garc{\'\i}a-Hern{\'a}ndez}, D.~A. and {Hayes}, C. and {Majewski}, S.~R. and {Masseron}, T. and {Schiavon}, R.~P. and {Weinberg}, D.~H. and {Beaton}, R.~L. and {Fern{\'a}ndez-Trincado}, J.~G. and {J{\"o}nsson}, H. and {Lane}, R.~R. and {Minniti}, D. and {Manchado}, A. and {Moni Bidin}, C. and {Nitschelm}, C. and {O'Connell}, J. and {Villanova}, S.},
        title = "{Exploring the S-process History in the Galactic Disk: Cerium Abundances and Gradients in Open Clusters from the OCCAM/APOGEE Sample}",
      journal = {\apj},
         year = 2022,
        month = feb,
       volume = {926},
       number = {2},
          eid = {154},
        pages = {154},
          doi = {10.3847/1538-4357/ac4254},
archivePrefix = {arXiv},
       eprint = {2112.02196},
 primaryClass = {astro-ph.GA},
       adsurl = {https://ui.adsabs.harvard.edu/abs/2022ApJ...926..154S}
}

@ARTICLE{BHP2016a,
       author = {{Beniamini}, Paz and {Hotokezaka}, Kenta and {Piran}, Tsvi},
        title = "{r-process Production Sites as Inferred from Eu Abundances in Dwarf Galaxies}",
      journal = {\apj},
         year = 2016,
        month = dec,
       volume = {832},
       number = {2},
          eid = {149},
        pages = {149},
          doi = {10.3847/0004-637X/832/2/149},
archivePrefix = {arXiv},
       eprint = {1608.08650},
 primaryClass = {astro-ph.HE},
       adsurl = {https://ui.adsabs.harvard.edu/abs/2016ApJ...832..149B}
}

@ARTICLE{Roederer2016,
       author = {{Roederer}, Ian U. and {Mateo}, Mario and {Bailey}, III, John I. and {Song}, Yingyi and {Bell}, Eric F. and {Crane}, Jeffrey D. and {Loebman}, Sarah and {Nidever}, David L. and {Olszewski}, Edward W. and {Shectman}, Stephen A. and {Thompson}, Ian B. and {Valluri}, Monica and {Walker}, Matthew G.},
        title = "{Detailed Chemical Abundances in the r-process-rich Ultra-faint Dwarf Galaxy Reticulum 2}",
      journal = {\aj},
         year = 2016,
        month = mar,
       volume = {151},
       number = {3},
          eid = {82},
        pages = {82},
          doi = {10.3847/0004-6256/151/3/82},
archivePrefix = {arXiv},
       eprint = {1601.04070},
 primaryClass = {astro-ph.SR},
       adsurl = {https://ui.adsabs.harvard.edu/abs/2016AJ....151...82R}
}

@ARTICLE{Honda2006ApJ,
       author = {{Honda}, S. and {Aoki}, W. and {Ishimaru}, Y. and {Wanajo}, S. and {Ryan}, S.~G.},
        title = "{Neutron-Capture Elements in the Very Metal Poor Star HD 122563}",
      journal = {\apj},
         year = 2006,
        month = jun,
       volume = {643},
       number = {2},
        pages = {1180-1189},
          doi = {10.1086/503195},
archivePrefix = {arXiv},
       eprint = {astro-ph/0602107},
 primaryClass = {astro-ph},
       adsurl = {https://ui.adsabs.harvard.edu/abs/2006ApJ...643.1180H}
}

@ARTICLE{Hansen2014ApJ,
       author = {{Hansen}, C.~J. and {Montes}, F. and {Arcones}, A.},
        title = "{How Many Nucleosynthesis Processes Exist at Low Metallicity?}",
      journal = {\apj},
         year = 2014,
        month = dec,
       volume = {797},
       number = {2},
          eid = {123},
        pages = {123},
          doi = {10.1088/0004-637X/797/2/123},
archivePrefix = {arXiv},
       eprint = {1408.4135},
 primaryClass = {astro-ph.SR},
       adsurl = {https://ui.adsabs.harvard.edu/abs/2014ApJ...797..123H}
}

@ARTICLE{Frebel2010,
       author = {{Frebel}, Anna and {Simon}, Joshua D. and {Geha}, Marla and {Willman}, Beth},
        title = "{High-Resolution Spectroscopy of Extremely Metal-Poor Stars in the Least Evolved Galaxies: Ursa Major II and Coma Berenices}",
      journal = {\apj},
         year = 2010,
        month = jan,
       volume = {708},
       number = {1},
        pages = {560-583},
          doi = {10.1088/0004-637X/708/1/560},
archivePrefix = {arXiv},
       eprint = {0902.2395},
 primaryClass = {astro-ph.GA},
       adsurl = {https://ui.adsabs.harvard.edu/abs/2010ApJ...708..560F}
}

@ARTICLE{Roederer2014,
       author = {{Roederer}, Ian U. and {Kirby}, Evan N.},
        title = "{Detailed abundance analysis of the brightest star in Segue 2, the least massive galaxy}",
      journal = {\mnras},
         year = 2014,
        month = may,
       volume = {440},
       number = {3},
        pages = {2665-2675},
          doi = {10.1093/mnras/stu491},
archivePrefix = {arXiv},
       eprint = {1403.2733},
 primaryClass = {astro-ph.GA},
       adsurl = {https://ui.adsabs.harvard.edu/abs/2014MNRAS.440.2665R}
}

@ARTICLE{Frebel2014,
       author = {{Frebel}, Anna and {Simon}, Joshua D. and {Kirby}, Evan N.},
        title = "{Segue 1: An Unevolved Fossil Galaxy from the Early Universe}",
      journal = {\apj},
         year = 2014,
        month = may,
       volume = {786},
       number = {1},
          eid = {74},
        pages = {74},
          doi = {10.1088/0004-637X/786/1/74},
archivePrefix = {arXiv},
       eprint = {1403.6116},
 primaryClass = {astro-ph.GA},
       adsurl = {https://ui.adsabs.harvard.edu/abs/2014ApJ...786...74F}
}

@ARTICLE{RoedererI+23,
       author = {{Roederer}, Ian U. and {Vassh}, Nicole and {Holmbeck}, Erika M. and {Mumpower}, Matthew R. and {Surman}, Rebecca and {Cowan}, John J. and {Beers}, Timothy C. and {Ezzeddine}, Rana and {Frebel}, Anna and {Hansen}, Terese T. and {Placco}, Vinicius M. and {Sakari}, Charli M.},
        title = "{Element abundance patterns in stars indicate fission of nuclei heavier than uranium}",
      journal = {Science},
         year = 2023,
        month = dec,
       volume = {382},
       number = {6675},
        pages = {1177-1180},
          doi = {10.1126/science.adf1341},
archivePrefix = {arXiv},
       eprint = {2312.06844},
 primaryClass = {astro-ph.SR},
       adsurl = {https://ui.adsabs.harvard.edu/abs/2023Sci...382.1177R}
}

@article{Astropy_2013,
    author = {{Astropy Collaboration} and Thomas P. Robitaille and Erik J. Tollerud and Perry Greenfield and Michael Droettboom and Erik Bray and Tom Aldcroft and Matt Davis and Adam Ginsburg and Adrian M. Price-Whelan and Wolfgang E. Kerzendorf and Alexander Conley and Neil Crighton and Kyle Barbary and Demitri Muna and Henry Ferguson and Frédéric Grollier and Madhura M. Parikh and Prasanth H. Nair and Hans M. Günther and Christoph Deil and Julien Woillez and Simon Conseil and Roban Kramer and James E.H. Turner and Leo Singer and Ryan Fox and Benjamin A. Weaver and Victor Zabalza and Zachary I. Edwards and K. Azalee Bostroem and D. J. Burke and Andrew R. Casey and Steven M. Crawford and Nadia Dencheva and Justin Ely and Tim Jenness and Kathleen Labrie and Pey Lian Lim and Francesco Pierfederici and Andrew Pontzen and Andy Ptak and Brian Refsdal and Mathieu Servillat and Ole Streicher},
    doi = {10.1051/0004-6361/201322068},
    issn = {00046361},
    journal = {Astronomy and Astrophysics},
    pages = {A33},
    title = {Astropy: A community Python package for astronomy},
    volume = {558},
    url = {https://ui.adsabs.harvard.edu/abs/2013A&A...558A..33A/abstract},
    year = {2013},
}

@ARTICLE{BisterzoS+14,
       author = {{Bisterzo}, S. and {Travaglio}, C. and {Gallino}, R. and {Wiescher}, M. and {K{\"a}ppeler}, F.},
        title = "{Galactic Chemical Evolution and Solar s-process Abundances: Dependence on the $^{13}$C-pocket Structure}",
      journal = {\apj},
         year = 2014,
        month = may,
       volume = {787},
       number = {1},
          eid = {10},
        pages = {10},
          doi = {10.1088/0004-637X/787/1/10},
archivePrefix = {arXiv},
       eprint = {1403.1764},
 primaryClass = {astro-ph.SR},
       adsurl = {https://ui.adsabs.harvard.edu/abs/2014ApJ...787...10B}
}

@article{Astropy_2018,
    author = {{Astropy Collaboration} and A. M. Price-Whelan and B. M. Sipőcz and H. M. Günther and P. L. Lim and S. M. Crawford and S. Conseil and D. L. Shupe and M. W. Craig and N. Dencheva and A. Ginsburg and J. T. VanderPlas and L. D. Bradley and D. Pérez-Suárez and M. de Val-Borro and T. L. Aldcroft and K. L. Cruz and T. P. Robitaille and E. J. Tollerud and C. Ardelean and T. Babej and Y. P. Bach and M. Bachetti and A. V. Bakanov and S. P. Bamford and G. Barentsen and P. Barmby and A. Baumbach and K. L. Berry and F. Biscani and M. Boquien and K. A. Bostroem and L. G. Bouma and G. B. Brammer and E. M. Bray and H. Breytenbach and H. Buddelmeijer and D. J. Burke and G. Calderone and J. L. Cano Rodríguez and M. Cara and J. V. M. Cardoso and S. Cheedella and Y. Copin and L. Corrales and D. Crichton and D. D'Avella and C. Deil and É. Depagne and J. P. Dietrich and A. Donath and M. Droettboom and N. Earl and T. Erben and S. Fabbro and L. A. Ferreira and T. Finethy and R. T. Fox and L. H. Garrison and S. L. J. Gibbons and D. A. Goldstein and R. Gommers and J. P. Greco and P. Greenfield and A. M. Groener and F. Grollier and A. Hagen and P. Hirst and D. Homeier and A. J. Horton and G. Hosseinzadeh and L. Hu and J. S. Hunkeler and Ž. Ivezić and A. Jain and T. Jenness and G. Kanarek and S. Kendrew and N. S. Kern and W. E. Kerzendorf and A. Khvalko and J. King and D. Kirkby and A. M. Kulkarni and A. Kumar and A. Lee and D. Lenz and S. P. Littlefair and Z. Ma and D. M. Macleod and M. Mastropietro and C. McCully and S. Montagnac and B. M. Morris and M. Mueller and S. J. Mumford and D. Muna and N. A. Murphy and S. Nelson and G. H. Nguyen and J. P. Ninan and M. Nöthe and S. Ogaz and S. Oh and J. K. Parejko and N. Parley and S. Pascual and R. Patil and A. A. Patil and A. L. Plunkett and J. X. Prochaska and T. Rastogi and V. Reddy Janga and J. Sabater and P. Sakurikar and M. Seifert and L. E. Sherbert and H. Sherwood-Taylor and A. Y. Shih and J. Sick and M. T. Silbiger and S. Singanamalla and L. P. Singer and P. H. Sladen and K. A. Sooley and S. Sornarajah and O. Streicher and P. Teuben and S. W. Thomas and G. R. Tremblay and J. E. H. Turner and V. Terrón and M. H. van Kerkwijk and A. de la Vega and L. L. Watkins and B. A. Weaver and J. B. Whitmore and J. Woillez and V. Zabalza and Astropy Contributors},
    doi = {10.3847/1538-3881/AABC4F},
    issn = {0004-6256},
    issue = {3},
    journal = {AJ},
    month = {9},
    pages = {123},
    publisher = {American Astronomical Society},
    title = {The Astropy Project: Building an Open-science Project and Status of the v2.0 Core Package},
    volume = {156},
    url = {https://ui.adsabs.harvard.edu/abs/2018AJ....156..123A/abstract},
    year = {2018},
}

@ARTICLE{Hotokezaka2015,
       author = {{Hotokezaka}, Kenta and {Piran}, Tsvi and {Paul}, Michael},
        title = "{Short-lived $^{244}$Pu points to compact binary mergers as sites for heavy r-process nucleosynthesis}",
      journal = {Nature Physics},
         year = 2015,
        month = dec,
       volume = {11},
       number = {12},
        pages = {1042-1042},
          doi = {10.1038/nphys3574},
archivePrefix = {arXiv},
       eprint = {1510.00711},
 primaryClass = {astro-ph.HE},
       adsurl = {https://ui.adsabs.harvard.edu/abs/2015NatPh..11.1042H}
}

@ARTICLE{Bauswein2014,
       author = {{Bauswein}, A. and {Ardevol Pulpillo}, R. and {Janka}, H.-T. and {Goriely}, S.},
        title = "{Nucleosynthesis Constraints on the Neutron Star-Black Hole Merger Rate}",
      journal = {\apjl},
         year = 2014,
        month = nov,
       volume = {795},
       number = {1},
          eid = {L9},
        pages = {L9},
          doi = {10.1088/2041-8205/795/1/L9},
archivePrefix = {arXiv},
       eprint = {1408.1783},
 primaryClass = {astro-ph.SR},
       adsurl = {https://ui.adsabs.harvard.edu/abs/2014ApJ...795L...9B}
}

@ARTICLE{OkadaH+25,
       author = {{Okada}, Hiroko and {Aoki}, Wako and {Tominaga}, Nozomu and {Honda}, Satoshi},
        title = "{SMSS J022423.27$-$573705.1: An Extremely Metal-Poor Star with the Most Pronounced Weak $r$-Process Signature}",
      journal = {arXiv e-prints},
         year = 2025,
        month = nov,
          eid = {arXiv:2512.00721},
        pages = {arXiv:2512.00721},
          doi = {10.48550/arXiv.2512.00721},
archivePrefix = {arXiv},
       eprint = {2512.00721},
 primaryClass = {astro-ph.SR},
       adsurl = {https://ui.adsabs.harvard.edu/abs/2025arXiv251200721O}
}

@article{Astropy_2022,
    author      = {{Astropy Collaboration} and Adrian M. Price-Whelan and Pey Lian Lim and Nicholas Earl and Nathaniel Starkman and Larry Bradley and David L. Shupe and Aarya A. Patil and Lia Corrales and C. E. Brasseur and Maximilian Nöthe and Axel Donath and Erik Tollerud and Brett M. Morris and Adam Ginsburg and Eero Vaher and Benjamin A. Weaver and James Tocknell and William Jamieson and Marten H. van Kerkwijk and Thomas P. Robitaille and Bruce Merry and Matteo Bachetti and H. Moritz Günther and Thomas L. Aldcroft and Jaime A. Alvarado-Montes and Anne M. Archibald and Attila Bódi and Shreyas Bapat and Geert Barentsen and Juanjo Bazán and Manish Biswas and Médéric Boquien and D. J. Burke and Daria Cara and Mihai Cara and Kyle E. Conroy and Simon Conseil and Matthew W. Craig and Robert M. Cross and Kelle L. Cruz and Francesco D'Eugenio and Nadia Dencheva and Hadrien A. R. Devillepoix and Jörg P. Dietrich and Arthur Davis Eigenbrot and Thomas Erben and Leonardo Ferreira and Daniel Foreman-Mackey and Ryan Fox and Nabil Freij and Suyog Garg and Robel Geda and Lauren Glattly and Yash Gondhalekar and Karl D. Gordon and David Grant and Perry Greenfield and Austen M. Groener and Steve Guest and Sebastian Gurovich and Rasmus Handberg and Akeem Hart and Zac Hatfield-Dodds and Derek Homeier and Griffin Hosseinzadeh and Tim Jenness and Craig K. Jones and Prajwel Joseph and J. Bryce Kalmbach and Emir Karamehmetoglu and Mikołaj Kałuszyński and Michael S. P. Kelley and Nicholas Kern and Wolfgang E. Kerzendorf and Eric W. Koch and Shankar Kulumani and Antony Lee and Chun Ly and Zhiyuan Ma and Conor MacBride and Jakob M. Maljaars and Demitri Muna and N. A. Murphy and Henrik Norman and Richard O'Steen and Kyle A. Oman and Camilla Pacifici and Sergio Pascual and J. Pascual-Granado and Rohit R. Patil and Gabriel I. Perren and Timothy E. Pickering and Tanuj Rastogi and Benjamin R. Roulston and Daniel F. Ryan and Eli S. Rykoff and Jose Sabater and Parikshit Sakurikar and Jesús Salgado and Aniket Sanghi and Nicholas Saunders and Volodymyr Savchenko and Ludwig Schwardt and Michael Seifert-Eckert and Albert Y. Shih and Anany Shrey Jain and Gyanendra Shukla and Jonathan Sick and Chris Simpson and Sudheesh Singanamalla and Leo P. Singer and Jaladh Singhal and Manodeep Sinha and Brigitta M. Sipőcz and Lee R. Spitler and David Stansby and Ole Streicher and Jani Šumak and John D. Swinbank and Dan S. Taranu and Nikita Tewary and Grant R. Tremblay and Miguel de Val-Borro and Samuel J. Van Kooten and Zlatan Vasović and Shresth Verma and José Vinícius de Miranda Cardoso and Peter K. G. Williams and Tom J. Wilson and Benjamin Winkel and W. M. Wood-Vasey and Rui Xue and Peter Yoachim and Chen Zhang and Andrea Zonca and {Astropy Project Contributors} },
    doi         = {10.3847/1538-4357/AC7C74},
    issn        = {0004-637X},
    issue       = {2},
    journal     = {ApJ},
    month       = {8},
    pages       = {167},
    publisher   = {American Astronomical Society},
    title       = {The Astropy Project: Sustaining and Growing a Community-oriented Open-source Project and the Latest Major Release (v5.0) of the Core Package},
    volume      = {935},
    url         = {https://ui.adsabs.harvard.edu/abs/2022ApJ...935..167A/abstract},
    year        = {2022}
}

@article{Hunter_2007,
    Author = {Hunter, J. D.},
    Title = {Matplotlib: A 2D graphics environment},
    Journal = {Computing in Science \& Engineering},
    Volume = {9},
    Number = {3},
    Pages = {90--95},
    publisher = {IEEE COMPUTER SOC},
    doi = {10.1109/MCSE.2007.55},
    year = 2007
}

@ARTICLE{KobayashiC+23,
       author = {{Kobayashi}, Chiaki and {Mandel}, Ilya and {Belczynski}, Krzysztof and {Goriely}, Stephane and {Janka}, Thomas H. and {Just}, Oliver and {Ruiter}, Ashley J. and {Vanbeveren}, Dany and {Kruckow}, Matthias U. and {Briel}, Max M. and {Eldridge}, Jan J. and {Stanway}, Elizabeth},
        title = "{Can Neutron Star Mergers Alone Explain the r-process Enrichment of the Milky Way?}",
      journal = {\apjl},
         year = 2023,
        month = feb,
       volume = {943},
       number = {2},
          eid = {L12},
        pages = {L12},
          doi = {10.3847/2041-8213/acad82},
archivePrefix = {arXiv},
       eprint = {2211.04964},
 primaryClass = {astro-ph.HE},
       adsurl = {https://ui.adsabs.harvard.edu/abs/2023ApJ...943L..12K}
}

@ARTICLE{Beniamini_Tsvi19,
       author = {{Beniamini}, Paz and {Piran}, Tsvi},
        title = "{The Gravitational waves merger time distribution of binary neutron star systems}",
      journal = {\mnras},
         year = 2019,
        month = aug,
       volume = {487},
       number = {4},
        pages = {4847-4854},
          doi = {10.1093/mnras/stz1589},
archivePrefix = {arXiv},
       eprint = {1903.11614},
 primaryClass = {astro-ph.HE},
       adsurl = {https://ui.adsabs.harvard.edu/abs/2019MNRAS.487.4847B}
}

@ARTICLE{CoteB+18,
       author = {{C{\^o}t{\'e}}, Benoit and {Fryer}, Chris L. and {Belczynski}, Krzysztof and {Korobkin}, Oleg and {Chru{\'s}li{\'n}ska}, Martyna and {Vassh}, Nicole and {Mumpower}, Matthew R. and {Lippuner}, Jonas and {Sprouse}, Trevor M. and {Surman}, Rebecca and {Wollaeger}, Ryan},
        title = "{The Origin of r-process Elements in the Milky Way}",
      journal = {\apj},
         year = 2018,
        month = mar,
       volume = {855},
       number = {2},
          eid = {99},
        pages = {99},
          doi = {10.3847/1538-4357/aaad67},
archivePrefix = {arXiv},
       eprint = {1710.05875},
 primaryClass = {astro-ph.GA},
       adsurl = {https://ui.adsabs.harvard.edu/abs/2018ApJ...855...99C}
}

@ARTICLE{Beniamini_Tsvi16,
       author = {{Beniamini}, Paz and {Piran}, Tsvi},
        title = "{Formation of double neutron star systems as implied by observations}",
      journal = {\mnras},
         year = 2016,
        month = mar,
       volume = {456},
       number = {4},
        pages = {4089-4099},
          doi = {10.1093/mnras/stv2903},
archivePrefix = {arXiv},
       eprint = {1510.03111},
 primaryClass = {astro-ph.HE},
       adsurl = {https://ui.adsabs.harvard.edu/abs/2016MNRAS.456.4089B}
}

@ARTICLE{TsujimotoT_BabaJ19,
       author = {{Tsujimoto}, Takuji and {Baba}, Junichi},
        title = "{Galactic r-process Abundance Feature Shaped by Radial Migration}",
      journal = {\apj},
         year = 2019,
        month = jun,
       volume = {878},
       number = {2},
          eid = {125},
        pages = {125},
          doi = {10.3847/1538-4357/ab22b3},
archivePrefix = {arXiv},
       eprint = {1905.08275},
 primaryClass = {astro-ph.GA},
       adsurl = {https://ui.adsabs.harvard.edu/abs/2019ApJ...878..125T}
}

@article{Harris_2020,
    title = {Array programming with {NumPy}},
    author = {Charles R. Harris and K. Jarrod Millman and St{\'{e}}fan J. van der Walt and Ralf Gommers and Pauli Virtanen and David Cournapeau and Eric Wieser and Julian Taylor and Sebastian Berg and Nathaniel J. Smith and Robert Kern and Matti Picus and Stephan Hoyer and Marten H. van Kerkwijk and Matthew Brett and Allan Haldane and Jaime Fern{\'{a}}ndez del R{\'{i}}o and Mark Wiebe and Pearu Peterson and Pierre G{\'{e}}rard-Marchant and Kevin Sheppard and Tyler Reddy and Warren Weckesser and Hameer Abbasi and Christoph Gohlke and Travis E. Oliphant},
    year = {2020},
    month = sep,
    journal = {Nature},
    volume = {585},
    number = {7825},
    pages = {357--362},
    doi = {10.1038/s41586-020-2649-2},
    publisher = {Springer Science and Business Media {LLC}},
    url = {https://doi.org/10.1038/s41586-020-2649-2}
}

@inproceedings{McKinney_2010,
    author = {Wes McKinney},
    title = {Data Structures for Statistical Computing in Python},
    booktitle = {Proceedings of the 9th Python in Science Conference},
    pages = {56 - 61},
    year = {2010},
    editor = {St\'efan van der Walt and Jarrod Millman},
    doi = {10.25080/Majora-92bf1922-00a}
}
\bibliographystyle{aasjournalv7}
\end{document}